\documentclass{aa}  

\usepackage[maxfloats=256]{morefloats}
\usepackage{graphicx}
\usepackage{mathrsfs}
\usepackage{CJKutf8}
\newcommand{\CNnames}[1]{{\begin{CJK}{UTF8}{gbsn}~(#1)~\end{CJK}}}
\usepackage{txfonts}
\usepackage{hyperref}
\usepackage{xcolor}
\newcommand{\starnumber}{13 }
\newcommand{\avalueexamplestar}{$a=0.74_{-0.18}^{+0.17}$}
\newcommand{\deltanuBexamplestar}{$\delta \nu_\mathrm{g}=0.104_{-0.019}^{+0.03}\,\mu\mathrm{Hz}$}

\newcommand{\seb}[1]{{#1}}
\newcommand{\fr}[1]{{#1}}
\newcommand{\comment}[1]{\textcolor{black}{#1}}
\newcommand{\kepler}{\textit{Kepler}}

\begin{document}

   \title{Internal magnetic fields in \starnumber red giants detected by asteroseismology}

   %\subtitle{implication of angular momentum transport mechanism}

   \author{Gang Li \CNnames{李刚}
          \inst{1,2}
          \and
          S\'{e}bastien Deheuvels\inst{1}%\fnmsep
          %\thanks{Just to show the usage of the elements in the author field}
          \and
          Tanda Li \CNnames{李坦达} \inst{3,4,5}
            \and
            J\'{e}r\^ome Ballot\inst{1}
            \and
            Fran\c{c}ois Ligni\`eres\inst{1}
          }

   \institute{IRAP, Université de Toulouse, CNRS, CNES, UPS, 31400 Toulouse, France\\
   \email{gang.li@kuleuven.be, sebastien.deheuvels@irap.omp.eu}
   \and
   Institute of Astronomy, KU Leuven, Celestijnenlaan 200D, 3001 Leuven, Belgium
   \and
   Institute for Frontiers in Astronomy and Astrophysics, Beijing Normal University, Beijing 102206, China
   \and
    Department of Astronomy, Beijing Normal University, Beijing 100875, China
    \and
    School of Physics and Astronomy, The University of Birmingham, UK, B15 2TT
         %\and
          %   , ...\\
         %    \email{c.ptolemy@hipparch.uheaven.space}
          %   \thanks{The university of heaven temporarily does not
          %           accept e-mails}
             }

   %\date{Received September 15, 1996; accepted March 16, 1997}

% \abstract{}{}{}{}{} 
% 5 {} token are mandatory
 
  \abstract
   {Magnetic fields affect stars at all evolutionary stages. \comment{While surface fields have been measured for stars across the HR diagram, internal magnetic fields remain largely unknown. The recent seismic detection of magnetic fields in the cores of several \kepler\ red giants has opened a new avenue to better understand the origin of magnetic fields and their impact on stellar structure and evolution.}}
   %While they are generated by the dynamo process in the convective zone, about 5\% to 10\% of stars with radiative envelopes still show surface magnetic fields observed by spectropolarimetry. The magnetic field is considered as a candidate for angular momentum transport to resolve the discrepancy between theoretical and observed internal rotation rates. However, due to the opacity of stellar material, observations of internal magnetic fields, especially in the radiative layer, are still lacking.}
   {The goal of our study is to use asteroseismology to systematically search for internal magnetic fields in red giant stars observed with the \kepler\ satellite, and to determine the strengths and geometries of these  fields.}
   {\comment{Magnetic fields are known to break the symmetry of rotational multiplets. In red giants, oscillation modes are mixed, behaving as pressure modes in the envelope and as gravity modes in the core. Magnetism-induced asymmetries are expected to be stronger for gravity-dominated modes than for pressure-dominated modes, and to decrease with frequency. Among \kepler\ red giants, we searched for stars that exhibit asymmetries satisfying these properties.}}  
   %To search for internal magnetic fields, we use recently developed mathematical frameworks which suggest that magnetic fields break the symmetry of rotational splittings, with the asymmetry being more significant for g-dominated modes than p-dominated modes, and decreasing with frequency. Using these theoretical predictions, we were able to search for targets that exhibit such asymmetries. }
   {After collecting a sample of $\sim$2500 Kepler red giant stars with clear mixed-mode patterns, we specifically searched for targets among $\sim$1200 stars with dipole triplets. We identified \starnumber stars exhibiting clear asymmetric multiplets and measured their parameters, especially the asymmetry parameter $a$ and the magnetic frequency shift $\delta\nu_\mathrm{g}$. By combining these estimates with best-fitting stellar models, we measured average core magnetic fields ranging from $\sim$20 to $\sim$150\,kG, corresponding to $\sim5\%$ to $\sim30\%$ of the critical field strengths. \comment{We showed that the detected core fields have various horizontal geometries, some of which significantly differ from a dipolar configuration.} We found that the field strengths decrease with stellar evolution, despite the fact that the cores of these stars are contracting. Additionally, even though these stars have strong internal magnetic fields, they display normal core rotation rates, suggesting no significantly different histories of angular momentum transport compared to other red giant stars. We also discuss the possible origin of the detected fields. }
   {}

   %Our observations thus suggest that these stars have not experienced significantly different histories of angular momentum transport compared to other red giant stars.


   \maketitle
%
%-------------------------------------------------------------------

\section{Introduction \label{sect_intro}}

Understanding the creation and evolution of magnetic fields is one of the main challenges faced by modern stellar physics. An important effect of magnetic fields on stellar evolution is that they are efficient at transporting angular momentum \citep{Cantiello2014, Rudiger2015,Fuller2019, Gouhier2022aa}. Thus, they influence the internal rotation of stars, and in turn the transport of chemical elements. Surface magnetic fields have been observed across the Hertzsprung-Russell diagram \citep{Landstreet1992, Donati2009}. These measurements, together with numerical simulations, establish the presence of dynamo-generated magnetic fields in convective regions \citep{Donati2009,auriere15}. A small fraction ($5-10 \%$) of intermediate- and high-mass stars with radiative envelopes harbours strong kilogauss surface fields that remain stable over decades \citep{Wade2012,Braithwaite2017}. These fields, which are thought to result from the star formation process, can subsist thanks to negligible ohmic diffusion and are potential progenitors for the magnetic white dwarfs and neutron stars \citep{Ferrario2020}. Another potentially widespread class of magnetic intermediate-mass stars has been identified with the detection of $\sim \!\!1$ Gauss fields in a few stars \citep{Lignieres09, Blazere2016}. Although it seems plausible that magnetic fields pervade much of stellar interior plasmas, the absence of direct measurements of internal fields has posed a significant obstacle to the study of their properties and their impact on stellar evolution. Fortunately, we have asteroseismology to help us explore the interior of stars \citep[e.g.][]{Aerts2010}.

Asteroseismology has yielded measurements of various physical processes within stars, one of which is stellar rotation at various stages of their evolution: on the main sequence (e.g., \citealt{Kurtz2014}, \citealt{Benomar2015}, \citealt{VanReeth2016}), in the subgiant and red giant branch (RGB) phases (\citealt{Beck2012Natur}, \citealt{Mosser2012}, \citealt{Deheuvels2014}, \citealt{Triana2017}, \citealt{Gehan2018}, \citealt{Deheuvels2020}, \citealt{kuszlewicz23}), in the core-He burning phase (\citealt{Mosser2012}, \citealt{Deheuvels2015}) and in white dwarfs (\citealt{hermes17}). In all of these phases, it was concluded that stars rotate more slowly than predicted by theoretical models \citep{Zahn1992, Eggenberger2012, Ceillier2013, Marques2013, Ouazzani2019, LiGang2020}. This shows that there must be additional not-yet-identified processes that efficiently carry angular momentum inside stars, beyond the classical hydrodynamic processes \citep{Cantiello2014, Fuller2014, Belkacem2015, Spada2016, Pincon2016, Eggenberger2017, Eggenberger2019I}. One of the main solutions proposed is the presence of magnetic fields in radiative cores.

%However, given the need for extremely high frequency resolution and photometric accuracy to discern modes of stellar oscillations, asteroseismologic observations were not well-developed before the era of space missions. 
The advent of space missions partly dedicated to asteroseismology has yielded the high frequency resolution and photometric accuracy required to characterise stellar oscillation modes.
Thanks to the \kepler\ mission \citep{Borucki2010Sci}, solar-like oscillators have been discovered in tens of thousands of stars \citep{Bedding2010ApJ, Yu2018}. Their oscillations are excited stochastically by the outer envelope convection similar to the Sun. Most of these solar-like oscillators are post-main-sequence stars that exhibit mixed dipole ($l=1$) modes, which arise from the coupling between the outer pressure modes and the interior gravity modes \citep{Bedding2014}. Mixed modes enable us to probe the physics from the stellar core to surface, for example, to distinguish evolutionary stages \citep{Bedding2011,mosser11b}, infer previous processes such as mergers or mass transfers  \citep{Deheuvels2022A&A, Rui2021MNRAS,li22_mass_transfer}, or measure the internal rotation, as mentioned above. Around 20\% of red giants exhibit suppressed \seb{dipole} mixed modes (\citealt{mosser12c}, \citealt{garcia14}, \citealt{Stello2016}). \seb{It was suggested that this phenomenon could be caused by} central magnetic fields exceeding the critical field intensity $B_{\rm c}$ above which magneto-gravity waves no longer propagate in the core \citep{Fuller2015,Stello2016, Rui2023MNRAS}. However, this \seb{interpretation} is still a topic of debate \citep{Mosser2017_dipole_modes,loi20}. 

From a theoretical perspective, it has been known for decades that magnetic fields impact stellar oscillations \citep{Gough1990, Hasan2005}. Similarly to rotation, they break the degeneracy of oscillation modes with \seb{same degree $l$ but} different azimuthal order $m$. Rotational effects alone produce multiplets (triplets for $l=1$ modes) that are \seb{generally} symmetric \seb{with respect to the central $m=0$ component} when the rotation rate is not too fast (\seb{that is, when} second-order effects are negligible). The effects of magnetic fields on mixed modes in red giants has been addressed, considering the simple case of dipolar fields with different radial profiles, either aligned with the rotation axis (\citealt{Gomes2020}, \citealt{Mathis2021}, \citealt{Bugnet2021}) or inclined \citep{Loi2021, Mathis2023A&A}. These studies showed that magnetic fields are expected to break the symmetry of rotational multiplets, and they produced estimates of the minimal field intensities required to detect magnetic asymmetries, defined as $\delta_\mathrm{asym}=\nu_{m=-1}+\nu_{m=1}-2\nu_{m=0}$ (\citealt{Deheuvels2017}).

Observation breakthrough has only been recently achieved. 
%compared to theoretical development. 
\cite{LiGang_2022_nature} reported the detection of clear asymmetries in the $l=1$ multiplets of three \kepler\ red giants, which they found could only be accounted for by the presence of strong magnetic fields in their cores. For this purpose, they extended previous theoretical works to \fr{magnetic fields} with arbitrary configurations and found that the magnetic asymmetries of multiplets can be either positive or negative depending on the field topology (all the configurations studied before them yielded positive asymmetries). They measured radial field intensities ranging from 30 to 130~kG in these stars, and placed constraints on their topology. \seb{If the magnetic field is strong enough, it can also significantly modify the regular spacing of g-mode period, which can be used to detect them (\citealt{LiGang_2022_nature, Bugnet2022}). \cite{Deheuvels2023} thus detected even stronger core fields in 11 \kepler\ red giants, with intensities that are comparable to the critical field strength $B_{\rm c}$. } 

These studies naturally raise the question of the prevalence of magnetic red giants, which can shed light on the origin of these fields. In this study, we systematically searched for asymmetries  in the rotational multiplets of \kepler\ red giants with detected oscillations.
The paper is organised as follows. In Sect.~\ref{sec:method}, we present the method we used to search for multiplet asymmetries among \kepler\ red giants and we list the underlying assumptions. This leads us to identify \starnumber targets with multiplet asymmetries that exhibit all the features expected in the presence of an internal magnetic field. In Sect.~\ref{sec:measure_field_strength}, we fit asymptotic expressions of mixed modes including rotational and magnetic perturbations to the observations for these stars. We thus obtain estimates of the average field strength in the core, as well as constraints on its horizontal topology.
In Sect.~\ref{sec:discussion}, we discuss the implications of these results for the origin and evolution of internal magnetic fields, and their impact on angular momentum transport.
Sect.~\ref{sec:conclusion} is dedicated to conclusions.

%--------------------------------------------------------------------
\section{Systematic search for multiplet asymmetries in Kepler data} \label{sec:method}

\subsection{Assumptions \label{sec:assumptions}}

\seb{The power spectra of mixed modes in red giants are complex. Specific methods were derived based on asymptotic expressions of mixed modes (\citealt{shibahashi79}, \citealt{Unno1989}) to identify the modes and recover general properties of p and g modes. In this study, we used methods that are derived from those prescribed by \cite{Mosser2015}, \cite{Vrard2016} and \cite{Gehan2018} in order to search for asymmetric multiplets within \kepler\ data. These methods assume that rotational multiplets are symmetric, so they needed to be adapted. They also make use of the regularity in the period spacings of asymptotic g modes. For field intensities comparable to those found by \cite{LiGang_2022_nature}, this assumption remains approximately correct. However, stronger fields can significantly modify this regularity, as already mentioned in Section~\ref{sect_intro}, and it is likely that the methods that we use in this study are ill-suited to detect such strong fields. This introduces an observational bias, which is further discussed in Section~\ref{subsec:biases}.}

We also assume that the effects of non-axisymmetry of the magnetic field on oscillations are small. If it is not the case, multiplets can be split into $(2l+1)^2$ components, instead of $(2l+1)$ in the axysymmetric case (\citealt{Gough1990},\citealt{Loi2021},\citealt{LiGang_2022_nature}). Dipole multiplets can thus have up to nine components instead of three. \cite{LiGang_2022_nature} have shown that this effect arises only if the ratio between the magnetic frequency shift and the rotational frequency shift exceeds unity. We note that this was not the case for the three red giants studied in \cite{LiGang_2022_nature} (we found ratios that do not exceed $\sim0.6$). For these stars, the magnetic perturbations of the oscillation frequencies are expected to be indistinguishable, whether the field is axisymmetric or not. In this study, we restrict our search to stars in the same regime, postponing the search for stars showing the seismic signature of non-axisymmetric magnetic fields to a future work. This also introduces an observational bias (see Section~\ref{subsec:biases}).

\seb{The following subsections describe the different steps of the method that we applied to search for multiplet asymmetries, in the framework exposed above.} The results obtained in this section serve as the basis for the measurements of magnetic field strengths and topology reported in Section~\ref{sec:measure_field_strength}. 

\subsection{Data reduction and sample selection} \label{subsubsec:data_reduction}

We used the \textit{Kepler} 4-yr long-cadence data to calculate the power spectra, which were downloaded from the Mikulski Archive for Space Telescopes (MAST, \url{https://archive.stsci.edu}). We calculated the power spectra density (PSD)  \citep{Lomb1976,Scargle1982,Kjeldsen1995}, the global asteroseismic parameters ($\nu_\mathrm{max}$, $\Delta\nu$), and the background properties following the processes used in the SYD pipeline \citep{Huber2009,Chontos2021}.

We visually inspected about 8000 red giant branch (RGB) stars reported by \cite{Yu2018} and \cite{Gehan2018} to select the stars with good patterns of $l=1$ mixed modes. 
During the visual inspection, our primary focus is on identifying stars that exhibit clear and distinct peaks between each $l=0$ and $l=2$ p mode. 
Some stars
%, however, 
do not display any discernible peaks in their mixed-mode regions, a phenomenon referred to as `suppressed $l=1$ mixed modes' as mentioned in the Introduction. 
%This suppression may be attributed to the presence of strong central magnetic fields \citep{Fuller2015, Stello2016, Rui2023MNRAS}. 
Additionally, 
%some stars only show a hump in their mixed-mode area, possibly due to either the relatively short lifetimes of mixed modes or the presence of a very small period spacing. 
the pattern of mixed modes becomes unclear for more evolved red giants because of their smaller asymptotic period spacing and the effects of radiative damping, which become large for g-dominated modes \citep[e.g.][]{Grosjean14}.
To ensure the selection of stars with well-defined mixed-mode features, 
we excluded red giant stars 
%that do not meet these criteria, 
in such unfavourable cases,
resulting in a cutoff for $\Delta \nu$ at approximately $7\,\mathrm{\mu Hz}$. Consequently, we have identified and retained around 2500 RGB stars for further investigation.

\subsection{Identification of azimuthal order $m$}\label{subsubsec:m_identification}

\subsubsection{Streched periods}

\seb{The first step of the method consists of identifying the azimuthal order $m$ of the detected modes. For this purpose, it is convenient to use the so-called ``stretched'' periods introduced by \cite{Mosser2015}. These authors have shown that the period spacing between consecutive dipole mixed modes can be expressed as $\Delta P = \zeta\Delta\Pi_1$, where $\zeta$ represents the fraction of the g-mode inertia over the whole inertia \citep{Goupil13} ($\zeta$ tends to unity for pure g modes and to zero for pure p modes), and $\Delta\Pi_1$ is the asymptotic period spacing of pure g modes. The frequencies $\nu$ are transformed into the so-called ``stretched periods'' $\tau$ using the differential equation
\begin{equation}
    \mathrm{d}\tau = \frac{1}{\zeta}\frac{\mathrm{d}\nu}{\nu^2}, \label{eq:stretched_differential_equation}
\end{equation}
so that the mixed modes are equally spaced by $\Delta\Pi_1$. When representing the stretched periods in an \'echelle diagram folded with $\Delta\Pi_1$, the modes of the same azimuthal order $m$ are expected to align nearly vertically in this diagram. }

\subsubsection{Asymptotic expression of $\zeta$}

To estimate the value of $\zeta$ for the detected modes, we used an asymptotic expression of this quantity, as is now commonly done (e.g., \citealt{Mosser2015}, \citealt{LiGang_2022_nature}). We briefly recall the steps of the procedure.
Following the work of \cite{shibahashi79} and \cite{Unno1989}, the implicit asymptotic relation of mixed modes is expressed as
\begin{equation}
    \tan \theta_\mathrm{p} = q \tan \theta_\mathrm{g}, \label{eq:tan_theta_p_q_tan_theta_g}
\end{equation}
where $q$ is the coupling factor between p and g components of the modes \citep[e.g.][]{Mosser2017_q}, and $\theta_\mathrm{p}$ and $\theta_\mathrm{g}$ are the phases for the pure p and g modes. 

The phase of the pure p modes is written as
\begin{equation}
\theta_\mathrm{p} = \pi \frac{\nu-\nu_\mathrm{p}}{\Delta\nu(n_\mathrm{p})},\label{eq:theta_p_unperturbed}
\end{equation}
where $\nu_\mathrm{p}$ is the pure p-mode frequency 
%which can be derived by $l=0$ and $l=2$ modes, and 
and $\Delta\nu(n_\mathrm{p})$ is the local frequency separation at the radial order $n_\mathrm{p}$ \citep{mosser15}.
\seb{We used the detected $l=0$ and $l=2$ mode frequencies \footnote{We fit a Lorentzian profile to each detected $l=0$ and $l=2$ modes to obtain their frequencies \citep{Anderson1990ApJ}.} to derive an expression of $\nu_\mathrm{p}$ for dipole mode frequencies. From asymptotic expressions, 
\begin{equation}
    \nu_\mathrm{p} = \left[n_\mathrm{p}+\frac{l}{2}+\varepsilon+\frac{\alpha}{2}(n_\mathrm{p}-n_\mathrm{max})^2\right]\Delta\nu - l(l+1)D, \label{eq:p_mode_asymptotic_relation}
\end{equation}
where $\varepsilon$ is the phase term and $n_\mathrm{max}=\nu_\mathrm{max}/\Delta\nu$ is the radial order at the max power frequency. The term $D$ describes the small separation $\delta\nu_{02}$ between $l=0$ and $2$ modes. We fit the expression given by Eq. \ref{eq:p_mode_asymptotic_relation} to the frequencies of the detected $l=0$ and $l=2$ modes.} The fit results are listed in Table~\ref{tab:p_param} for \seb{the stars that are discussed in the following sections of the paper}. 
Since the pure $l=1$ p modes are \seb{not observable}, and the small separation ratio $\delta\nu_{02}/\delta\nu_{01}$ deviate from the solar value (three) with stellar evolution \citep{Lund2017}, we add an extra free parameter $f_\mathrm{shift}$ to the expression of $\nu_\mathrm{p}$ given in Eq. \ref{eq:p_mode_asymptotic_relation} only for dipole mode frequencies, \textcolor{black}{that is:
\begin{equation}
    \nu_{\mathrm{p}, l=1} = \nu_\mathrm{p} + f_\mathrm{shift}.
\end{equation}
We use $\nu_{\mathrm{p}, l=1}$ to compute the dipole p-mode phase $\theta_\mathrm{p}$ in eq.~\ref{eq:theta_p_unperturbed}. The initial guess of $f_\mathrm{shift}$ is 0.6\,$\mu\mathrm{Hz}$ and it will be set to be a free parameter in further fitting steps. }

The phase of the pure g modes is
\begin{equation}
    \theta_\mathrm{g} = \frac{\pi}{\Delta P_\mathrm{g}}\left(\frac{1}{\nu}-P_\mathrm{g}\right),\label{eq:theta_g_unperturbed}
\end{equation}
in which $\Delta P_\mathrm{g}$ is the local period spacing and $P_\mathrm{g}$ is the pure g-mode period \citep{mosser15}. Without considering any perturbations by rotation and magnetism, the pure g-mode period is equally spaced in period, shown as
\begin{equation}
    P_\mathrm{g} = \Delta\Pi_1\left(n_\mathrm{g}+\varepsilon_\mathrm{g}\right) \label{eq:pure_g_mode}
\end{equation}
for $l=1$ g modes, where $n_\mathrm{g}$ is the g-mode radial order and $\varepsilon_\mathrm{g}$ is the g-mode phase.

%We plot the stretched \'{e}chelle diagram for red giants to identify the azimuthal order $m$. The frequency $\nu$ is transformed to be the so-called `stretched period' $\tau$ by the differential equation
%\begin{equation}
 %   \mathrm{d}\tau = \frac{1}{\zeta}\frac{\mathrm{d}\nu}{\nu^2}, \label{eq:stretched_differential_equation}
%\end{equation}
%so that the mixed modes are equally spaced in unit of the stretched period. The $\zeta$ function represents the fraction of the g-mode inertia over the whole inertia \citep{Goupil13} and its asymptotic expression derived by \cite{mosser15} and \cite{hekker17} is
\seb{As shown by \cite{mosser15} and \cite{hekker17}, the $\zeta$ function can then be expressed as}
\begin{equation}
    \zeta = \left[1+\frac{\nu^2}{q}\frac{\Delta \Pi_1}{\Delta\nu}\frac{1}{\frac{1}{q}\sin^2\theta_\mathrm{p}+\cos^2\theta_\mathrm{p}}\right]^{-1}. \label{eq:zeta_function}
\end{equation}
%The modes with different $m$ are not exactly equally spaced at the period spacing $\Delta \Pi_1$. They show slightly different spacings in the stretched period diagram \citep[see eq.21 in][]{Mosser2015}, 
%\begin{equation}
%\Delta \tau_m=\Delta \Pi_1\left(1+2m\frac{\mathscr{N}\left(\nu_\mathrm{max}\right)}{\mathscr{N}\left(\nu_\mathrm{max}\right)+1}\frac{\delta\nu_\mathrm{rot, core}}{\nu_\mathrm{max}}\right), \label{eq:delta_tau_m} 
%\end{equation}
%with $\mathscr{N}=\Delta\nu/(\Delta \Pi_1\nu^2)$, where $\delta\nu_\mathrm{rot, core}$ is the half of the core rotation frequency. 
%hence we can use this nature to identify $m$ values for the peaks.

\subsubsection{Identification of $m$ with stretched \'echelle diagrams}

\seb{For all the stars of our sample, we selected }the peaks with signal-to-noise ratio larger than ten to plot the initial stretched \'{e}chelle diagram (eqs.~\ref{eq:stretched_differential_equation} and \ref{eq:zeta_function}), which is used to measure the asymptotic period spacing $\Delta\Pi_1$ and identify azimuthal orders. %The y-axis of the stretched \'{e}chelle diagram is frequency while the x-axis is the stretched period modulo $\Delta \Pi_1$. 
In this step, we set $q=0.15$, as a typical value for hydrogen-shell-burning (HSB) stars \citep{Mosser2017_q} and varied $\Delta \Pi_1$ to produce vertical ridges in the stretched \'{e}chelle diagram. The modes with different $m$ are not exactly equally spaced by the period spacing $\Delta \Pi_1$. As shown by \cite{Mosser2015}, they have slightly different spacings $\Delta\tau_m$ in the stretched period diagram
\begin{equation}
\Delta \tau_m=\Delta \Pi_1\left(1+2m\frac{\mathcal{N}}{\mathcal{N}+1}\frac{\delta\nu_\mathrm{rot, core}}{\nu_\mathrm{max}}\right), \label{eq:delta_tau_m} 
\end{equation}
with $\mathcal{N}=\Delta\nu/(\Delta \Pi_1\nu_{\rm max}^2)$, where $\delta\nu_\mathrm{rot, core}$ is the rotational splitting of pure g modes, that is, half the core rotation frequency. 
Multiple ridges may thus appear.
%because they have different azimuthal orders. 
We measured their period spacing $\Delta\tau_m$ by slightly changing $\Delta \Pi_1$ and identified the $m$ value as described by Eq. 21 in \cite{Mosser2015} ($m=1$ or $-1$ for doublets, and $m=1$, $0$, $-1$ for triplets). About 1200 stars with clear triplets were used for the subsequent analysis, and we also obtained about 800 stars with doublets (their asymmetries cannot be measured due to the lack of $m=0$ modes). The rest $\sim$500 stars do not show splittings. 

\begin{figure}
    \centering
    \includegraphics[width=\linewidth]{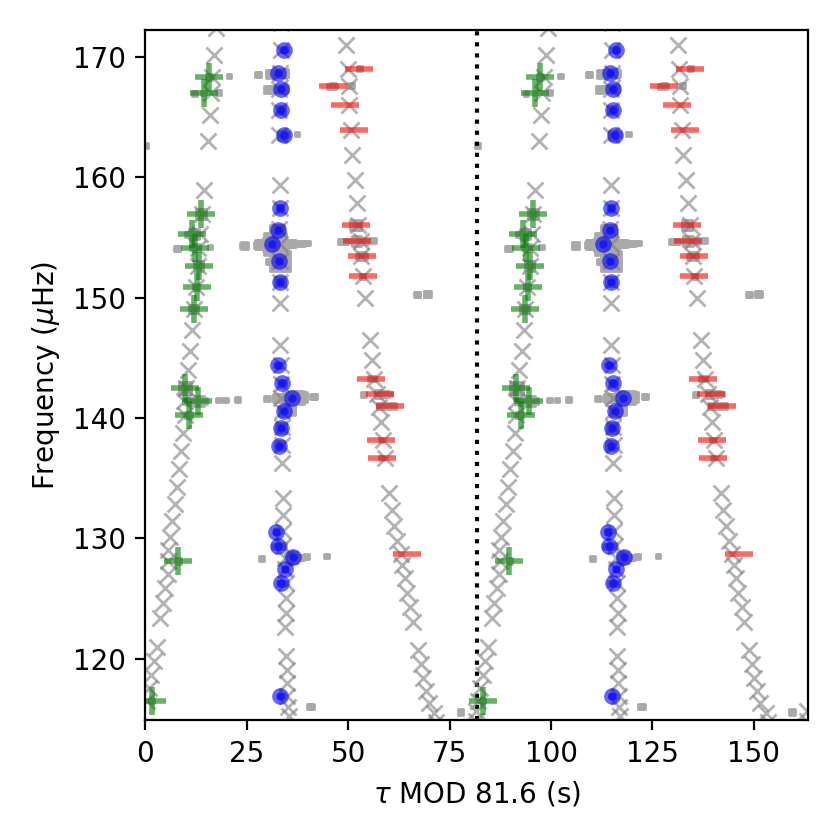}
    \caption{Stretched \'{e}chelle diagram of KIC\,5792889 that does not show any magnetism-induced perturbation. The x-axis is the stretched periods $\tau$ modulo $\Delta \Pi_1\approx 81.6\,\mathrm{s}$. The peaks with S/N>10 are shown by the grey points ($l=0$ and $l=2$ modes have been removed). The green `+' stands for the $m=1$ modes. The red `-' stands for the $m=-1$ modes. The blue `$\bullet$' shows the $m=0$ modes. The best-fitting results are plotted by the cross.}
    \label{fig:stretched_KIC5792889}
\end{figure}

Figure~\ref{fig:stretched_KIC5792889} shows the stretched \'{e}chelle diagram of KIC\,5792889, which is selected randomly from our sample. Three distinct ridges are visible, each marked by a different symbol denoting their azimuthal order (refer to the caption of Fig.~\ref{fig:stretched_KIC5792889}). However, this star does not show magnetism-induced asymmetries, as a result, the ridge corresponding to $m=0$ appears halfway between the $m=1$ and $m=-1$ ridges. 

In the presence of a core magnetic field, the oscillation modes undergo a frequency shift that depends on $|m|$ (see Sect. \ref{sec:measure_field_strength}). If the core field is strong enough, we anticipate that it can modify the ordering of the $m$-components in a multiplet. This would invalidate our identification of $m$ base on Eq. \ref{eq:delta_tau_m}. For all the stars where multiplet asymmetries were detected, we thus investigated alternate identifications of $m$, assigning $m=0$ to the external components of the multiplet. In all these cases, these alternate identifications led to poor fits to the observations, so we ruled out the possibility of having a different ordering of the components in the multipltets.

%\textbf{In some cases it is difficult to identify the azimuthal orders for three ridges in the stretched \'{e}chelle diagram due to their curvature caused by the magnetic effect. As a result, their $\Delta \tau_m$ values may not be clearly defined. In this case, we assume that the magnetic shift is not significant enough for the $m=0$ components to be located outside the $m=-1$ and $m=1$ modes. To check this assumption, we have attempted different mode identifications for $m$ (such as assigning the rightmost peaks in triplets to $m=0$), but we have not found a satisfactory fit. These findings confirm the validity of our assumption.  }
%\comment{(I modified a bit the organisation of Section \ref{subsubsec:m_identification}, trying to make it more reader-friendly (giving the reader the keys to understand why you are doing things you describe). I still have a question. You mentioned that you identify $m$ for all three ridges based on the ordering of the $\Delta\tau_m$, but if the magnetic shift is strong enough that the $m=0$ component is no longer between the $m=\pm1$ components, you will get the wrong answer with this procedure, no?) Gang's reply: see the new paragraph in front}

\subsection{Identification of the rotational multiplets}\label{subsubsec:splitting_identification}

\begin{figure}
    \centering
    \includegraphics[width=\linewidth]{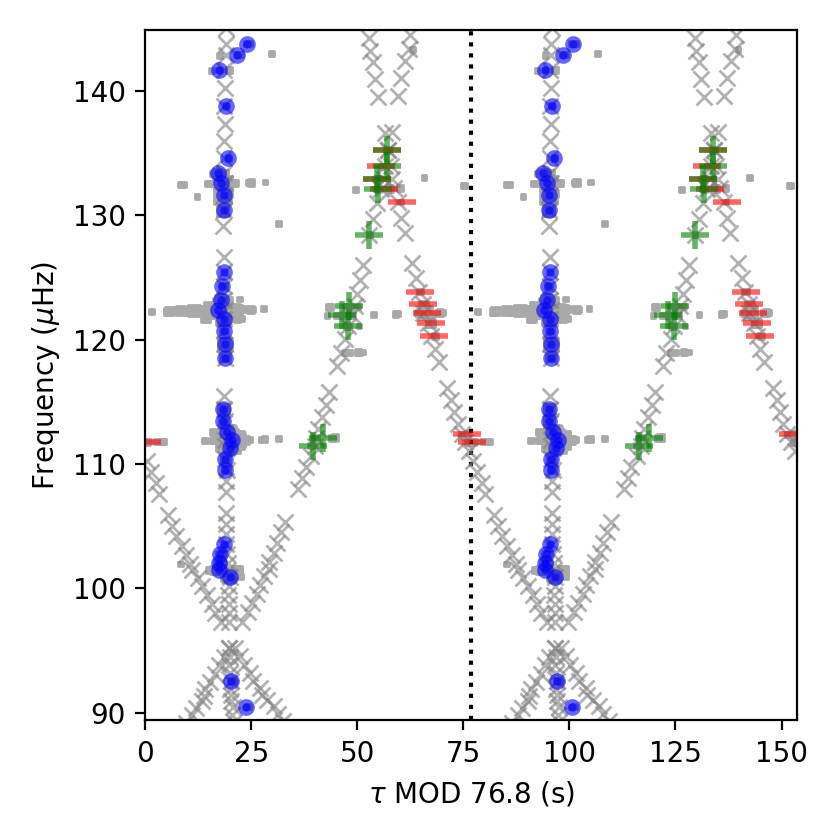}
    \caption{Stretched \'{e}chelle diagram of KIC\,9467102, which does not show any magnetism-induced perturbation. The symbols are the same as fig.~\ref{fig:stretched_KIC5792889}. We show this star as an example because its splittings overlap seriously. }
    \label{fig:stretched_echelle_diagram_7749842}
\end{figure}

\seb{In order to measure multiplet asymmetry, we then needed to identify modes that belong to a same multiplet, that is, modes that share common values of $l$ and $n$ but have different values of $m$. This step can be complicated when the rotational splitting is comparable to or larger than the frequency spacing between modes of consecutive radial order $n$.}
%The goal of this section is to identify rotational splittings, which are modes with the same $l$ and $n$ but different $m$. These splittings will be used in the next step to automatically measure splitting asymmetry. Identifying the splittings can be challenging because they may overlap and exceed the local frequency spacing, as illustrated in Figure~\ref{fig:stretched_echelle_diagram_7749842} for KIC\,7749842. 
In such cases, the nearest three modes no longer form a single multiplet, making the identification more difficult, as illustrated in Figure~\ref{fig:stretched_echelle_diagram_7749842} for KIC\,7749842. 

\seb{To identify rotational multiplets, we fit an asymptotic expression of mixed modes including rotational effects to the detected modes. For given values of $\Delta\Pi_1$, $\varepsilon_{\rm g}$, $q$, and $f_{\rm shift}$, the frequencies of the unperturbed modes are given by solving Eq. \ref{eq:tan_theta_p_q_tan_theta_g}. We then add rotational splittings to these modes.}
%To calculate rotational splittings properly, we calculate the frequencies of $m=0$ mixed modes first (using eq.\ref{eq:tan_theta_p_q_tan_theta_g}), then we add rotational splittings on $m=0$ modes. 
The rotational splittings $\delta \nu_\mathrm{R}$ \seb{can be expressed as}
\begin{equation}
    2\pi \delta \nu_\mathrm{R} = 0.5\Omega_\mathrm{core}\zeta + \Omega_\mathrm{env}\left(1-\zeta \right), \label{eq:rotational_splittings}
\end{equation}
where $\Omega_\mathrm{core}$ and $\Omega_\mathrm{env}$ are the mean rotation rates (in unit of angular frequency) in the core and the outer envelope \citep{Goupil13, Deheuvels2014}. 
%Therefore, the splittings of mixed modes show a linear relation with $\zeta$, whose slope is $0.5\Omega_\mathrm{core}-\Omega_\mathrm{env}$ and intercept is $\Omega_\mathrm{env}$. 
\seb{This expression leads to symmetric rotational multiplets, contrary to what we generally expect for red giants harbouring core magnetic fields. However, if the effects of non-axisymmetry of the magnetic field on the frequency shifts are negligible (as we have assumed in Sect. \ref{sec:assumptions}), the mode frequencies of azimuthal order $m=\pm 1$ are affected in the same way. In this case, the frequency spacing between these two components remains equal to $2\delta \nu_\mathrm{R}$, as in the non-magnetic case. We thus used only the $m=\pm 1$ modes to perform the fit.} 
%\comment{(I think it makes more sense to only use $m=\pm1$ modes at this stage, knowing that if the multiplets are asymmetric, you expect Eq. \ref{eq:rotational_splittings} to be incorrect. Tell me if this formulation correctly represents what you actually did.) Gang's reply: no problem.}

We ran a Markov chain Monte Carlo (MCMC) method to optimise the parameters using the python package \textsc{emcee} \citep{Foreman-Mackey2013PASP}. %The frequencies of $m=0$ $l=1$ mixed modes were calculated by eq.~\ref{eq:tan_theta_p_q_tan_theta_g}. The splittings were calculated by eq.~\ref{eq:rotational_splittings} with eq.~\ref{eq:zeta_function}. 
There are in total six parameters that will be optimised. We applied uniform priors to the parameters and defined their ranges for optimisation in the MCMC as follows: 
%\comment{(you should add a justification for the choice of boundaries of the priors for each parameter) Gang's reply: see below}

\begin{enumerate}
    \item Asymptotic period spacing $\Delta\Pi_1$. $[\Delta \Pi_\mathrm{1,init}-0.5\,\mathrm{s}, \Delta \Pi_\mathrm{1,init}+0.5\,\mathrm{s}]$ where $\Delta \Pi_\mathrm{1,init}$ is the initial guess of the period spacing \textcolor{black}{from the stretched \'{e}chelle diagram}. 
    %\comment{(explain how these initial guesses are obtained)}
    \item Coupling factor $q$. [0.08, 0.25]. 
    \item Phase of g mode $\varepsilon_g$. [-0.1, 1.1]. 
    \item Dipole ($l=1$) frequency shift $f_\mathrm{shift}$. [0.3$\,\mathrm{\mu Hz}$, 1.0$\,\mathrm{\mu Hz}$].
    \item Core angular frequency $\Omega_\mathrm{core}$. [0$\,\mathrm{\mu Hz}$, 20$\,\mathrm{\mu Hz}$].
    \item Envelope angular frequency $\Omega_\mathrm{env}$. [-1$\,\mathrm{\mu Hz}$, 1$\,\mathrm{\mu Hz}$].
\end{enumerate}
%the asymptotic period spacing $\Delta\Pi_1$, the coupling factor $q$, the g-mode phase $\varepsilon_g$, the $l=1$ frequency shift $f_\mathrm{shift}$, the core angular frequency $\Omega_\mathrm{core}$, and the envelope angular frequency $\Omega_\mathrm{env}$. %In the MCMC optimisation, the best-fitting result also provides the identification of the rotational splittings, which might be difficult if the splittings overlap each other, e.g. the splittings are larger than the local frequency separations. 
\textcolor{black}{The prior ranges of these parameters were determined based on several previous analyses of large samples of red giant stars \citep{Mosser2017_q,mosser18, Gehan2018, Triana2017}.} The MCMC algorithm maximises the likelihood function defined as follows:
\begin{equation}
        \ln L = -\frac{1}{2}\sum_{m=1, 0, -1} \sum_i\left[\frac{(\nu_{m, i}^\mathrm{obs}-\nu_{m, i}^\mathrm{cal})^2}{\sigma_{m, i}^2} +\ln \left(2\pi\sigma_{m, i}^2\right)\right],\label{eq:likelihood}
\end{equation}
where $\nu_{m, i}^\mathrm{obs}$ is the $i^\mathrm{th}$ observed frequency with azimuthal order of $m$ and $\nu_{m, i}^\mathrm{cal}$ is the calculated frequency. 
At \seb{this stage}, the observed frequencies are estimated as the mean of the nearby points in the PSD whose signal-to-noise ratio (S/N) is greater than 10, which is more conservative than the criterion applied in previous studies, such as \cite{mosser15}. 
%\textbf{In some previous studies, a S/N threshold greater than eight has been used \citep[e.g.][]{mosser15}. Nevertheless, it is worth noting that our approach considers the overall shape in the stretched \'echelle diagram, rather than focusing on individual peaks. Consequently, different choices of the S/N criteria do not substantially impact the mode identification and fitting process. }

Currently, we assume that the uncertainty in the observed frequency $\sigma_{m, i}$ is 0.02\,$\mu$Hz. More proper estimates of the mode frequencies and their uncertainties are obtained in section~\ref{subsubsec:asymmetry_measurement} by fitting Lorentzian profiles to the PSD.

We ran 14 parallel chains with length of 5000 steps. The first 50\% samples are discarded. Finally, we obtained the best-fitting results that give the identification \seb{of the multiplets} and allow us to run an automated algorithm (in section~\ref{subsubsec:asymmetry_measurement}) to measure the asymmetric splittings 
%\comment{(For non-specialist, the results of the procedure described in this subsection are unclear, I think. Perhaps adding a plot with the PSD of KIC9467102 showing the identified multiplets would make this clearer?)}. 
\textcolor{black}{We show the best-fitting results of KIC\,9467102 in Fig.~\ref{apdxfig:echelle_diagram_9467102}. The identified triplets are marked by the horizontal red lines in each panel. Even though there is significant overlap between multiplets, we still can distinguish them.}
We also find that the splitting identification works well even for the stars with asymmetric splittings.

\subsection{Asymmetry measurement} \label{subsubsec:asymmetry_measurement}

We measured the asymmetries of the identified \seb{multiplets} by fitting three Lorentzian profiles, whose initial locations are given by the MCMC algorithm in section~\ref{subsubsec:splitting_identification}. \textcolor{black}{In this fitting for the Lorentzian profile, we can determine the following parameters: frequencies of components, linewidths assuming identical for three components, amplitudes, and inclinations.} The amplitudes of the three components were determined by a relation that was modified by inclination \citep{Gizon2003},
%\comment{ (You mean that the inclination is a free parameter of the fit? This would deserve to be clarified.)}
and the best-fitting results were obtained by maximising the likelihood function defined by \cite{Anderson1990ApJ}. We allowed the $m=0$ component to shift freely to reproduce the asymmetry.

\begin{figure}
    \centering
    \includegraphics[width = 0.9\linewidth]{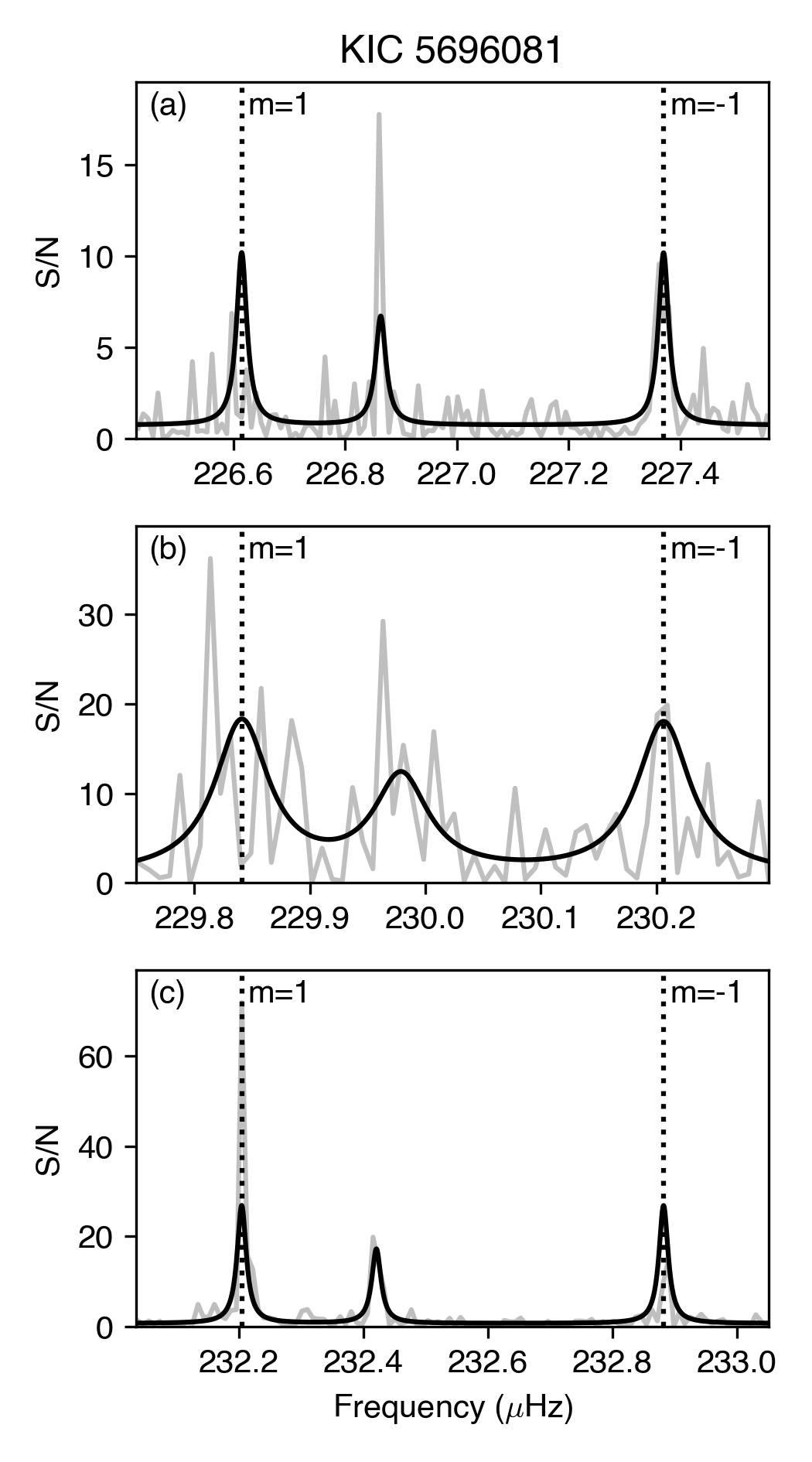}
    \caption{Three continuous asymmetric splittings in KIC\,5696081. The top and bottom panels show g-dominated modes while the middle panel displays a p-dominated one. Note that the x-axes of three panels are aligned by the $m=1$ and $-1$ mode frequencies.}
    \label{fig:KIC5696081_asymmtry}
\end{figure}

\begin{figure}
    \centering
    \includegraphics[width=1\linewidth]{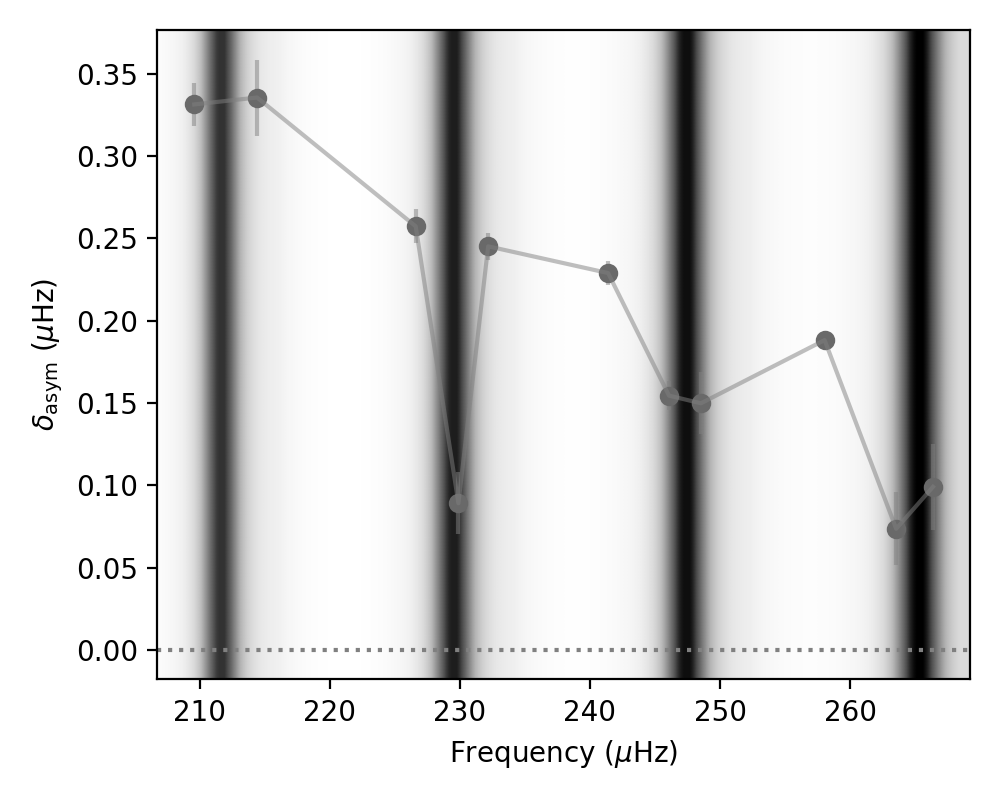}
    \caption{The asymmetries as a function of frequency in KIC\,5696081. The dark fringes show where the p-dominated modes are, while the white background shows the locations of g-dominated modes.}
    \label{fig:KIC5696081_asymmtry_vs_freq}
\end{figure}

Among the stars showing significant multiplet asymmetries, we selected those that share the features expected for magnetic asymmetries, namely:
\begin{itemize}
    \item \seb{They should have the same sign (either positive or negative) for all multiplets in a given star.}
    \item \seb{They} should decrease with frequency (in absolute value).
    \item \seb{They should be larger for g-dominated modes than for p-dominated modes.}
\end{itemize}
Finally, \starnumber stars \seb{were} found to show magnetism-induced asymmetries, \textcolor{black}{including the three stars reported by \cite{LiGang_2022_nature}.}
%\comment{(Perhaps mention that three of them are from the Nature paper)} 
Here we show KIC\,5696081 as an example. The three panels in Fig.~\ref{fig:KIC5696081_asymmtry} show three continuous asymmetric splittings, whose x-axes are aligned by the $m=1$ and $m=-1$ mode frequencies. The modes in the top and the bottom panels are g-dominated, so they show narrower linewidths and larger asymmetries. While the mode in the middle panel is p-dominated, hence it has wider linewidth and smaller asymmetry. Although the asymmetries vary with different modes, all the asymmetries keep positive. 
Figure~\ref{fig:KIC5696081_asymmtry_vs_freq} shows the \seb{variations in the} asymmetry as a function of \seb{the} frequency in KIC\,5696081. We find that they follow all the %predictions in eq.~\ref{eq:asymmetry_vs_freq}: 
characteristics expected from a magnetic perturbation (as listed in Sect. \ref{sect_intro}): the asymmetries are all positive in this star, decrease with frequency, and are smaller for p-dominated modes. We show all the asymmetry measurements in Appendix~\ref{appendixsection:all_figures}. 

\section{Magnetic perturbation and field strength measurement}\label{sec:measure_field_strength}

\subsection{Magnetic perturbation}

\seb{For the stars identified as showing multiplet asymmetries of magnetic origin in Sect. \ref{subsubsec:asymmetry_measurement}, we estimated the properties of the field that could reproduce the observations. For this purpose, we followed the approach that we proposed in \cite{LiGang_2022_nature}. We again solved Eq. \ref{eq:tan_theta_p_q_tan_theta_g} to obtain the asymptotic expression of mixed mode frequencies, but here the frequencies of pure p and g modes include perturbations arising from rotation and magnetic field.}

%In this section, we introduce the method to measure the internal magnetic fields. 
In the case of axisymmetric fields, or the non-axisymmetric effects are negligible, the frequency perturbation of pure g modes caused by both magnetism and rotation is given by
\begin{equation}
    \delta \nu_{\mathrm{g~mode}, m=0} = \left(1-a\right)\delta\nu_\mathrm{g}\left(\frac{\nu_\mathrm{max}}{\nu}\right)^3 \label{eq:g_perturb_m_0}
\end{equation}
for $m=0$ modes, and 
\begin{equation}
    \delta \nu_{\mathrm{g~mode}, m=\pm1} = \left(1+\frac{a}{2}\right)\delta\nu_\mathrm{g}\left(\frac{\nu_\mathrm{max}}{\nu}\right)^3 \mp \frac{\Omega_\mathrm{core}}{4\pi}  \label{eq:g_perturb_m_pm_1}
\end{equation}
for $m=\pm1$ modes \citep{LiGang_2022_nature}. %(non-axisymmetric effect of Br is ignored. no $\zeta$ function) 
In Eqs.~\ref{eq:g_perturb_m_0} and \ref{eq:g_perturb_m_pm_1}, $\delta\nu_\mathrm{g}$ is the magnetic shift for pure g modes 
%($\zeta=1$) 
at the frequency at maximum power of the oscillations. 
%\comment{(For clarity, I suggest replacing $\delta\nu_\mathrm{g}$ by $\delta\nu_0$ and $\delta \nu_{\mathrm{g~mode}, m}$ by $\delta \nu_{\mathrm{g}, m}$) Gang's reply: $\delta \nu_g$ is consistent with the Nature paper. I think it is clear here since we use different subscript. }%, which is linked to the asymmetry of splittings together with $a$,
%\begin{equation}
%    \delta_\mathrm{asym} = 6\pi a\delta\nu_\mathrm{B}
%\end{equation}
The asymmetry parameter $a$ is a dimensionless average of $B_r^2$ in the oscillation cavity 
%\seb{depends on} the horizontal \seb{average} of the field 
\seb{weighted by the second order Legendre polynomial $P_2(\cos\theta)$}:

\begin{equation}
a = \frac{ \displaystyle \int_{r_\mathrm{i}}^{r_\mathrm{o}} K(r) \iint B_r^2 P_2(\cos\theta) \sin\theta \,\hbox{d}\theta\hbox{d}\phi \,\hbox{d}r} {\displaystyle \int_{r_\mathrm{i}}^{r_\mathrm{o}} K(r) \iint B_r^2  \sin\theta \,\hbox{d}\theta\hbox{d}\phi \,\hbox{d}r}.
\label{eq_coef_a}
\end{equation}
It verifies $-0.5 < a < 1$, its exact value depending on the latitudinal distribution of $B_r^2$ in the oscillation cavity \citep{LiGang_2022_nature}.
%disFor a dipolar field aligned with the rotation axis $a=2/5$. Generally, positive $a$ occurs when $B_r^2$ is concentrated near the pole, while negative values indicate a field concentrated towards the equator.}
%The range of possible values for $a$ 
%%causes a phenomenon that the magnetism-modified asymmetry can be negative if $a<0$, and it is also possible that no asymmetry is seen when $a=0$. 
%shows that the multiplet asymmetries induced by magnetic fields can be negative, while previous studies limited to a particular field geometry all led to positive asymmetries (\citealt{Hasan2005}, \citealt{Bugnet2021}). 
%It is also possible that the asymmetry vanishes with specific field configurations, such as a dipolar field inclined by about 55$^\circ$ with respect to the rotation axis or if the latitudinal variations of $B_r^2$ only occur at lengthscales much smaller than the star radius.
%\comment{(Not sure this sentence is needed, as all this was said in the Nature paper) Gang's reply: not everybody read our Nature paper so it is ok to mention it again. }

The pure g-mode periods in Eq.~\ref{eq:pure_g_mode} can then be rewritten as
\begin{equation}
    P'_{\mathrm{g}, m} = \left( \frac{1}{P_\mathrm{g}} + \delta\nu_{\mathrm{g~mode}, m} \right)^{-1}. \label{eq:pure_g_mode_with_perturbation}
\end{equation}

For pure p modes, the perturbation only arises from rotation, since the magnetic field being buried inside the star acts mainly on the g-mode parts of mixed modes. The effect on p modes is still negligible even if the magnetic field extends to the p-mode cavity \citep{Mathis2021}.Therefore, the pure p-mode frequency in Eq.~\ref{eq:p_mode_asymptotic_relation} is rewritten as
\begin{equation}
    \nu'_{\mathrm{p}, m} = \nu_\mathrm{p} - m \frac{\Omega_\mathrm{env}}{2\pi}. \label{eq:pure_p_mode_with_perturbation}
\end{equation}

%We perturbed the pure g- and p-mode frequencies using Eqs.~\ref{eq:pure_g_mode_with_perturbation} and \ref{eq:pure_p_mode_with_perturbation}, then re-calculated $\theta_\mathrm{g}$ and $\theta_\mathrm{p}$ using Eqs.~\ref{eq:theta_g_unperturbed} and \ref{eq:theta_p_unperturbed}.
\seb{We then} solved the asymptotic expressions for $m=1$, $0$, and $-1$ respectively,
\begin{equation}
    \tan \theta'_{\mathrm{p},m} = q \tan \theta'_{\mathrm{g},m}~\mathrm{for}~m=1,~0,~-1,\label{eq:magnetic_asymptotic_expressions}
\end{equation}
where $\theta'_{\mathrm{p},m}$ and $\theta'_{\mathrm{g},m}$ are the perturbed phases for pure p and g modes. \seb{Their expressions are similar to those of the unperturbed phases $\theta_\mathrm{p}$ and $\theta_\mathrm{g}$ given by Eqs.~\ref{eq:theta_p_unperturbed} and \ref{eq:theta_g_unperturbed}, but the frequencies of pure p and g modes are now replaced with the perturbed expressions given}
by Eqs.~\ref{eq:g_perturb_m_0}, \ref{eq:g_perturb_m_pm_1}, \ref{eq:pure_g_mode_with_perturbation}, and \ref{eq:pure_p_mode_with_perturbation}. 
%eq.~\ref{eq:magnetic_asymptotic_expressions} is that it %avoids using the $\zeta$ function (Eq.~\ref{eq:zeta_function}). %, \seb{which} 
%because $\zeta$ 
%is hard to define here \comment{(This would deserve a more detailed comment) Gang's reply: could you say more about that?}.
\seb{Eq.~\ref{eq:magnetic_asymptotic_expressions} can then be solved to obtain the perturbed mixed mode frequencies for any set of parameters characterising $(\Delta\Pi_1, q, \varepsilon_{\rm g}, f_\mathrm{shift}, \Omega_\mathrm{core},\Omega_\mathrm{env},a,\delta\nu_\mathrm{g})$. } 

\subsection{Fit to the observations}

\seb{In order to fit our asymptotic expression of mixed modes including rotational and magnetic perturbations to the observations, }we ran an MCMC algorithm \seb{similar to the one used in Sect. \ref{subsubsec:splitting_identification} } 
%to search for the best-fitting parameters. 
The priors for the first six parameters ($\Delta\Pi_1, q, \varepsilon_{\rm g}, f_\mathrm{shift}, \Omega_\mathrm{core},\Omega_\mathrm{env}$) are the same as in Sect. \ref{subsubsec:splitting_identification}, and for the 
%, which contains the six parameters introduced in Section~\ref{subsubsec:splitting_identification} and 
two additional parameters $a$ and $\delta\nu_\mathrm{g}$, 
%listed as follows (and their priors): 
\seb{we considered} 
%\comment{(The priors for the first six parameters are the same as before, so no need to repeat them) Gang's reply: good}
\begin{itemize}
    \item Asymmetry parameter $a$: uniform prior [-0.5, 1.0].
    \item Magnetic shift $\delta\nu_\mathrm{g}$: uniform prior [0$\,\mathrm{\mu Hz}$, 1$\,\mathrm{\mu Hz}$].
\end{itemize}
%The priors for each parameter are uniform within the ranges shown above. 
The same likelihood function as eq.~\ref{eq:likelihood} was used with the uncertainty defined as
the quadratic summation of model and observation errors:
\begin{equation}
    %\sigma_{m, i}^2 = \left( 0.02{\mu\mathrm{Hz}} \right)^2+\left(\sigma_{m, i}^\mathrm{obs}\right)^2.
    \sigma_{m, i}^2 = \left(\sigma_{m, i}^\mathrm{mod} \right)^2+\left(\sigma_{m, i}^\mathrm{obs}\right)^2.
\end{equation}
When fitting asymptotic expressions of mixed modes to the observations, we found that the optimal solutions
have a typical residual spread of 0.02\,$\mathrm{\mu Hz}$, which we consider as some residual spread caused by the WKB approximation since the asymptotic expression is not expected to be exact. Hence we included $\sigma_{m, i}^\mathrm{mod}=0.02\,\mathrm{\mu Hz}$ as additional white noise into the frequency uncertainties to account for the model uncertainty, assuming they are completely uncorrelated. The uncertainties of the mode frequencies ($\sigma_{m, i}^\mathrm{obs}$) were obtained in Section~\ref{subsubsec:asymmetry_measurement}. 

\begin{figure}
    \centering
    \includegraphics[width=\linewidth]{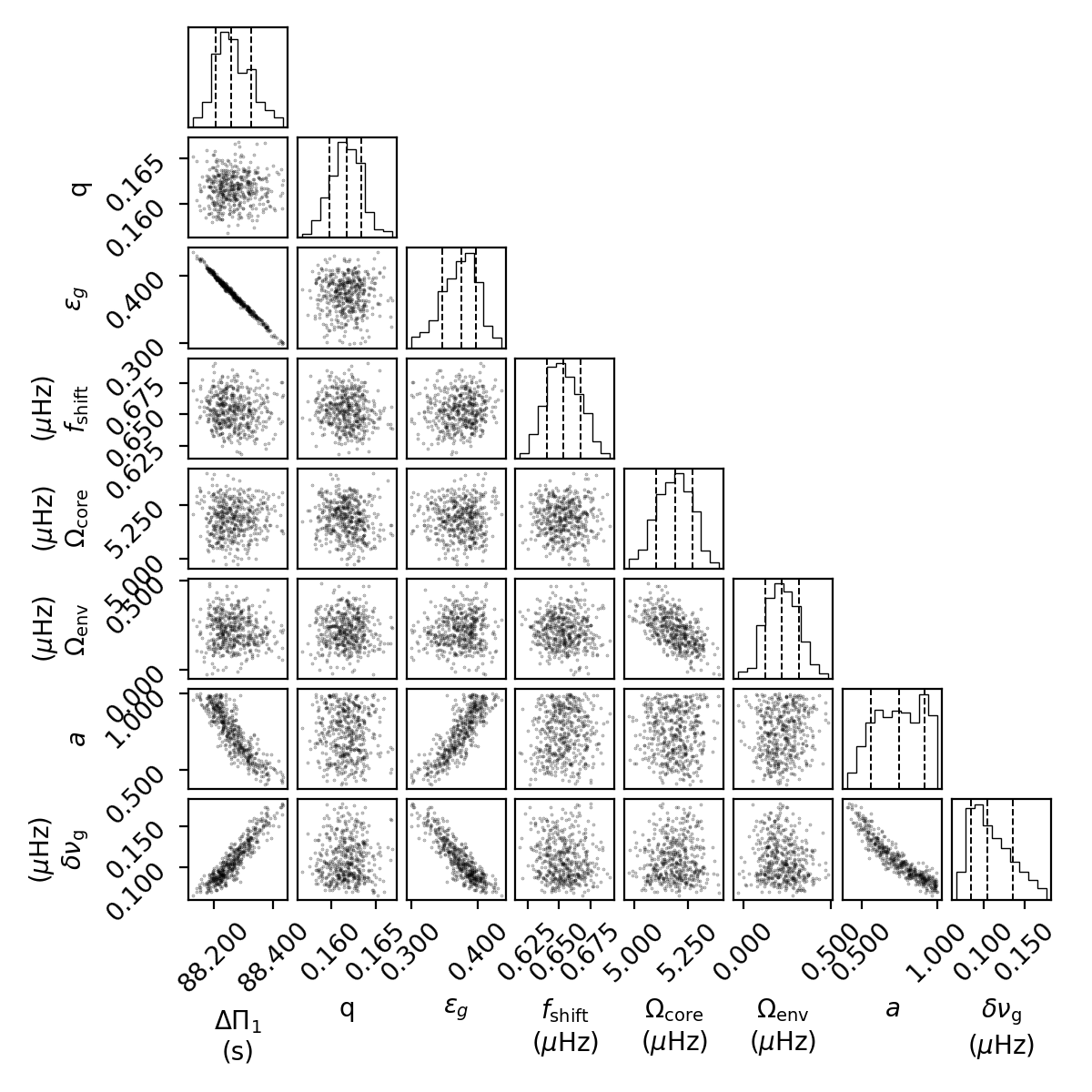}
    \caption{The corner diagram of the magnetic fitting of KIC\,5696081. The vertical dashed lines mark the median values and $\pm1\sigma$ ranges. }
    \label{fig:corner}
\end{figure}

\begin{figure}
    \centering
    \includegraphics[width= \linewidth]{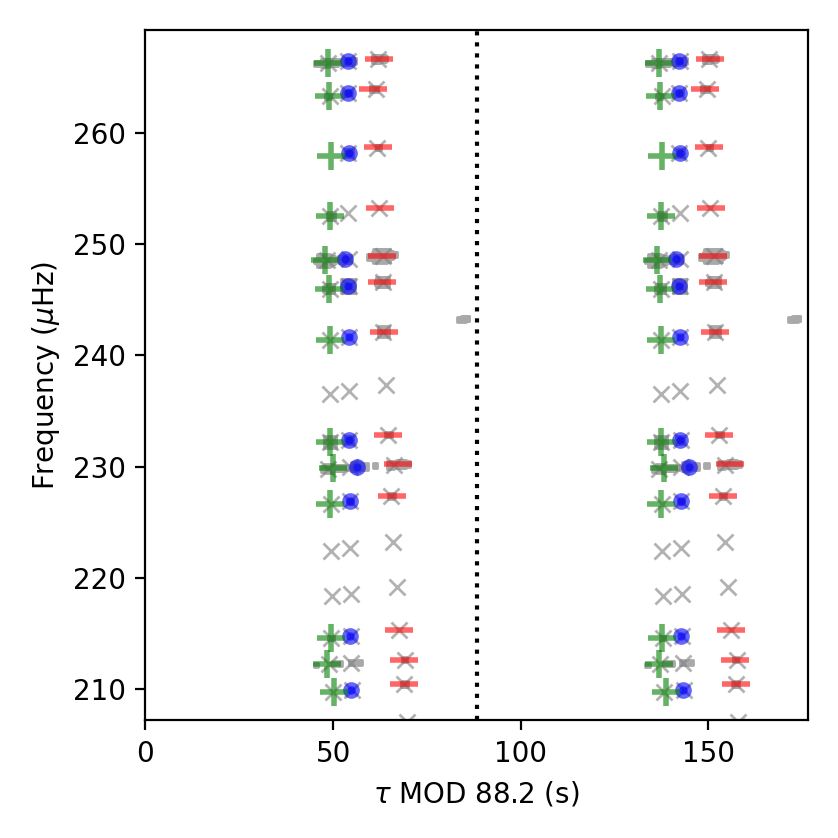}
    \caption{Same as fig.~\ref{fig:stretched_KIC5792889} but for KIC\,5696081. Note that this star shows clear asymmetries by magnetic field. }
    \label{fig:stretched_echelle_5696081}
\end{figure}

In the MCMC algorithm, 18 parallel chains \seb{were} used with length of 4000 steps, and we discarded the first 50\% sample results. The posterior distributions of the eight parameters are shown in Fig.~\ref{fig:corner}. We obtained good agreement between the formula including both rotational and magnetic perturbation and the observed asymmetric splittings. 
\seb{To illustrate this,} Figure~\ref{fig:stretched_echelle_5696081} displays the stretched \'{e}chelle diagram of KIC\,5696081. The \seb{dipole} triplets form three \seb{nearly} vertical ridges, and the $m=0$ ridge does not fall exactly halfway between $m=1$ and $m=-1$ ridges due to the magnetic perturbations. Our best-fitting frequencies (shown by the crosses) follow the observations well, showing a good agreement between the theory and observations. We display the 
%normal and 
stretched \'{e}chelle diagrams of the \starnumber stars in Appendix~\ref{appendixsection:all_figures}. They 
%all prove 
\seb{show} that our fits 
%show 
\seb{are in} good agreement with the observations and reproduce the asymmetries well.

The best-fitting parameters for all the \starnumber stars are listed in Table~\ref{tab:g_para}. 
Here we \seb{show} KIC\,5696081 as an example. The corner diagram of the MCMC result is shown in Fig.~\ref{fig:corner}, where the best-fitting result is found with \avalueexamplestar ~and \deltanuBexamplestar. %The mean angular frequency of the core rotation is \Omegacoreovertwopiexamplestar ~while the envelope rotation rate is \Omegaenvelopeovertwopiexamplestar. Therefore, we find a non-zero envelope rotation rate for KIC\,5696081 and the core-to-envelope rotation ratio is \Omegaratioexamplestar. 
In Fig.~\ref{fig:corner}, as well as the corner diagrams for the other stars in Appendix~\ref{appendixsection:all_figures}, we identify correlations between several parameters. \seb{As in the case of non-magnetic red giants, the measurement of $\Delta \Pi_1$ is strongly anti-correlated with the measurement of $\varepsilon_\mathrm{g}$. This is a direct result from the linear relation given by Eq. \ref{eq:pure_g_mode}. We also find a clear correlation between $\Delta \Pi_1$ and $\delta\nu_\mathrm{g}$. This can be understood as follows: starting from a best-fit solution, if we increase the value of $\delta\nu_\mathrm{g}$ (that is, if we increase the intensity of the field), the magnetic frequency perturbations increase, which tends to decrease the period spacing between consecutive g modes. Therefore, to correctly reproduce the observations, the asymptotic period spacing of unperturbed g modes $\Delta \Pi_1$ needs to be increased. Also, the parameters $\delta\nu_\mathrm{g}$ and $a$ are found to be strongly anti-correlated. Again, this was expected. Indeed, using Eqs. \ref{eq:g_perturb_m_0} and \ref{eq:g_perturb_m_pm_1}, we find that the asymmetry of g-mode multiplets near $\nu_{\rm max}$ corresponds to $3a\delta\nu_\mathrm{g}$. Therefore, if $\delta\nu_\mathrm{g}$ increases, one needs to decrease $a$ to reproduce the observed asymmetries. Finally, we observe a slight anti-correlation between the measurements of $\Omega_\mathrm{core}$ and $\Omega_\mathrm{env}$. This can be understood from Eq. \ref{eq:rotational_splittings} (even though this relation has not been used in our fits here): in this linear relation, $\Omega_\mathrm{core}$ and $\Omega_\mathrm{env}$ are related to the slope and the intercept, respectively, and the measurements of these quantities are not independent.}
%\textbf{, such as $\Delta \Pi_1$ and $\varepsilon_\mathrm{g}$, $\Omega_\mathrm{core}$ and $\Omega_\mathrm{env}$, $\Delta \Pi_1$ and $a$, and $a$ and $\delta\nu_\mathrm{g}$. 

The asymmetry parameter $a$ generally has larger uncertainties and broader distributions within its prior range (from -0.5 to 1) compared to the other parameters, suggesting that the constraints on $a$ are weaker. The parameter $\delta\nu_\mathrm{g}$ often had an asymmetric distribution, while the other parameters have symmetric distributions and generally tighter constraints. 

These stars display normal values for the other parameters of the fits ($\Delta \Pi_1$, $q$, $\varepsilon_\mathrm{g}$, and $f_\mathrm{shift}$), as shown in Appendix~\ref{appendixsec:other_paramters}. Despite the fact that we used non-informative priors for $\varepsilon_\mathrm{g}$, we obtained results that are in line with typical values for other red giants \citep[$0.28\pm0.08$][]{takata16, mosser18}.

%Figure~\ref{fig:stretched_echelle_5696081} displays the stretched \'{e}chelle diagram of KIC\,5696081. The rotational triplets form three vertical ridges, and the $m=0$ ridge does not fall exactly halfway between $m=1$ and $m=-1$ ridges due to the magnetic perturbations. Our best-fitting frequencies (shown by the crosses) follow the observations well, showing a good agreement between the theory and observations.  

%We display all the normal and stretched \'{e}chelle diagrams of \starnumber stars in appendix~\ref{appendixsection:all_figures}. They all prove that our fits show good agreement with the observations and reproduce the asymmetries well.

\subsection{Stellar model and field strength}

After obtaining the magnetic shift $\delta \nu_\mathrm{g}$, we can calculate the magnetic field strength in the stellar interior. What we can measure about the magnetic field strength is a weighted integral of the horizontal average of the squared radial magnetic field, $\overline{B_r^2}$, given by \citep{LiGang_2022_nature}:
\begin{equation}
    \langle B_r^2\rangle = \int_{r_\mathrm{i}}^{r_\mathrm{o}} K\left(r\right)\overline{B_r^2}\mathrm{d}r= \frac{16\pi^4\mu_0 \delta\nu_\mathrm{g} \nu_{\rm max}^3}{\mathcal{I}}, \label{eq:square_field_strength}
\end{equation}
with $K(r)$ the weight function and where \seb{$r_\mathrm{i}$ and $r_\mathrm{o}$ are the inner and outer turning points of the g-mode cavity, respectively, and} $\mu_0$ is the vacuum permeability. The core factor $\mathcal{I}$ is determined by the internal structure of the star:
\begin{equation}
    \mathcal{I} = \frac{\int_{r_\mathrm{i}}^{r_\mathrm{o}} \left(\frac{N}{r}\right)^3 \frac{\mathrm{d}r}{\rho}}{\int_{r_\mathrm{i}}^{r_\mathrm{o}} \left(\frac{N}{r}\right) \,\mathrm{d}r}, \label{eq:core_factor_I}
\end{equation}
where $N$ is the buoyancy frequency and $\rho$ is the local density. The weighted function,
\begin{equation}
    K\left(r\right) = \frac{\frac{1}{\rho}\left(\frac{N}{r}\right)^3}{\int_{r_\mathrm{i}}^{r_\mathrm{o}}\left(\frac{N}{r}\right)^3\frac{\mathrm{d}r}{\rho}},\label{eq:kernel}
\end{equation}
sharply peaks at the hydrogen-burning shell, \seb{with a much lower sensitivity in the layers below.}
%hence in fact we \seb{essentially} measure the magnetic field strength in the vicinity of hydrogen burning shell. 

To \seb{estimate the intensity of the detected magnetic fields, we needed to} characterise the internal structures of these stars (such as the buoyancy frequency $N$ and the density profile $\rho$). For this purpose, we adopted the seismology-modelling pipeline and the model grid introduced by \citet{Li2022keplerRG} to search for the best-fitting models. The input constraints were based on global parameters, such as effective temperature, luminosity, and metallicity, which were reported by \cite{Berger2020} and listed in table~\ref{tab:strength_table}. We also used the observed radial-mode frequencies given by \citet{Li2022keplerRG}, and the asymptotic period spacing of g modes ($\Delta \Pi_{1}$) measured by this work as additional constraints.
By applying the pipeline, we determined the best-fitting structural model, whose parameters (masses, ages, and radii) are listed in Table~\ref{tab:strength_table}.

Using our best-fit stellar models, we could derive estimates of the core factor $\mathcal{I}$, and we were thus able to obtain measurements of the average radial magnetic fields $\langle B_r^2\rangle^{0.5}$ using Eq. \ref{eq:square_field_strength} (see Table~\ref{tab:strength_table}). We found field strengths ranging from about 20~kG to 150~kG.

%In table~\ref{tab:strength_table}, we list our results of the core factor $\mathcal{I}$ and the measured magnetic field strengths $\langle B_r^2\rangle^{0.5}$. 
Since this work and \cite{LiGang_2022_nature} used different approaches in the stellar modelling and the fitting of the asymptotic expression of mixed mode frequencies to the observations, there are slight differences in the inferred field strengths for the three stars that they have in common. We compared the two sets of results and found that the field strengths derived by both works are generally consistent, with slightly larger strengths obtained in this work. For KIC\,7518143 and KIC\,8684542, they have consistent field strengths within 1-$\sigma$ ranges. While for KIC\,11515377, the field strength obtained by this work is approximately 30\% times larger than the strength by \cite{LiGang_2022_nature}, though still within the 2-$\sigma$ range. The ratios between the field strength and the critical field strength show good agreement between the two sets of results, likely due to this ratio being more sensitive to the stellar structure rather than the observed frequencies or the fitting strategies.

\begin{figure}
    \centering
    \includegraphics[width=\linewidth]{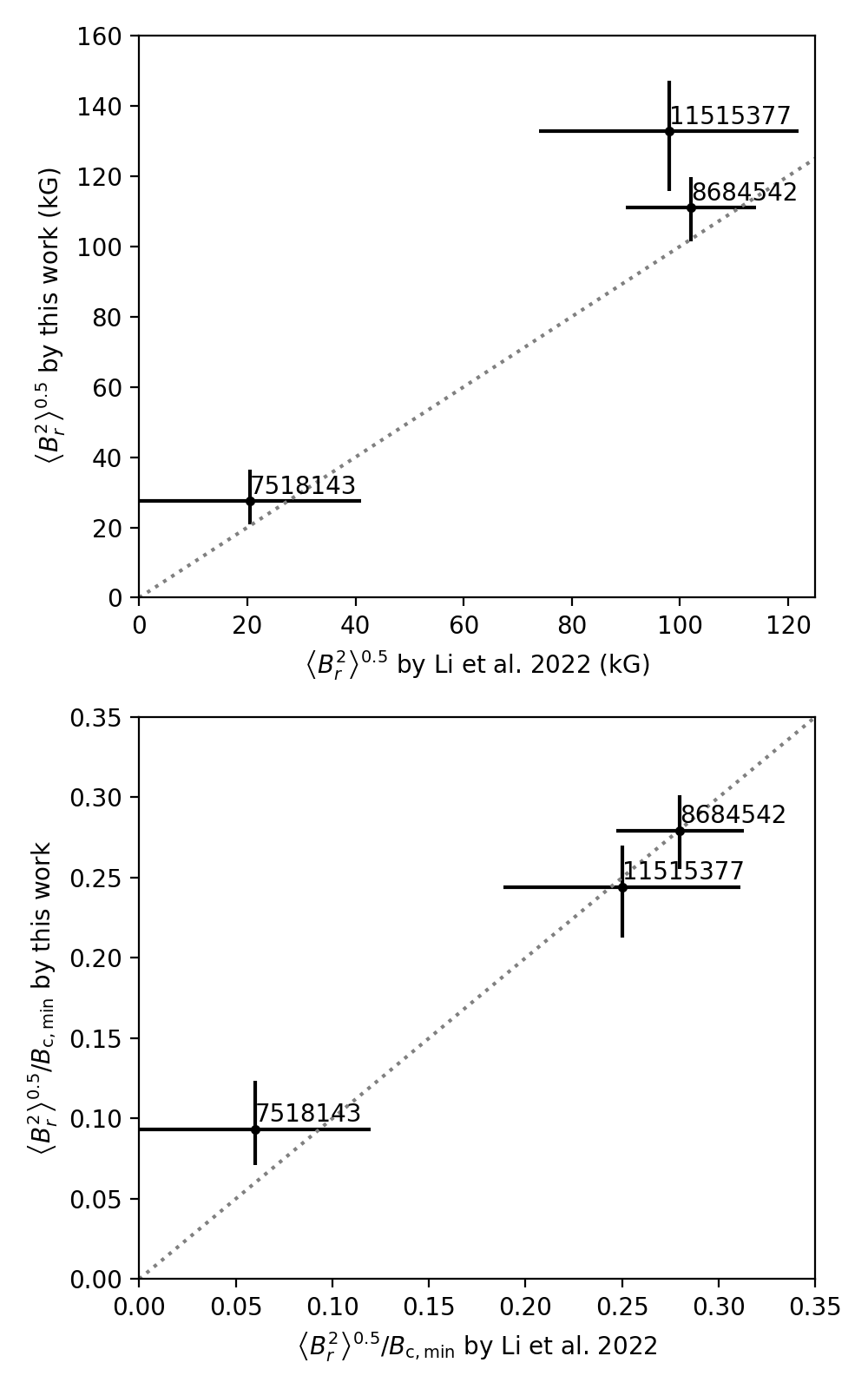}
    \caption{Top panel: measured field strengths by this work and by \cite{LiGang_2022_nature}. Bottom: the ratio between the measured field strengths and the critical strengths. The dotted lines show 1:1 relations. Note that the value of the field strength of KIC\,7518143 reported by \cite{LiGang_2022_nature} is smaller than 41\,kG, while we use the median here (20.5\,kG). }
    \label{fig:strength_comparison}
\end{figure}

In Appendix~\ref{appendixsec:relation_between_I_and_Delta_Pi_1}, we also offer a linear relation between $\mathcal{I}$ and $\Delta\Pi_1$. This relation can be used for model-independent calculations of field strength, particularly when a best-fitting stellar model is unavailable.

\begin{table*}[]
%\scriptsize
    \caption{The fit results of the p-mode asymptotic relations in eq.~\ref{eq:p_mode_asymptotic_relation}. }
    \label{tab:p_param}
    \centering
    \begin{tabular}{lrrrrr}
    \hline
    \multicolumn{1}{c}{KIC} & \multicolumn{1}{c}{$\nu_\mathrm{max}$, $\mu$Hz} & \multicolumn{1}{c}{$\Delta\nu$, $\mu$Hz} & \multicolumn{1}{c}{$\varepsilon$} & \multicolumn{1}{c}{$\alpha$} & \multicolumn{1}{c}{$D$, $\mu\mathrm{Hz}$} \\
    \hline
4458118 & $207.6\pm1.6$ & $16.442\pm0.004$$\phantom{0}$ & $0.326\pm0.003$$\phantom{0}$ & $0.0057\pm0.0005$$\phantom{0}$ & $0.3620\pm0.0015$\\ 
5196300 & $207.6\pm1.1$ & $15.138\pm0.003$$\phantom{0}$ & $0.340\pm0.003$$\phantom{0}$ & $0.00406\pm0.00018$ & $0.3215\pm0.0011$\\ 
5696081 & $248.0\pm1.3$ & $17.901\pm0.004$$\phantom{0}$ & $0.362\pm0.003$$\phantom{0}$ & $0.00115\pm0.00024$ & $0.3544\pm0.0012$\\ 
6936091 & $94.3\pm0.4$ & $8.731\pm0.004$$\phantom{0}$ & $0.174\pm0.006$$\phantom{0}$ & $0.0057\pm0.0005$$\phantom{0}$ & $0.2090\pm0.0011$\\ 
7009365 & $210.5\pm1.2$ & $15.6082\pm0.0029$ & $0.3332\pm0.0027$ & $0.00679\pm0.00019$ & $0.3147\pm0.0011$\\ 
7518143 & $159.2\pm0.5$ & $12.394\pm0.003$$\phantom{0}$ & $0.216\pm0.003$$\phantom{0}$ & $0.00388\pm0.00021$ & $0.2555\pm0.0012$\\ 
8540034 & $195.4\pm0.7$ & $15.108\pm0.003$$\phantom{0}$ & $0.2517\pm0.0029$ & $0.00563\pm0.00025$ & $0.3285\pm0.0012$\\ 
8619145 & $130.0\pm0.4$ & $10.896\pm0.004$$\phantom{0}$ & $0.163\pm0.004$$\phantom{0}$ & $0.0097\pm0.0004$$\phantom{0}$ & $0.2428\pm0.0012$\\ 
8684542 & $179.4\pm0.7$ & $13.4921\pm0.0027$ & $0.2741\pm0.0028$ & $0.00345\pm0.00020$ & $0.282\pm0.0010$\\ 
9202471 & $218.5\pm1.3$ & $15.946\pm0.004$$\phantom{0}$ & $0.358\pm0.003$$\phantom{0}$ & $-0.0005\pm0.0003$$\phantom{0}$ & $0.3138\pm0.0012$\\ 
9589420 & $109.9\pm0.5$ & $9.473\pm0.004$$\phantom{0}$ & $0.187\pm0.005$$\phantom{0}$ & $-0.0015\pm0.0005$$\phantom{0}$ & $0.1993\pm0.0012$\\ 
10801792 & $199.7\pm1.0$ & $15.0092\pm0.0028$ & $0.3358\pm0.0026$ & $0.00244\pm0.00018$ & $0.326\pm0.0010$\\ 
11515377 & $194.6\pm0.5$ & $14.739\pm0.005$$\phantom{0}$ & $0.340\pm0.005$$\phantom{0}$ & $0.0011\pm0.0003$$\phantom{0}$ & $0.3204\pm0.0012$\\

\hline
    \end{tabular}

\end{table*}

\begin{table*}[]
    \caption{The best-fitting results of the $l=1$ mixed mode frequencies considering the magnetism-induced perturbations in eqs.~\ref{eq:g_perturb_m_0}, \ref{eq:g_perturb_m_pm_1}, and \ref{eq:pure_p_mode_with_perturbation}. The parameters are: period spacing $\Delta \Pi_1$, the coupling factor $q$, g-mode phase $\varepsilon_\mathrm{g}$, the frequency correction of $l=1$ pure p mode $ f_\mathrm{shift}$, core and envelope rotation rates $\Omega_\mathrm{core}/2\pi$ and $\Omega_\mathrm{env}/2\pi$, the asymmetry parameter $a$, and the magnetic shift $\delta\nu_\mathrm{g}$. }
    \label{tab:g_para}
    \centering
    \small
    \begin{tabular}{lllllllll}
    \hline
    \multicolumn{1}{c}{KIC} & $\Delta \Pi_1$ (s) & \multicolumn{1}{c}{$q$} & \multicolumn{1}{c}{$\varepsilon_\mathrm{g}$} & $ f_\mathrm{shift}$ ($\mu$Hz) & $\Omega_\mathrm{core}/2\pi$ ($\mu$Hz) & $\Omega_\mathrm{env}/2\pi$ ($\mu$Hz) & \multicolumn{1}{c}{$a$} & $\delta\nu_\mathrm{g}$ ($\mu$Hz) \\
    \hline
4458118 & $88.71_{-0.06}^{+0.06}$ & $0.1617_{-0.0025}^{+0.0025}$ & $0.27_{-0.03}^{+0.03}$ & $0.734_{-0.015}^{+0.016}$ & $0.867_{-0.020}^{+0.019}$ & $\phantom{-}$$0.019_{-0.022}^{+0.020}$ & $\phantom{-}$$0.42_{-0.26}^{+0.3}$ & $\phantom{-}$$0.023_{-0.012}^{+0.022}$\\ 
5196300 & $82.99_{-0.05}^{+0.07}$ & $0.1304_{-0.0020}^{+0.0019}$ & $0.306_{-0.04}^{+0.028}$ & $0.620_{-0.013}^{+0.014}$ & $1.251_{-0.012}^{+0.012}$ & $\phantom{-}$$0.031_{-0.014}^{+0.014}$ & $\phantom{-}$$0.71_{-0.22}^{+0.20}$ & $\phantom{-}$$0.080_{-0.018}^{+0.029}$\\ 
5696081 & $88.26_{-0.05}^{+0.07}$ & $0.1617_{-0.0019}^{+0.0017}$ & $0.375_{-0.029}^{+0.022}$ & $0.653_{-0.013}^{+0.014}$ & $0.826_{-0.014}^{+0.013}$ & $\phantom{-}$$0.035_{-0.014}^{+0.015}$ & $\phantom{-}$$0.74_{-0.18}^{+0.17}$ & $\phantom{-}$$0.104_{-0.019}^{+0.03}$\\ 
6936091 & $75.29_{-0.13}^{+0.16}$ & $0.132_{-0.008}^{+0.007}$ & $0.39_{-0.24}^{+0.21}$ & $0.550_{-0.024}^{+0.023}$ & $0.362_{-0.012}^{+0.013}$ & $-0.01_{-0.03}^{+0.03}$ & $\phantom{-}$$0.30_{-0.13}^{+0.23}$ & $\phantom{-}$$0.064_{-0.029}^{+0.04}$\\ 
7009365 & $83.298_{-0.027}^{+0.029}$ & $0.1377_{-0.0016}^{+0.0017}$ & $0.270_{-0.018}^{+0.017}$ & $0.600_{-0.012}^{+0.012}$ & $0.546_{-0.012}^{+0.013}$ & $\phantom{-}$$0.023_{-0.012}^{+0.012}$ & $\phantom{-}$$0.83_{-0.19}^{+0.12}$ & $\phantom{-}$$0.036_{-0.006}^{+0.010}$\\ 
7518143 & $78.50_{-0.03}^{+0.04}$ & $0.1209_{-0.0022}^{+0.0025}$ & $0.29_{-0.04}^{+0.03}$ & $0.611_{-0.013}^{+0.016}$ & $0.315_{-0.011}^{+0.013}$ & $\phantom{-}$$0.068_{-0.018}^{+0.018}$ & $\phantom{-}$$0.49_{-0.23}^{+0.3}$ & $\phantom{-}$$0.020_{-0.008}^{+0.015}$\\ 
8540034 & $84.25_{-0.03}^{+0.03}$ & $0.1524_{-0.0022}^{+0.0024}$ & $0.308_{-0.022}^{+0.020}$ & $0.612_{-0.011}^{+0.014}$ & $1.138_{-0.010}^{+0.011}$ & $\phantom{-}$$0.009_{-0.014}^{+0.012}$ & $\phantom{-}$$0.57_{-0.27}^{+0.29}$ & $\phantom{-}$$0.020_{-0.009}^{+0.012}$\\ 
8619145 & $76.49_{-0.14}^{+0.19}$ & $0.126_{-0.005}^{+0.005}$ & $0.54_{-0.20}^{+0.15}$ & $0.645_{-0.019}^{+0.020}$ & $0.920_{-0.013}^{+0.013}$ & $\phantom{-}$$0.020_{-0.018}^{+0.023}$ & $\phantom{-}$$0.58_{-0.22}^{+0.29}$ & $\phantom{-}$$0.11_{-0.04}^{+0.06}$\\ 
8684542 & $80.55_{-0.09}^{+0.08}$ & $0.1213_{-0.0021}^{+0.0023}$ & $0.22_{-0.06}^{+0.06}$ & $0.680_{-0.013}^{+0.014}$ & $0.663_{-0.016}^{+0.016}$ & $\phantom{-}$$0.016_{-0.018}^{+0.017}$ & $\phantom{-}$$0.38_{-0.05}^{+0.05}$ & $\phantom{-}$$0.21_{-0.03}^{+0.03}$\\ 
9202471 & $84.66_{-0.09}^{+0.14}$ & $0.1351_{-0.0019}^{+0.0016}$ & $0.26_{-0.07}^{+0.05}$ & $0.684_{-0.013}^{+0.015}$ & $0.950_{-0.018}^{+0.019}$ & $\phantom{-}$$0.014_{-0.016}^{+0.015}$ & $\phantom{-}$$0.21_{-0.09}^{+0.19}$ & $\phantom{-}$$0.10_{-0.04}^{+0.07}$\\ 
9589420 & $74.44_{-0.08}^{+0.08}$ & $0.130_{-0.004}^{+0.006}$ & $0.28_{-0.12}^{+0.12}$ & $0.612_{-0.018}^{+0.018}$ & $0.645_{-0.008}^{+0.010}$ & $-0.008_{-0.018}^{+0.017}$ & $\phantom{-}$$0.37_{-0.12}^{+0.17}$ & $\phantom{-}$$0.070_{-0.018}^{+0.020}$\\ 
10801792 & $83.17_{-0.10}^{+0.13}$ & $0.1416_{-0.0019}^{+0.0020}$ & $0.25_{-0.08}^{+0.06}$ & $0.640_{-0.015}^{+0.016}$ & $0.918_{-0.016}^{+0.015}$ & $\phantom{-}$$0.028_{-0.016}^{+0.016}$ & $\phantom{-}$$0.37_{-0.12}^{+0.16}$ & $\phantom{-}$$0.12_{-0.04}^{+0.06}$\\ 
11515377 & $83.7_{-0.10}^{+0.10}$ & $0.1542_{-0.0021}^{+0.0020}$ & $0.21_{-0.06}^{+0.06}$ & $0.615_{-0.014}^{+0.013}$ & $0.649_{-0.014}^{+0.014}$ & $\phantom{-}$$0.050_{-0.016}^{+0.016}$ & $-0.16_{-0.05}^{+0.03}$ & $\phantom{-}$$0.18_{-0.04}^{+0.04}$\\

    \hline
    \end{tabular}

\end{table*}

\begin{table*}[]
    \centering
        \caption{In this table, we list the values and their uncertaintiers of $T_\mathrm{eff}$, $\log L/L_\odot$, and [Fe/H] given by \cite{Berger2020} that were used to constrain the theoretical models. Then we list the model-inferred masses, radii, and ages. We also provide the core structure factor $\mathcal{I}$, the measured field strengths $\left<B_\mathrm{r}^2\right>^{0.5}$, and the model-inferred critical strengths at the hydrogen-burning shells $B_\mathrm{c}$ of the \starnumber stars in this work.}
    \label{tab:strength_table}
    \begin{tabular}{cllllllllc}
\hline
KIC & $T_\mathrm{eff}$ & $\log L/L_\odot$ & [Fe/H] & Mass & radius & Age  & $\mathcal{I}$  & $\langle B_\mathrm{r}^2\rangle^{0.5}$ & $B_\mathrm{c}$\\
   & K & & dex & $\mathrm{M_\odot}$ & $\mathrm{R_\odot}$ & Gyr &$10^{-23}\frac{\mathrm{m}}{\mathrm{s^2kg}}$ & kG & kG \\
\hline
4458118 & $4900_{-80}^{+90}$ & $0.97_{-0.15}^{+0.11}$ & $-0.18_{-0.15}^{+0.13}$ & $1.077_{-0.04}^{+0.021}$ & $4.10_{-0.05}^{+0.03}$ & $8.4_{-0.4}^{+1.5}$ & 1.1732 & $\phantom{0}58_{-19}^{+24}$ & 704.7\\ 
5196300 & $4840_{-80}^{+90}$ & $1.08_{-0.03}^{+0.03}$ & $\phantom{-}0.20_{-0.12}^{+0.11}$ & $1.44_{-0.03}^{+0.08}$ & $4.78_{-0.05}^{+0.09}$ & $3.6_{-0.6}^{+0.7}$ & 1.7617 & $\phantom{0}89_{-12}^{+16}$ & 564.0\\ 
5696081 & $4910_{-80}^{+80}$ & $0.93_{-0.28}^{+0.26}$ & $-0.02_{-0.11}^{+0.15}$ & $1.34_{-0.04}^{+0.04}$ & $4.19_{-0.05}^{+0.04}$ & $4.5_{-0.6}^{+0.6}$ & 1.3404 & $152_{-16}^{+23}$ & 957.4\\ 
6936091 & $4840_{-90}^{+90}$ & $1.24_{-0.18}^{+0.22}$ & $-0.17_{-0.15}^{+0.12}$ & $1.04_{-0.03}^{+0.04}$ & $6.13_{-0.09}^{+0.10}$ & $8.1_{-1.2}^{+1.7}$ & 2.0698 & $\phantom{0}23_{-6}^{+7}$ & 107.9\\ 
7009365 & $4880_{-80}^{+80}$ & $1.04_{-0.17}^{+0.14}$ & $\phantom{-}0.10_{-0.12}^{+0.14}$ & $1.396_{-0.021}^{+0.06}$ & $4.64_{-0.03}^{+0.07}$ & $3.8_{-0.6}^{+0.8}$ & 1.7098 & $\phantom{0}62_{-6}^{+9}$ & 585.2\\ 
7518143 & $4780_{-80}^{+80}$ & $1.148_{-0.006}^{+0.028}$ & $\phantom{-}0.15_{-0.14}^{+0.10}$ & $1.40_{-0.05}^{+0.06}$ & $5.41_{-0.08}^{+0.08}$ & $3.8_{-0.8}^{+1.0}$ & 2.1362 & $\phantom{0}27_{-6}^{+9}$ & 294.1\\ 
8540034 & $4940_{-80}^{+80}$ & $1.06_{-0.11}^{+0.08}$ & $-0.12_{-0.15}^{+0.11}$ & $1.20_{-0.03}^{+0.08}$ & $4.52_{-0.06}^{+0.09}$ & $5.0_{-0.3}^{+1.3}$ & 1.5042 & $\phantom{0}44_{-11}^{+12}$ & 545.0\\ 
8619145 & $4850_{-80}^{+80}$ & $1.19_{-0.04}^{+0.07}$ & $-0.03_{-0.14}^{+0.12}$ & $1.26_{-0.04}^{+0.04}$ & $5.68_{-0.06}^{+0.06}$ & $5.6_{-1.1}^{+0.6}$ & 2.1872 & $\phantom{0}47_{-9}^{+12}$ & 197.3\\ 
8684542 & $4850_{-80}^{+100}$ & $1.112_{-0.06}^{+0.014}$ & $\phantom{-}0.19_{-0.15}^{+0.12}$ & $1.40_{-0.06}^{+0.07}$ & $5.11_{-0.08}^{+0.09}$ & $3.6_{-0.6}^{+0.7}$ & 1.9372 & $111_{-9}^{+9}$ & 397.6\\ 
9202471 & $4720_{-80}^{+80}$ & $0.97_{-0.17}^{+0.14}$ & $\phantom{-}0.36_{-0.11}^{+0.12}$ & $1.40_{-0.04}^{+0.03}$ & $4.57_{-0.05}^{+0.05}$ & $4.0_{-0.4}^{+1.3}$ & 1.5490 & $110_{-30}^{+40}$ & 663.3\\ 
9589420 & $4780_{-80}^{+80}$ & $1.31_{-0.18}^{+0.21}$ & $-0.01_{-0.15}^{+0.11}$ & $1.30_{-0.03}^{+0.06}$ & $6.27_{-0.08}^{+0.08}$ & $5.6_{-1.2}^{+0.6}$ & 2.6778 & $\phantom{0}26_{-4}^{+4}$ & 124.7\\ 
10801792 & $4890_{-80}^{+80}$ & $1.08_{-0.05}^{+0.010}$ & $-0.06_{-0.16}^{+0.15}$ & $1.34_{-0.05}^{+0.04}$ & $4.68_{-0.06}^{+0.05}$ & $4.21_{-0.28}^{+1.2}$ & 1.7296 & $102_{-19}^{+24}$ & 535.6\\ 
11515377 & $4960_{-80}^{+80}$ & $1.11_{-0.08}^{+0.07}$ & $-0.21_{-0.12}^{+0.15}$ & $1.30_{-0.07}^{+0.06}$ & $4.70_{-0.10}^{+0.06}$ & $4.2_{-0.9}^{+0.4}$ & 1.4839 & $133_{-17}^{+14}$ & 544.3\\ 

\hline
    \end{tabular}
\end{table*}

\section{Discussion}\label{sec:discussion}

\subsection{Asymmetry parameter and magnetic shift}
%We list the best-fitting results of the p-mode asymptotic relation in Table~\ref{tab:p_param}, and the fit results of the magnetism-induced perturbations in Table~\ref{tab:g_para}. 

\begin{figure}
    \centering
    \includegraphics[width=\linewidth]{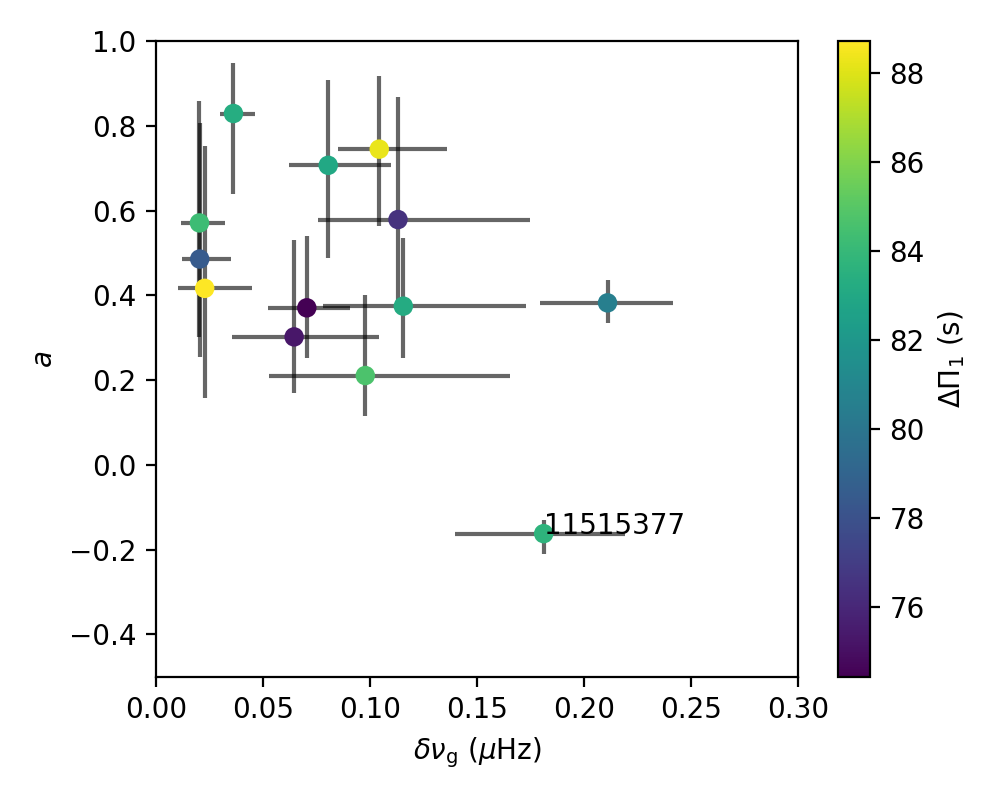}
    \caption{Relation between the asymmetry parameter $a$ and the magnetic shift $\delta\nu_\mathrm{g}$. KIC\,11515377 is the only star that shows negative asymmetries. }
    \label{fig:a_vs_magshift}
\end{figure}

\begin{figure}
    \centering
    \includegraphics[width=\linewidth]{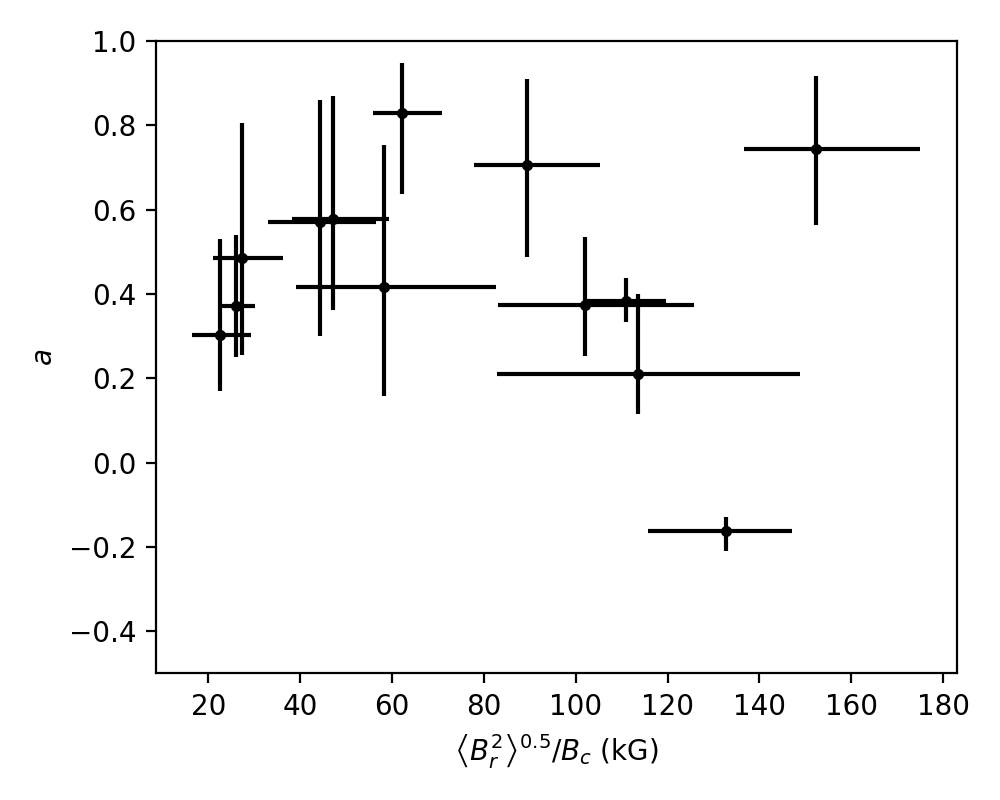}
    \caption{Relation between the asymmetry parameter $a$ and the field strength.}
    \label{fig:a_vs_field_strength}
\end{figure}

Figure~\ref{fig:a_vs_magshift} displays the 
%correlation 
\seb{relation} between the asymmetry parameter $a$ and the magnetic shift $\delta \nu_\mathrm{g}$.
%\comment{(Perhaps it would make more sense to show the relation between the asymmetry parameter and the field strength, instead of $\delta\nu_{\rm g}$. It doesn't show any correlation either.) Gang's reply: I show it in the new figure}
We find that there is no obvious correlation between these two parameters and there is also no correlation with the period spacing $\Delta \Pi_1$. \textcolor{black}{Figure~\ref{fig:a_vs_field_strength} shows the relation between the asymmetry parameter $a$ and the field strength of our sample, and we still do not find any correlation between them, meaning that the field strength does not correlate with the field topology.}

The asymmetry parameter $a$ reaches 1 when the field is \seb{entirely} concentrated to the poles \seb{and} it reaches $-0.5$ when the field is concentrated to the equator. Dipolar fields have values of $a$ ranging from $-0.2$ (corresponding to a dipolar field aligned with the equator) to $0.4$ (corresponding to a dipolar field aligned with the rotation axis). The asymmetry parameter can also vanish, 
%\seb{value of $a$}, 
even in the presence of a strong field, for example if a dipolar field is inclined by about 55$^\circ$ with respect to the rotation axis or if the latitudinal variations of $B_r^2$ only occur at length scales much smaller than the star radius \citep{LiGang_2022_nature}. 
Most of the stars exhibit $a$ values between 0.2 and 0.8. KIC\,11515377 is the only star that shows negative asymmetries, which lead to $a$ close to $-0.2$. This suggests that this configuration might be rarer among red giants. We note the diversity of the values obtained for the asymmetry parameter $a$, which clearly shows that the core fields of red giants have various horizontal geometries. 
In some stars, the posterior probabilities for the asymmetry parameter $a$ nearly vanish below $0.4$, which means that for these stars the magnetic fields are more sharply concentrated on the poles than a purely dipolar field aligned with the rotation axis. Our measurements of $a$ will be useful to constrain future models of red giant core magnetic fields.

\subsection{Core and envelope rotation rates}

\begin{figure}
    \centering
    \includegraphics[width=\linewidth]{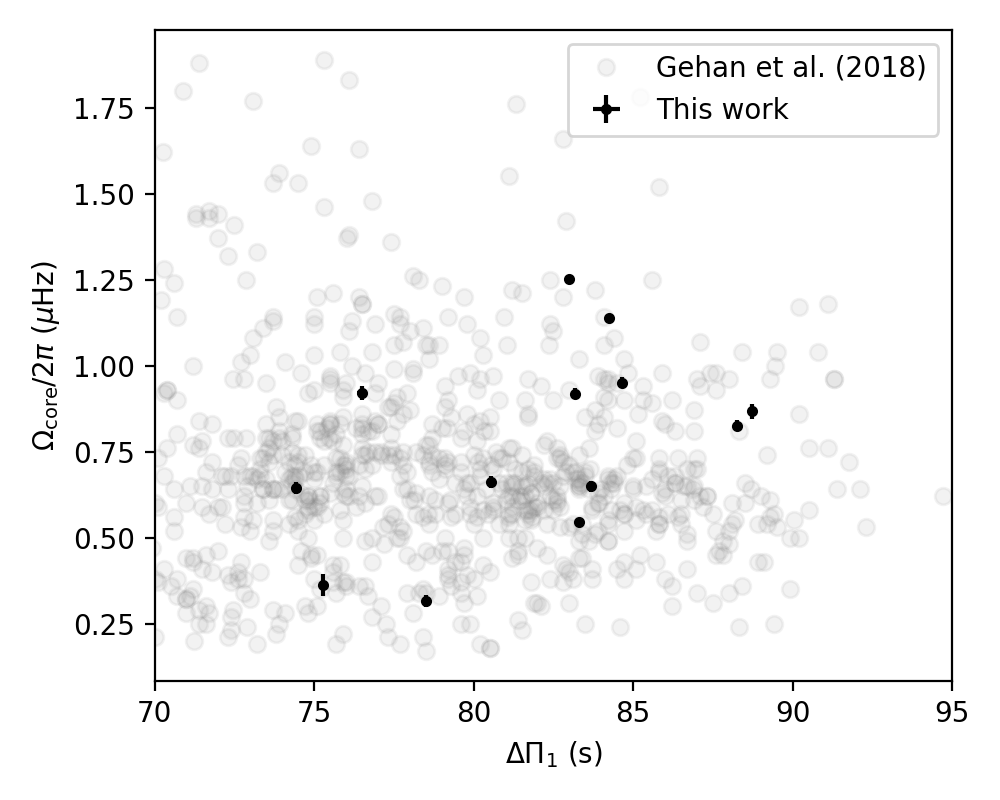}
    \caption{The core rotation rates $\Omega_\mathrm{g}$ as a function of $\Delta \Pi_1$. The black dots are the stars in this work, and the grey circles are reported by \cite{Gehan2018}.  }
    \label{fig:core_rotation}
\end{figure}

We display the relation between $\Omega_\mathrm{core}$ and $\Delta \Pi_1$ in Fig.~\ref{fig:core_rotation}. The figure shows that the rotation rates of our stars are typical for red giants and are consistent with those of other studies \citep[shown as the grey circles by][]{Gehan2018}. Our observations thus suggest that these stars have not experienced significantly different histories of angular momentum transport compared to other red giant stars. We note that this does not contradict the hypothesis that magnetic fields could be responsible for angular momentum transport in red giants because we cannot exclude that magnetic fields currently escaping detection might exist in other red giants.

We compared our measurements of the core rotation rates with those of \cite{Gehan2018} for the five stars in our sample that were also in their study. We find that our results are consistent in four stars (KIC\,7518143, 6936091, 11515377, and 8540034), while a large discrepancy appears in KIC\,8684542. The reason for the consistency is that the magnetism-induced perturbation does not change the \seb{frequency separation} between $m=1$ and $m=-1$ modes, hence it does not affect the measurements of splittings if the mode identification is correct. However, the discrepancy for KIC\,8684542 arises because only symmetric rotational splittings were considered in the previous work, which resulted in an incorrect identification of the splitting of $m=1$ and $-1$ modes \citep[see the online peer review file of][]{LiGang_2022_nature}.

\begin{figure}
    \centering
    \includegraphics[width=\linewidth]{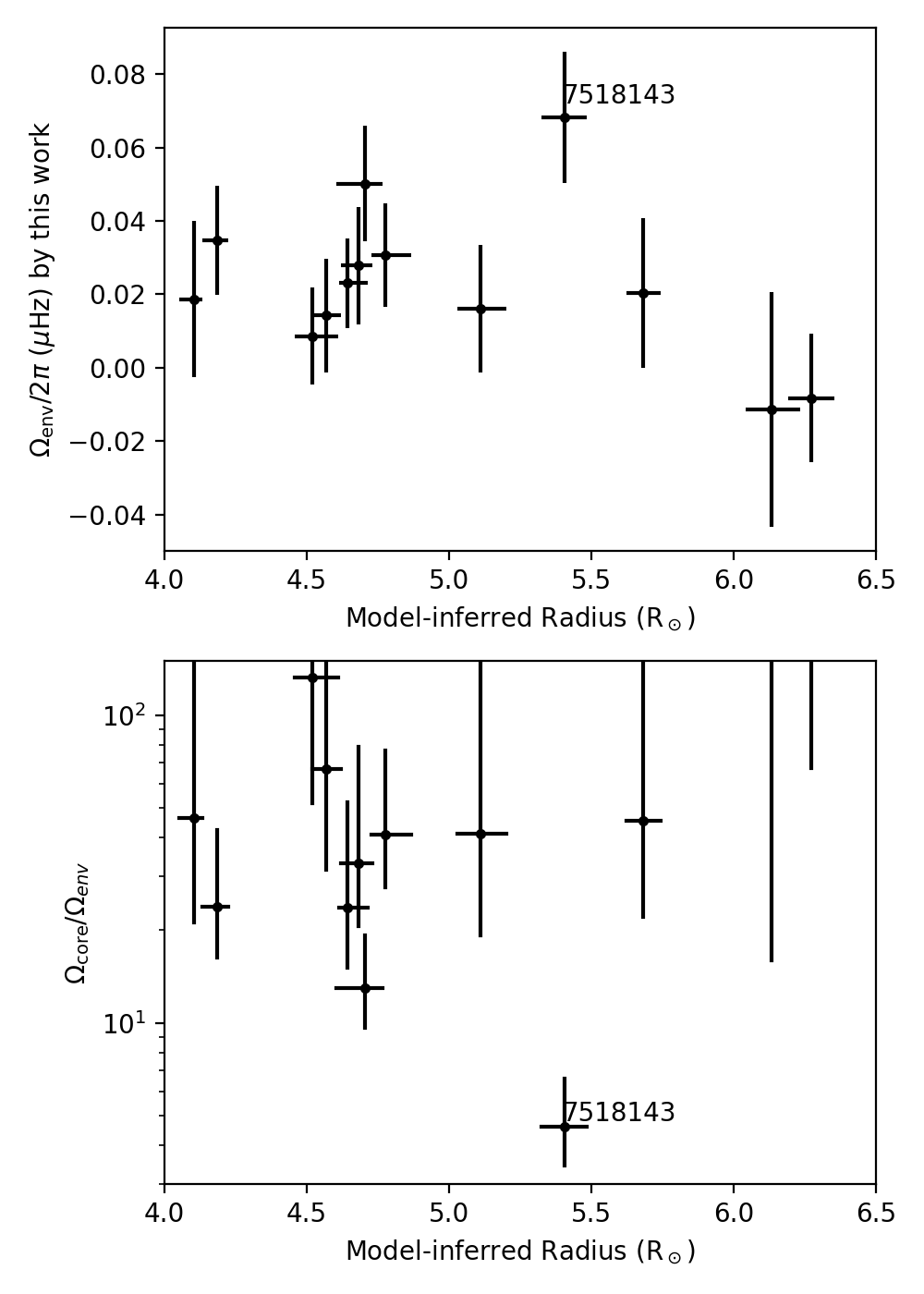}
    \caption{Top panel: surface rotation rates as a function of model-inferred radii. Bottom panel: %the differential rotations
    variations in the ratio $\Omega_\mathrm{core}/\Omega_\mathrm{env}$ with stellar radii.}
    \label{fig:surface_rotation_vs_radii}
\end{figure}

The top panel of fig.~\ref{fig:surface_rotation_vs_radii} displays the measurements of the envelope rotation rates as a function of their model-inferred radii. The method used in section~\ref{sec:measure_field_strength} does not require the use of the $\zeta$ function, which is different from the method \seb{used} by \cite{LiGang_2022_nature} \seb{to measure the internal rotation.} Consequently, the envelope rotation rates obtained using the current method exhibit differences compared to the results reported by \cite{LiGang_2022_nature}. Due to the expansion of the envelopes, red giant stars show \seb{very} slow envelope rotations. Most of the stars have surface rotation rates around $0.02\,\mathrm{\mu Hz}$, equivalent to about $\sim 600\,\mathrm{d}$. KIC\,7518143 has the fastest surface rotation rate \seb{with a rotation period of} about 170 days. For the two stars KIC\,6936091 and KIC\,9589420, with relatively large radii inferred from the modelling (larger than six solar radii), the rotation rates are slower than the detection limit of the seismic signal, resulting in surface rotation rates that are %equal to 
\seb{compatible with} zero within $1\sigma$ ranges.

The bottom panel of fig.~\ref{fig:surface_rotation_vs_radii} shows the ratio between the core and the envelope rotations ($\Omega_\mathrm{core}/\Omega_\mathrm{env}$), which represents the level of differential rotation.
%\comment{(Two questions about this figure: the two stars with a radius of about $6\,R_\odot$ seem to have different radii in the top and the bottom panel, which is strange. Secondly, how can you obtain positive ratios for these stars knowing the inferred envelope rotation is negative? You should give lower limits of the ratio for these stars.) Gang's reply: I used different model radii, one is the radii from the median posterior, another is the radii from the max posterior. Now I use the first one for both panels. The upper limit for the two stars (radii larger than 6) is positive, I use that values and lower limit, as plotted in the figure. }. %However, 
For the two stars with near-zero envelope rotations (with radii larger than 6 solar radii), %we cannot determine the ratio. 
\seb{we can only obtain a lower limit to this ratio.} Among the remaining stars, we find that the rotation ratio %falls 
\seb{lies} between 10 and 100. Interestingly, KIC\,7518143, the star with the fastest envelope rotation, also shows a slow core-rotation rate, which leads to the rotation ratio $\Omega_\mathrm{core}/\Omega_\mathrm{env}$ is only around 4. This finding suggests that a strong transfer of angular momentum occurs inside the star, but the relation with magnetic field is unclear, as the measured field strength of this star is not the strongest.

%\subsection{Comparison between measured field strengths and critical field}
\subsection{Comparison with critical field strength} \label{subsec:evolution_of_the_fields}

\cite{Fuller2015} showed that \seb{magnetic fields that exceed a critical value $B_{\rm c}$} can \seb{prevent} the propagation of gravity waves. \seb{This phenomenon was invoked as a possible explanation for } the suppression of $l=1$ mixed modes in \seb{a fraction of} red giant stars \citep[see also in][]{Stello2016}. \seb{\cite{Fuller2015} suggested that when the g-mode cavity harbours a magnetic field stronger than $B_{\rm c}$, all the mode energy that reaches the core is dissipated, leading to modes that have a pure p-like behaviour. This interpretation was challenged by \cite{Mosser2017_dipole_modes}, who found that for red giants that show only partially-suppressed dipole modes, the modes still have a g-like character. This question is currently under debate. Even though we do not yet have a clear picture of how global oscillation modes are affected by a magnetic field exceeding $B_{\rm c}$, it is clear that it will have an impact. We thus compared the measured field strength to the critical field strength, which we computed using our best-fit stellar models. In practice, this critical field} varies as a function of radius. The minimum appears at the hydrogen-burning shell (HBS), as given by
\begin{equation}
    B_\mathrm{c, min} = \left(\frac{16\pi^4\mu_0 \rho_\mathrm{hbs} r_\mathrm{hbs}^2\nu_\mathrm{max}^4}{8N_\mathrm{hbs}^2} \right)^{0.5},
\end{equation}
where $\rho_\mathrm{hbs}$, $r_\mathrm{hbs}$, and $N_\mathrm{hbs}$ are the density, radius, and the Brunt-Väisälä frequency at the hydrogen-burning shell. Using our optimal stellar models, we computed $B_\mathrm{c, min}$ for the \starnumber stars of our sample. The obtained values ore listed in Table~\ref{tab:strength_table}.

%The bottom panel of Fig.~\ref{fig:field_strength} shows 
Using the values listed in Table~\ref{tab:strength_table}, we can calculate the ratio between the measured and critical field strengths as a function of $\Delta\Pi_1$. The ratios range from approximately 0.1 to 0.3. It is important to note that the critical field strength varies with radius, and we use the minimum value, which occurs at the hydrogen-burning shell and is consistent with the layer where we measure the field strength. Additionally, the measured field strength is the average over the weight 
%kernel 
function $K(r)$, which means that some contribution from the field strength inside the hydrogen-burning shell is also included \cite[see extended data Figure 1 in][]{LiGang_2022_nature}. Therefore, the reported ratios may not be representative of the fields ratio at the location of hydrogen-burning shell.

\begin{figure*}
    \centering
    \includegraphics[width=1\linewidth]{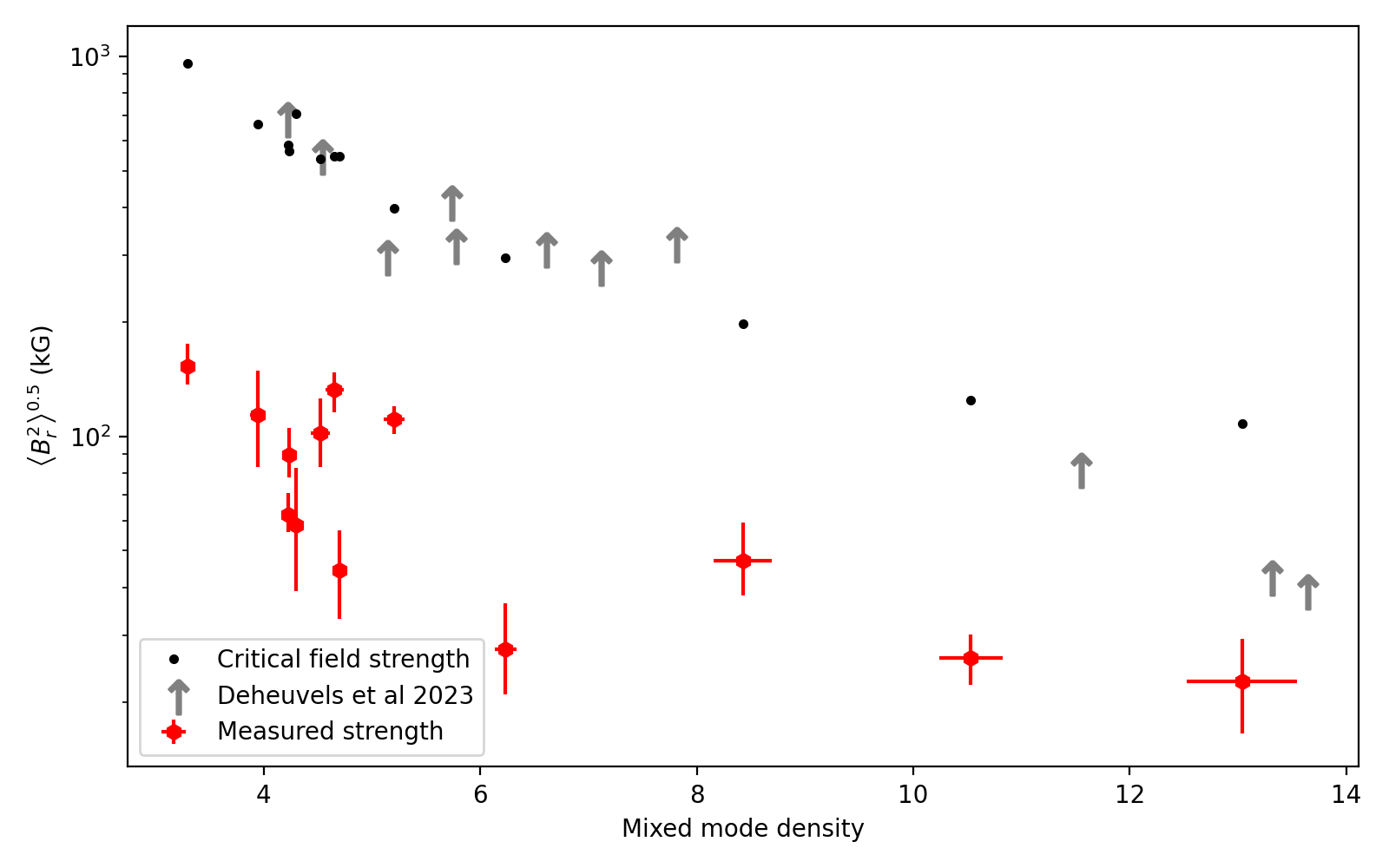}
    \caption{Field strength as a function of mixed mode density. The red hexagons with errorbars are the field strengths of the \starnumber stars reported by this work, and the black dots are their critical field strengths. The grey vertical arrows show the lower limits of the field strengths reported by \cite{Deheuvels2023}, where the stars do not show any asymmetric splittings, but only $m=0$ curved ridges in their stretched \'{e}chelle diagrams. }
    \label{fig:strength_vs_mixed_mode_density}
\end{figure*}

%\comment{moved from Sect. 4.3} We thereafter measure the field strengths of all the \starnumber stars and show the results in Fig.~\ref{fig:field_strength}. The top panel shows that the critical field strengths tend to decrease as $\Delta\Pi_1$ decreases \citep[also shown in ][]{Fuller2015}, indicating that an evolved star tends to have a smaller critical field strength. We also observe a correlation (albeit with relatively large scatter) between the measured field strength and $\Delta\Pi_1$, as shown by the black dots in the top panel of Fig.\ref{fig:field_strength}. This is in contrast to the theoretical prediction that field strength should increase with evolution, assuming the field flux is conserved while the core contracts. However, since the field strength cannot exceed the critical field strength, we observe a decrease in the field strength with evolution rather than an increase. \comment{There is repetition between this paragraph and Sect. 4.6. You should decide where to put this content and avoid duplicating it.) Gang's reply: I think it is ok to say it twice since a reader would like to know the explanation immediately.}

Fig.~\ref{fig:strength_vs_mixed_mode_density} shows the observed and critical field strength with evolution, indicated by the mixed mode density $\mathcal{N}=\Delta\nu/(\nu_\mathrm{max}^2\Delta\Pi_1)$ \citep{Gehan2018}. 
\seb{It is evident from Fig.~\ref{fig:strength_vs_mixed_mode_density} that the critical field strengths tend to decrease as stars evolve along the red giant branch \citep[this was already shown by ][]{Fuller2015}.
We also observe an overall decrease in the observed field strengths with the evolution (albeit with relatively large scatter).}
%We find a clear decrease trend in both the critical and observed field strengths. The decrease in observed field strength may seem counter-intuitive since magnetic flux is conserved, and as the core contracts with evolution, the field strength should increase. 
\seb{This is in contrast to the theoretical prediction that the field strength should increase with evolution, assuming the magnetic flux is conserved while the core contracts.}
However, if the field strength exceeds the critical field strength, the energy of gravity waves is \seb{thought to be} completely transferred to Alfv\'en waves, and we cannot observe any dipole mixed modes \citep{Fuller2015} %\comment{(repetition with what was said in Sect. 4.3) Gang's reply: I think it is ok to say it again so that reader can find answer easily and directly.}. 
Moreover, a strong field leads to a curvature in the stretched \'{e}chelle diagram \citep{LiGang_2022_nature, Bugnet2022, Deheuvels2023}, which could obscure the observations. Therefore, only a field that is $\sim10\%$ to $\sim 30\%$ of the critical field strength can generate observable asymmetric splittings.
%, and in this case, the field decreases with evolution due to the constraint of critical field strength. 
This could explain why the decrease in the detected field strength with evolution seems to follow the decrease in the critical field $B_{\rm c}$ (Fig. \ref{fig:strength_vs_mixed_mode_density}). 

\cite{Deheuvels2023} reported 11 stars showing curved stretched \'{e}chelle diagrams caused by their central magnetic fields. In this case, only the lower limit of the field strengths are derived. We plot the results by the grey vertical arrows in Fig.~\ref{fig:strength_vs_mixed_mode_density}, and find that they also follow a decreasing trend with evolution but with much stronger field strengths. Figure~\ref{fig:strength_vs_mixed_mode_density} reveals a large gap of field strength between the stars exhibiting asymmetric splittings, as seen in this work, and those with curved stretched \'{e}chelle diagrams, as observed by \cite{Deheuvels2023}. The search for such power spectra may be hindered by an observational bias, as explained in Sect.~\ref{subsec:biases}.

\subsection{Origin of the fields}

\begin{figure}
    \centering
    \includegraphics[width=\linewidth]{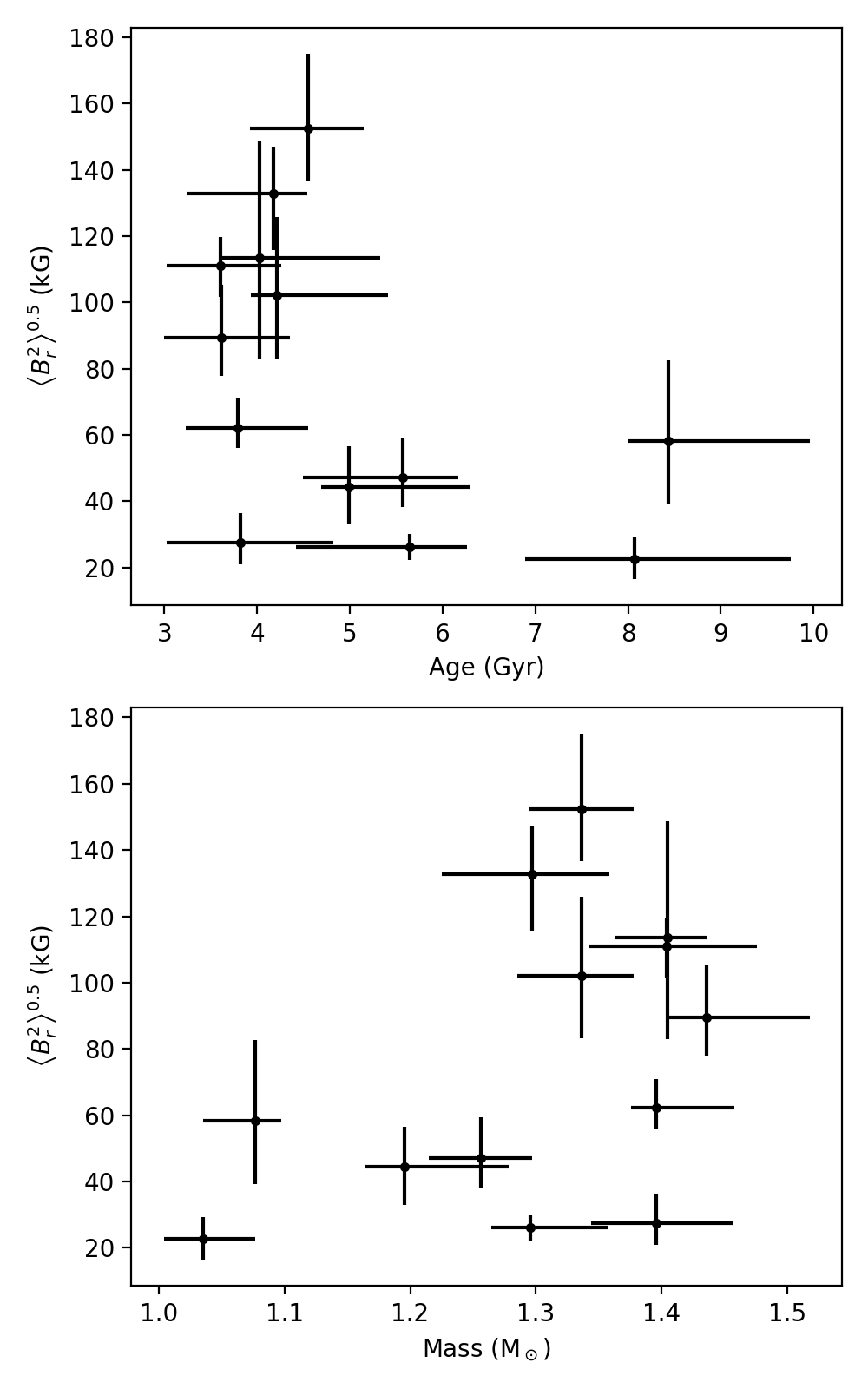}
    \caption{Top panel: field strengths with model-inferred ages. Bottom panel: field strengths but with model-inferred masses.}
    \label{fig:strength_vs_age_and_mass}
\end{figure}

Figure~\ref{fig:strength_vs_age_and_mass} depicts the measured field strengths plotted against model-inferred age and mass. In the top panel, most stars have ages smaller than 6\,Gyr and the field strengths of these stars show large scatter. However, for the two stars with longer ages (between 8 and 9\,Gyr), the field strengths are small. 
As stellar age is strongly related to mass, we examined the correlation between field strength and mass in the bottom panel of Fig.~\ref{fig:strength_vs_age_and_mass}. We find that most stars have masses larger than 1.3\,$\mathrm{M_\odot}$, meaning that they had convective cores in their main-sequence stages and can generate strong central fields by the dynamo processes. \seb{Since the Ohmic timescale, over which magnetic field dissipate, is longer than the evolution timescale (\citealt{cantiello16}), such fields could survive until the red-giant phase, where they may relax into stable mixed poloidal-toroidal configurations (\citealt{Braithwaite2004}).} 

However, we observed two stars with lower masses (1.08\,$\mathrm{M_\odot}$ for KIC\,4458118 and 1.04\,$\mathrm{M_\odot}$ for KIC\,6936091). 
%\textcolor{red}{Both the masses and metallicity} of the two stars suggest that they may not have had convective cores}
\seb{Our best-fit models for these stars had a radiative core} during \seb{the bulk of }their main sequence, 
%and their central fields may not have been 
\seb{which challenges the interpretation that their central fields may have been}
generated by the core dynamo.
%\comment{(There is still the question of the metallicity of these stars. You said you would show the atmospheric observables that were used to do the modelling, and in particular the metallicity, but I don't see them in the paper. I think this information is important and should be discussed here.) Gang's reply: I list [Fe/H] in Table 3.}.
To further study the origin of the field, we calculated the mass and the radius of the convective core at the beginning of stellar life. %Figure~\ref{fig:convective_core_size} gives an example from KIC\,6936091. 
For the two low-mass stars (KIC\,4458118 and 6936091), we found that they %still 
show convective cores with masses of $\sim$12\% of the total stellar masses at the very beginning of its evolution ($\sim30$\,Myr), owing to the burning of $^3\mathrm{He}$ and $^{12}\mathrm{C}$ outside of equilibrium \citep{Deheuvels2010}. However, the convective core was too small to reach \seb{the shell that is currently burning hydrogen} in red-giant phase (which is \seb{located at} about $\sim$20\% of the total stellar masses). 
%The minimum distance from the ancient convective core to the current hydrogen-burning shell is about $3\times10^{9}\,\mathrm{cm}$, containing $\sim 0.09\,\mathrm{M_\odot}$ masses of the star. 
\seb{Even though the hydrogen-burning shell in these two stars was never convective, }
we cannot rule out the dynamo origin
\seb{because the weight function $K(r)$ involved in the expression of the measured field has a contribution from the deeper layers of the star. If we assume that the magnetic field is confined to the layers that were convective at the beginning of the main sequence }
(i.e. the layers whose mass is smaller than 12\% of the total mass), we find that fields of $\sim$300\,kG for KIC\,4458118 and $\sim$200\,kG for KIC\,6936091 are needed to reproduce the observations. These values are several times larger than the values given in Table \ref{tab:strength_table}. However, we cannot exclude that such fields might result from a dynamo process in main-sequence cores because these field strengths  remain compatible with the amplitudes predicted by numerical simulations of core convection (\citealt{Brun2005ApJ}) and dynamo scaling laws (\citealt{Bugnet2021}). 
%\comment{(The estimates of the field strength that you obtain here are about one order of magnitude larger than the previous ones. Since the fields measured in the Nature paper were one order of magnitude smaller than those expected to result from core dynamo models, I think that these values are in fact well in line with this hypothesis, even for the lower-mass stars.) Gang's reply: that's great}
%However, we cannot rule out the dynamo origin, because the field generated by the convective core still can be included by the weighted function (eq.~\ref{eq:kernel}). \textbf{To estimate the contribution of the weighted function in the core area (e.g. the layers whose mass is smaller than 12\% of the total mass), we assume a constant magnetic field strength inside the 12\%-mass layers of the star, and find that the assumed uniform field strength should be several times larger than the current field strength ($\sim$300\,kG for KIC\,4458118 and $\sim$200\,kG for KIC\,6936091).}
In addition, the determination of convective core size still shows an uncertainty due to the poorly-understanding internal chemical mixing process \citep[e.g.][]{Johnston2021}. \seb{Another} possibility for the origin of the magnetic field is that it is inherited from a fossil field.
%since the hydrogen-burning shell is never convective. 

%\begin{figure}
%    \centering
%    \includegraphics[width=\linewidth]{./figures/KIC006936091_size_of_convective_cores.png}
%    \caption{Mass (top) and radius (bottom) of the convective core before the main sequence as a function of the age of KIC\,6936091, which has the smallest mass in our sample. The dashed lines mark the mass and radius of the hydrogen-burning shell at the current age, tracing back to the pre-main-sequence stage. The solid lines show the convective cores. }
%    \label{fig:convective_core_size}
%\end{figure}

\subsection{Ongoing dynamo or stable fields?}
\begin{figure}
    \centering
    \includegraphics[width=\linewidth]{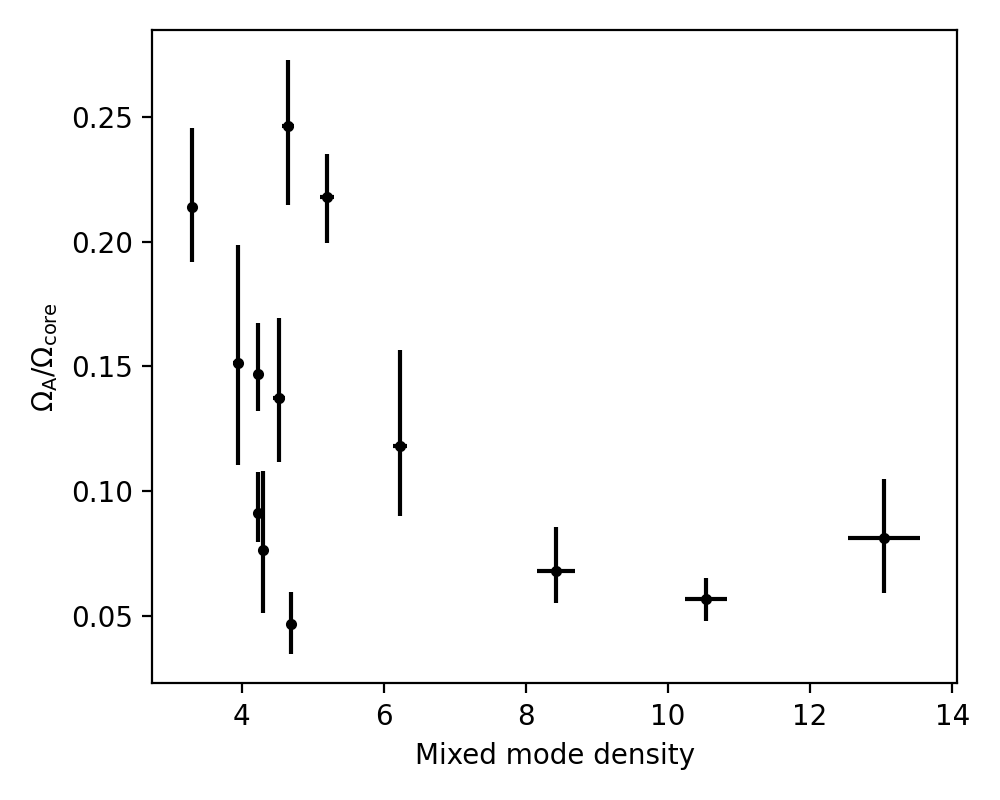}
    \caption{Ratio of the Alfv\`en frequency $\Omega_{\rm A} = \sqrt{\frac{<B_r^2>}{\mu_0 \rho r^2}}$
to the core rotation rate $\Omega_{\rm core}$ as a function of mixed mode density. The density $\rho$ and the radius $r$ have been computed at the HBS of the star models. }
    \label{fig:alfven_Omega_g_vs_mixed_mode_density}
\end{figure}

The ratio between the Alfv\`en frequency $\Omega_{\rm A} = \sqrt{\frac{<B_r^2>}{\mu_0 \rho r^2}}$
and the rotation rate $\Omega$ is
an important parameter of the dynamical interactions between magnetic field and rotation. Its value can provide some
clues about the nature of the detected magnetism. We computed this ratio at the HBS of the \starnumber red giants and, as shown in Fig. ~\ref{fig:alfven_Omega_g_vs_mixed_mode_density}, found
that it is comprised between $0.05$ to $0.25$.

A first consequence is that a Tayler-Spruit dynamo is probably not presently at work in these layers. Indeed,
the \cite{Fuller2019} model of the Tayler-Spruit dynamo finds that
$\Omega_{\rm A}/\Omega$ should scale with $(\Omega/N)^{5/3}$ where $N$, the Brunt-Vaisala
frequency, measures the strength of stable stratification in radiative zones.
As $(\Omega/N)^{5/3} \sim 10^{-7}$ at the HBS of a typical 1.2 M$_\sun$ and 4 R$_\sun$ star
at the base of the red giant branch \cite{Fuller2019},
the predicted $\Omega_{\rm A}/\Omega$ are extremely small and thus incompatible with our measurements.

The observed ratios are rather compatible with a stable magnetic field configuration.
Indeed, as anticipated in \cite{Spruit99}, numerical studies of a poloidal magnetic field embedded in a differentially rotating radiative zone
have shown that, if the ratio $\Omega_{\rm A}/ \Omega_{\rm core}$ exceeds a certain value, the field
is not affected by instabilities and evolves towards stable configurations while being only subjected to ohmic dissipation \citep{Jouve15,Jouve20,Gouhier2022aa}. The values obtained in numerical simulations
vary between $10^{-3}$ for a free differential rotation \citep{Jouve20} and $10^{-2}$ for a differential rotation forced by a radial flow simulating a star contraction
\citep{Gouhier2022aa}.
Another argument in favour of a stable magnetic field configuration comes from the magnetic fields observed at the surface of intermediate-mass and
massive main-sequence stars. In this mass range, the so-called fossil magnetic fields are stable at least over decades and their $\Omega_A/\Omega$ ratio, computed at the star's surface, is always greater than $\sim1$ \citep{Auriere2007}.

%Beyond the argument about the stability of the detected red giant internal fields, it is tempting to go further in the comparison with the magnetism of intermediate mass stars. In this mass range, the population of stable and strong (> 100 G) fossil magnetic fields has a low incidence (5-10\%), while another possibly widespread class of weakly magnetic stars has been revealed by the  detection of $\sim1$ G fields at the surface of a few very bright intermediate-mass stars \citep{Lignieres09, Blazere2016}. Extrapolated to red giant stars, a similar dichotomy between weak and strong fields could potentially explain the low incidence of the seismically detected fields. For intermediate-mass stars, it has been suggested that the dichotomy arises from a separation between high $\Omega_{\rm A}/\Omega$ stable fields and low $\Omega_{\rm A}/\Omega$ unstable fields as contraction forces differential rotation during the pre-main-sequence \citep{Auriere2007,Lignieres2014,Gouhier2022aa}. A similar process could be at work when the convective core dynamo fields go through the post-main-sequence contraction/expansion phases \citep{Gouhier2022aa}.

\subsection{Observational biases}\label{subsec:biases}

Our method of analysis of \kepler\ data presented in Sect. \ref{sec:method} leads to several observational biases that need to be acknowledged.

\subsubsection{Limitation on the measurable field strength \label{subsubsec:field_strength}}

\seb{As mentioned in Sect. \ref{sec:assumptions}, we have assumed that the magnetic field is not strong enough to significantly alter the regularity of g-mode period spacings. \cite{Deheuvels2023} estimated the magnetic intensity threshold $B_{\rm th}$ above which deviations from the regular period spacing of pure g modes become detectable (see their Appendix B). They found that $B_{\rm th}$ follows a similar decreasing trend with evolution as that of the critical field $B_{\rm c}$. Overall, a magnetic field that exceeds about 40\% of the critical field should produce significant deviations. We can thus expect that our method may start failing for field strengths above $0.4\,B_{\rm th}$. This could at least partly explain the gap that we found between the field measurements in this work and those obtained by \cite{Deheuvels2023}.}

\comment{Additionally, we assumed in Sect. \ref{subsubsec:m_identification} that the magnetic asymmetries are not large enough to push the $m=0$ component of the multiplets outside the interval formed by $m=\pm1$ components. This phenomenon occurs if $\nu_{\mathrm{g~mode}, m=0} < \nu_{\mathrm{g~mode}, m=+1}$ when $a>0$, and if $\nu_{\mathrm{g~mode}, m=0} > \nu_{\mathrm{g~mode}, m=-1}$ when $a<0$. Using Eq. \ref{eq:g_perturb_m_0} and \ref{eq:g_perturb_m_pm_1}, one finds that near $\nu_{\rm max}$, this condition is equivalent to
\begin{equation}
\delta\nu_{\rm g}\frac{3|a|}{2}\frac{4\pi} {\Omega_{\rm core}} > 1.
\label{eq:condition}
\end{equation}
For the \starnumber stars of our sample, the quantity on the left-hand-side of Eq. \ref{eq:condition} takes values between 0.03 and 0.36. It is not straightforward to translate the condition given by Eq. \ref{eq:condition} in terms of a constraint on the field intensity because it depends on properties that vary from star to star (namely $a$, $\mathcal{I}$, $\nu_{\rm max}$, and $\Omega_{\rm core}$). If we consider the values of these parameters that were obtained for the stars of our sample, we find that a change in the ordering of components in a multiplet would occur for these stars if the field strengths were multiplied by a factor ranging from 1.6 to 5.8. This corresponds to field intensities that are intermediate between the measured values and the critical field strengths. We thus conclude that our assumption that the $m=0$ component lies between the $m=\pm1$ components could partly explain the lack of red giants with measured fields closer to the critical field strength.}

% Firstly, the magnetic shift should not be too large so that the power spectra are still clear for identification. Secondly, the co-existence of asymmetric splittings and curved \'{e}chelle diagram in one star can make mode identifications complicated, potentially leading to discard some stars. Lastly, non-axisymmetric effect of magnetic fields may generates multiplets with more than three components, which also challenges the identification. Therefore, we only discover the stars with either asymmetric splittings, or a single curved ridge in the stretched \'{e}chelle diagram. 

\subsubsection{Non-axisymmetric effects}

Equations~\ref{eq:g_perturb_m_0} and \ref{eq:g_perturb_m_pm_1} hold when the non-axisymmetric effects of magnetic fields are negligible.
\comment{This is true (i) if $B_r^2$ is axisymmetric, but also (ii) if the ratio $b$ between the magnetic frequency and the rotational splitting ($b=4\pi\delta \nu_\mathrm{g}/\Omega_\mathrm{core}$) is smaller than $\sim 1$ \citep[see the supplementary material S2.7 in][]{LiGang_2022_nature}. Our results in Table~\ref{tab:g_para}  show that the ratio $b$ is much smaller than one. The stars with the largest values of $b$ are KIC\,8684542 ($b = 0.64\pm0.09$) and KIC\,11515377 ($b=0.56\pm0.12$), the other stars having $b$ values below $\sim0.35$. Hence, we state that case (ii) is valid for our stars, so that even if $B_r^2$ is non-axisymmetric, we still cannot see any significant non-axisymmetric effect in the oscillation spectra.}

As mentioned in Sect. \ref{sec:assumptions}, when $b\gtrsim 1$, non-axisymmetric magnetic fields can produce up to nine components for dipole modes (\citealt{LiGang_2022_nature}). Dedicated methods need to be devised and applied to search for such features in the oscillation spectra of red giants. This work is currently undertaken by our team and will be the subject of a next publication. Meanwhile, our assumption that non-axisymmetric effects are weak limits the strength of detectable fields. Similarly to Sect. \ref{subsubsec:field_strength}, the condition $b\gtrsim 1$ cannot directly be translated in terms of field strength. We thus estimated the minimal field strengths that the stars of our sample should have in order to produce significant non-axisymmetric effects. We found that the measured fields would need to be multiplied by a factor ranging from 1.2 to 5.4. Again, this corresponds to field intensities that lie between the measurements obtained in this study and those found by \cite{Deheuvels2023}. Taking non-axisymmetric effects into account could thus populate the gap observed in Fig. \ref{fig:strength_vs_mixed_mode_density}.

%, that is (i) if the ratio $b=4\pi\delta \nu_\mathrm{g}/\Omega_\mathrm{core}$ is smaller than $\sim 1$; (ii) or if $B_r^2$ is axisymmetric \citep[see the supplementary material S2.7 in][]{LiGang_2022_nature}. Currently, we cannot measure if $B_r^2$ is axisymmetric or not. However, our results in table~\ref{tab:g_para} indeed show that the ratio $b=4\pi\delta \nu_\mathrm{g}/\Omega_\mathrm{core}$ is much smaller than one. \textcolor{red}{The star that has the largest $b$ is KIC\,8684542 with $b = 0.64\pm0.09$. The star with the second largest $b$ is KIC\,11515377 with $b=0.56\pm0.12$, and the rest $b$ values are all below $\sim0.35$.} Hence, we state that case (i) is valid for our stars, and in this case even if $B_r^2$ is non-axisymmetric, we still can not see any non-axisymmetric effect in the power spectra.

%-------------------------------------- Two column figure (place early!)
  
%-----------------------------------------------------------------

\section{Conclusions} \label{sec:conclusion}
%Recently, \cite{LiGang_2022_nature} reported three red giant stars whose rotational splittings are asymmetric due to the central magnetic fields. The magnetism-induced asymmetries are predicted to be all positive or all negative, and should decrease with increasing frequency. The g-dominated mixed modes should show larger asymmetries since the field is in the stellar interior.

In this study, we conducted a systematic search for magnetism-induced asymmetric splittings in \kepler\ data. We successfully identified \starnumber stars (including the three stars previously reported by \citealt{LiGang_2022_nature}) exhibiting clear multiplet asymmetries with properties matching those expected in the presence of a core magnetic field. Notably, we found that only one star (KIC\,11515377) displayed negative asymmetries, while the remaining stars exhibited positive asymmetries. By fitting an asymptotic expression of mixed mode frequencies including rotational and magnetic effects to the observations, we were able to measure the magnetic frequency shift $\delta\nu_\mathrm{g}$ (which is related to the field strength), and the asymmetry parameter $a$ (which yields constraints on the field topology).
%By measuring the asymmetry parameter $a$, we were able to obtain constraints on the field topology, and we also determined the magnetic shift $\delta\nu_\mathrm{g}$, which allowed us to calculate the field strength.

%We collected the best-fitting stellar evolutionary models by using the radial mode frequencies, $\Delta\Pi_1$, effective temperature, luminosity, and metallicity. We calculated the core factor $\mathcal{I}$ (eq.~\ref{eq:core_factor_I}) and found that it shows a linear relation with $\Delta\Pi_1$, hence the magnetic field can be calculated with model independence in future work. 

Using the best-fitting stellar structure model, we were able to measure the average radial field strength in the core ($\langle B_r^2\rangle^{0.5}$) for the \starnumber stars. These field strengths were found to lie between approximately 20 and 150 kG, with maximal sensitivity in the vicinity of the hydrogen-burning shells. These values represent about 10\% to 30\% of the critical field strength above which gravity modes are no longer expected to propagate in the core (\citealt{Fuller2015}). 
%Additionally, we observed a decrease in field strength with evolution.

We also obtained estimates of the asymmetry parameter $a$, which provides a horizontal average of $B_r^2$ weighted by the second-order Legendre polynomial (see Eq. \ref{eq_coef_a}). For the \starnumber~stars, we found values of $a$ between about $-0.2$ and $0.95$, nearly spanning the entire possible range for this parameter ($-0.5\leqslant a\leqslant 1$). We recall that large negative values of $a$ are reached for fields that are concentrated near the equator, while large positive values of $a$ correspond to fields concentrated on the poles. Our results thus show that the core fields of red giants have various horizontal geometries. For some stars, we find values of $a$ that significantly exceed 0.4, which is the highest value that can be obtained with a dipolar magnetic field. This means that for these stars, the fields are more sharply concentrated on the poles than a pure dipolar field aligned with the rotation axis. The fact that a negative value of $a$ was obtained for only one star (KIC\,11515377) suggests that this configuration might be rare among red giants.

In agreement with the results of \cite{Deheuvels2023}, we found that the magnetic field strength in the core of red giants decreases with evolution (see Fig. \ref{fig:strength_vs_mixed_mode_density}). This is in contradiction with the general expectation that the contraction of the core should increase the magnetic field, if we assume a conservation of the magnetic flux. As was already pointed out in  \cite{Deheuvels2023}, the observed decrease seems to follow the overall decrease of the critical field strength $B_{\rm c}$ with evolution. One possible interpretation is thus that for a given star, the core magnetic field increases with evolution until it reaches $B_{\rm c}$. At this point, gravity waves no longer propagate in the core so that dipole modes do not have a mixed behaviour, and therefore core magnetic fields can no longer be detected. This could account for the observed decrease in the measured field strength, although further work is clearly needed to test this interpretation.

This work presents a larger sample of red giant stars with central magnetic fields, comprising \starnumber stars out of approximately 1200 red giant stars with triplets. The prevalence of such magnetic fields is thus currently found to be only 1\%. \comment{%much lower than 
\cite{Stello2016} claimed that about 5 to 15\% of red giants in the mass range of our sample are magnetised based on the assumption that suppressed dipole mixed modes are caused by magnetic fields exceeding the critical field strength. So far, this assumption remains challenged (\citealt{Mosser2017_dipole_modes}), but if it were correct, the stars of \cite{Stello2016} (which lie in a different range of field intensities compared to this work) could be added to the list of magnetic red giants.
%the previous study that claimed that 20\% of red giant stars are magnetised based on the suppression of their dipole mixed modes \citep{Fuller2015, Stello2016}. 
The low prevalence of magnetic giants in our sample may be partly due to observational biases related to the analysis method that we adopted in this study. We assumed that the field strength was not strong enough to significantly alter the regularity in the g-mode period spacings, modify the ordering of the components within dipole multiplets, or produce detectable effects related to the non-axisymmetric component of the magnetic field. We estimated that field intensities exceeding the measured field strengths by a factor of a few would be enough to make at least one of these assumptions invalid. Therefore, the present study might miss stars with stronger core fields. We also stress that in this study, magnetic giants have been identified by searching for multiplet asymmetries. This method can thus not detect magnetic fields with horizontal geometries that correspond to vanishing values of $a$.}
%due to the requirement that the field strength should be strong enough to generate significant asymmetric splittings, but not so strong as to destroy the stretched \'{e}chelle diagram or exceed the critical field strength, which is estimated to be within a range of no more than one order of magnitude.

\comment{Regarding the origin of the detected fields, one of the main scenarios is that they were produced by a dynamo in the main-sequence convective core. After the end of the main sequence, these fields would have relaxed into stable configurations (\citealt{Braithwaite2004}), undergoing only weak Ohmic diffusion (\citealt{cantiello16}). In this study, we found magnetic fields in two low-mass stars ($1.04_{-0.03}^{+0.04}\,M_\odot$ for KIC~6936091, and $1.08_{-0.04}^{+0.02}\,M_\odot$ for KIC~4458118), who had a small convective core only at the very beginning of the main sequence, owing to nuclear reactions outside of equilibrium. These convective cores never reached the layers where our magnetic field measurements have maximal sensitivity, namely the hydrogen-burning shell. Assuming that the core magnetic field is confined to the layers that were once convective, we found that field strengths of 200 to 300~kG need to be invoked. These intensities remain compatible with order-of-magnitude predictions of field strengths produced by dynamo in main-sequence convective cores (\citealt{Brun2005ApJ}, \citealt{Bugnet2021}). Another possible interpretation is that the detected fields might be inherited from fossil magnetic fields.}

Internal magnetic fields have been proposed as a candidate to provide additional transport of
%to carry 
angular momentum inside stars.
%very efficiently. 
%However, although the central fields are significant in these stars, they do not show any weird core rotation rates. 
\comment{In this study, we were also able to measure average core rotation rates for the \starnumber stars. We found values that are in line with the typical rotation rates of red giant cores, as obtained by \cite{Gehan2018}. This suggests that the stars of our sample do not undergo enhanced angular momentum redistribution compared to other red giants. This does not rule out magnetic fields as the origin of the angular momentum transport in red giants, as other red giants may harbour core magnetic fields that were not detected, either because of our observational biases or because they are below the detection threshold.}
%We also find the magnetic fields in two low-mass stars, whose convective cores never reach the observed layers in pre-main-sequence stages. Therefore, these fields might be inherited from fossil fields or an extension of inner dynamo-generated fields. 
The relatively high ratios between the Alfv\`en frequency and the rotation rate rule out that the observed fields are fed by an ongoing  Tayler-Spruit dynamo. They rather point towards fields that have settled into stable configurations.

This work constrains the strength and topology of the magnetic fields, hence it can be used in future numerical simulations and provides the possibility to further investigate the evolution of magnetic fields and their interaction with stellar rotation.

\begin{acknowledgements}
The authors acknowledge support from the project BEAMING ANR-18-CE31-0001 of
the French National Research Agency (ANR) and from the Centre
National d’Etudes Spatiales (CNES). 
Gang Li received funding from the KU Leuven Research Council under grant C16/18/005: PARADISE. TL acknowledges support from the Joint Research Fund in Astronomy (U2031203) under cooperative agreement between the National Natural Science Foundation of China (NSFC) and Chinese Academy of Sciences (CAS), NSFC grants (12090040, 12090042), and the European Research Council (ERC) under the European Union’s Horizon 2020 research and innovation programme (CartographY GA. 804752). 
This paper includes data collected by the Kepler mission and obtained from the MAST data archive at the Space Telescope Science Institute (STScI). Funding for the Kepler mission is provided by the NASA Science Mission Directorate. STScI is operated by the Association of Universities for Research in Astronomy, Inc., under NASA contract NAS 5–26555.
\end{acknowledgements}

   \bibliographystyle{aa} % style aa.bst
   \bibliography{main_paper_file} % your references Yourfile.bib

\begin{thebibliography}{95}
\expandafter\ifx\csname natexlab\endcsname\relax\def\natexlab#1{#1}\fi

\bibitem[{{Aerts} {et~al.}(2010){Aerts}, {Christensen-Dalsgaard}, \& {Kurtz}}]{Aerts2010}
{Aerts}, C., {Christensen-Dalsgaard}, J., \& {Kurtz}, D.~W. 2010, {Asteroseismology} (Springer, Dordrecht)

\bibitem[{{Anderson} {et~al.}(1990){Anderson}, {Duvall}, \& {Jefferies}}]{Anderson1990ApJ}
{Anderson}, E.~R., {Duvall}, Thomas~L., J., \& {Jefferies}, S.~M. 1990, \apj, 364, 699

\bibitem[{{Auri{\`e}re} {et~al.}(2015){Auri{\`e}re}, {Konstantinova-Antova}, {Charbonnel}, {Wade}, {Tsvetkova}, {Petit}, {Dintrans}, {Drake}, {Decressin}, {Lagarde}, {Donati}, {Roudier}, {Ligni{\`e}res}, {Schr{\"o}der}, {Landstreet}, {L{\`e}bre}, {Weiss}, \& {Zahn}}]{auriere15}
{Auri{\`e}re}, M., {Konstantinova-Antova}, R., {Charbonnel}, C., {et~al.} 2015, \aap, 574, A90

\bibitem[{{Auri{\`e}re} {et~al.}(2007){Auri{\`e}re}, {Wade}, {Silvester}, {Ligni{\`e}res}, {Bagnulo}, {Bale}, {Dintrans}, {Donati}, {Folsom}, {Gruberbauer}, {Hui Bon Hoa}, {Jeffers}, {Johnson}, {Landstreet}, {L{\`e}bre}, {Lueftinger}, {Marsden}, {Mouillet}, {Naseri}, {Paletou}, {Petit}, {Power}, {Rincon}, {Strasser}, \& {Toqu{\'e}}}]{Auriere2007}
{Auri{\`e}re}, M., {Wade}, G.~A., {Silvester}, J., {et~al.} 2007, \aap, 475, 1053

\bibitem[{{Beck} {et~al.}(2012){Beck}, {Montalban}, {Kallinger}, {De Ridder}, {Aerts}, {Garc{\'\i}a}, {Hekker}, {Dupret}, {Mosser}, {Eggenberger}, {Stello}, {Elsworth}, {Frandsen}, {Carrier}, {Hillen}, {Gruberbauer}, {Christensen-Dalsgaard}, {Miglio}, {Valentini}, {Bedding}, {Kjeldsen}, {Girouard}, {Hall}, \& {Ibrahim}}]{Beck2012Natur}
{Beck}, P.~G., {Montalban}, J., {Kallinger}, T., {et~al.} 2012, \nat, 481, 55

\bibitem[{{Bedding}(2014)}]{Bedding2014}
{Bedding}, T.~R. 2014, in Asteroseismology, ed. P.~L. {Pall{\'e}} \& C.~{Esteban}, 60

\bibitem[{{Bedding} {et~al.}(2010){Bedding}, {Huber}, {Stello}, {Elsworth}, {Hekker}, {Kallinger}, {Mathur}, {Mosser}, {Preston}, {Ballot}, {Barban}, {Broomhall}, {Buzasi}, {Chaplin}, {Garc{\'\i}a}, {Gruberbauer}, {Hale}, {De Ridder}, {Frandsen}, {Borucki}, {Brown}, {Christensen-Dalsgaard}, {Gilliland}, {Jenkins}, {Kjeldsen}, {Koch}, {Belkacem}, {Bildsten}, {Bruntt}, {Campante}, {Deheuvels}, {Derekas}, {Dupret}, {Goupil}, {Hatzes}, {Houdek}, {Ireland}, {Jiang}, {Karoff}, {Kiss}, {Lebreton}, {Miglio}, {Montalb{\'a}n}, {Noels}, {Roxburgh}, {Sangaralingam}, {Stevens}, {Suran}, {Tarrant}, \& {Weiss}}]{Bedding2010ApJ}
{Bedding}, T.~R., {Huber}, D., {Stello}, D., {et~al.} 2010, \apjl, 713, L176

\bibitem[{{Bedding} {et~al.}(2011){Bedding}, {Mosser}, {Huber}, {Montalb{\'a}n}, {Beck}, {Christensen-Dalsgaard}, {Elsworth}, {Garc{\'\i}a}, {Miglio}, {Stello}, {White}, {De Ridder}, {Hekker}, {Aerts}, {Barban}, {Belkacem}, {Broomhall}, {Brown}, {Buzasi}, {Carrier}, {Chaplin}, {di Mauro}, {Dupret}, {Frandsen}, {Gilliland}, {Goupil}, {Jenkins}, {Kallinger}, {Kawaler}, {Kjeldsen}, {Mathur}, {Noels}, {Silva Aguirre}, \& {Ventura}}]{Bedding2011}
{Bedding}, T.~R., {Mosser}, B., {Huber}, D., {et~al.} 2011, \nat, 471, 608

\bibitem[{{Belkacem} {et~al.}(2015){Belkacem}, {Marques}, {Goupil}, {Sonoi}, {Ouazzani}, {Dupret}, {Mathis}, {Mosser}, \& {Grosjean}}]{Belkacem2015}
{Belkacem}, K., {Marques}, J.~P., {Goupil}, M.~J., {et~al.} 2015, \aap, 579, A30

\bibitem[{{Benomar} {et~al.}(2015){Benomar}, {Takata}, {Shibahashi}, {Ceillier}, \& {Garc{\'\i}a}}]{Benomar2015}
{Benomar}, O., {Takata}, M., {Shibahashi}, H., {Ceillier}, T., \& {Garc{\'\i}a}, R.~A. 2015, \mnras, 452, 2654

\bibitem[{{Berger} {et~al.}(2020){Berger}, {Huber}, {van Saders}, {Gaidos}, {Tayar}, \& {Kraus}}]{Berger2020}
{Berger}, T.~A., {Huber}, D., {van Saders}, J.~L., {et~al.} 2020, \aj, 159, 280

\bibitem[{{Blaz{\`e}re} {et~al.}(2016){Blaz{\`e}re}, {Petit}, {Ligni{\`e}res}, {Auri{\`e}re}, {Ballot}, {B{\"o}hm}, {Folsom}, {Gaurat}, {Jouve}, {Lopez Ariste}, {Neiner}, \& {Wade}}]{Blazere2016}
{Blaz{\`e}re}, A., {Petit}, P., {Ligni{\`e}res}, F., {et~al.} 2016, \aap, 586, A97

\bibitem[{{Borucki} {et~al.}(2010){Borucki}, {Koch}, {Basri}, {Batalha}, {Brown}, {Caldwell}, {Caldwell}, {Christensen-Dalsgaard}, {Cochran}, {DeVore}, {Dunham}, {Dupree}, {Gautier}, {Geary}, {Gilliland}, {Gould}, {Howell}, {Jenkins}, {Kondo}, {Latham}, {Marcy}, {Meibom}, {Kjeldsen}, {Lissauer}, {Monet}, {Morrison}, {Sasselov}, {Tarter}, {Boss}, {Brownlee}, {Owen}, {Buzasi}, {Charbonneau}, {Doyle}, {Fortney}, {Ford}, {Holman}, {Seager}, {Steffen}, {Welsh}, {Rowe}, {Anderson}, {Buchhave}, {Ciardi}, {Walkowicz}, {Sherry}, {Horch}, {Isaacson}, {Everett}, {Fischer}, {Torres}, {Johnson}, {Endl}, {MacQueen}, {Bryson}, {Dotson}, {Haas}, {Kolodziejczak}, {Van Cleve}, {Chandrasekaran}, {Twicken}, {Quintana}, {Clarke}, {Allen}, {Li}, {Wu}, {Tenenbaum}, {Verner}, {Bruhweiler}, {Barnes}, \& {Prsa}}]{Borucki2010Sci}
{Borucki}, W.~J., {Koch}, D., {Basri}, G., {et~al.} 2010, Science, 327, 977

\bibitem[{{Braithwaite} \& {Spruit}(2004)}]{Braithwaite2004}
{Braithwaite}, J. \& {Spruit}, H.~C. 2004, \nat, 431, 819

\bibitem[{{Braithwaite} \& {Spruit}(2017)}]{Braithwaite2017}
{Braithwaite}, J. \& {Spruit}, H.~C. 2017, Royal Society Open Science, 4, 160271

\bibitem[{{Brun} {et~al.}(2005){Brun}, {Browning}, \& {Toomre}}]{Brun2005ApJ}
{Brun}, A.~S., {Browning}, M.~K., \& {Toomre}, J. 2005, \apj, 629, 461

\bibitem[{{Bugnet}(2022)}]{Bugnet2022}
{Bugnet}, L. 2022, arXiv e-prints, arXiv:2208.14954

\bibitem[{{Bugnet} {et~al.}(2021){Bugnet}, {Prat}, {Mathis}, {Astoul}, {Augustson}, {Garc{\'\i}a}, {Mathur}, {Amard}, \& {Neiner}}]{Bugnet2021}
{Bugnet}, L., {Prat}, V., {Mathis}, S., {et~al.} 2021, \aap, 650, A53

\bibitem[{{Cantiello} {et~al.}(2016){Cantiello}, {Fuller}, \& {Bildsten}}]{cantiello16}
{Cantiello}, M., {Fuller}, J., \& {Bildsten}, L. 2016, \apj, 824, 14

\bibitem[{{Cantiello} {et~al.}(2014){Cantiello}, {Mankovich}, {Bildsten}, {Christensen-Dalsgaard}, \& {Paxton}}]{Cantiello2014}
{Cantiello}, M., {Mankovich}, C., {Bildsten}, L., {Christensen-Dalsgaard}, J., \& {Paxton}, B. 2014, \apj, 788, 93

\bibitem[{{Ceillier} {et~al.}(2013){Ceillier}, {Eggenberger}, {Garc{\'\i}a}, \& {Mathis}}]{Ceillier2013}
{Ceillier}, T., {Eggenberger}, P., {Garc{\'\i}a}, R.~A., \& {Mathis}, S. 2013, \aap, 555, A54

\bibitem[{{Chontos} {et~al.}(2021){Chontos}, {Sayeed}, \& {Huber}}]{Chontos2021}
{Chontos}, A., {Sayeed}, M., \& {Huber}, D. 2021, in Posters from the TESS Science Conference II (TSC2), 189

\bibitem[{{Deheuvels} {et~al.}(2015){Deheuvels}, {Ballot}, {Beck}, {Mosser}, {{\O}stensen}, {Garc{\'\i}a}, \& {Goupil}}]{Deheuvels2015}
{Deheuvels}, S., {Ballot}, J., {Beck}, P.~G., {et~al.} 2015, \aap, 580, A96

\bibitem[{{Deheuvels} {et~al.}(2020){Deheuvels}, {Ballot}, {Eggenberger}, {Spada}, {Noll}, \& {den Hartogh}}]{Deheuvels2020}
{Deheuvels}, S., {Ballot}, J., {Eggenberger}, P., {et~al.} 2020, \aap, 641, A117

\bibitem[{{Deheuvels} {et~al.}(2022){Deheuvels}, {Ballot}, {Gehan}, \& {Mosser}}]{Deheuvels2022A&A}
{Deheuvels}, S., {Ballot}, J., {Gehan}, C., \& {Mosser}, B. 2022, \aap, 659, A106

\bibitem[{{Deheuvels} {et~al.}(2014){Deheuvels}, {Do{\u{g}}an}, {Goupil}, {Appourchaux}, {Benomar}, {Bruntt}, {Campante}, {Casagrande}, {Ceillier}, {Davies}, {De Cat}, {Fu}, {Garc{\'\i}a}, {Lobel}, {Mosser}, {Reese}, {Regulo}, {Schou}, {Stahn}, {Thygesen}, {Yang}, {Chaplin}, {Christensen-Dalsgaard}, {Eggenberger}, {Gizon}, {Mathis}, {Molenda-{\.Z}akowicz}, \& {Pinsonneault}}]{Deheuvels2014}
{Deheuvels}, S., {Do{\u{g}}an}, G., {Goupil}, M.~J., {et~al.} 2014, \aap, 564, A27

\bibitem[{{Deheuvels} {et~al.}(2023){Deheuvels}, {Li}, {Ballot}, \& {Ligni{\`e}res}}]{Deheuvels2023}
{Deheuvels}, S., {Li}, G., {Ballot}, J., \& {Ligni{\`e}res}, F. 2023, \aap, 670, L16

\bibitem[{{Deheuvels} {et~al.}(2010){Deheuvels}, {Michel}, {Goupil}, {Marques}, {Mosser}, {Dupret}, {Lebreton}, {Pichon}, \& {Morel}}]{Deheuvels2010}
{Deheuvels}, S., {Michel}, E., {Goupil}, M.~J., {et~al.} 2010, \aap, 514, A31

\bibitem[{{Deheuvels} {et~al.}(2017){Deheuvels}, {Ouazzani}, \& {Basu}}]{Deheuvels2017}
{Deheuvels}, S., {Ouazzani}, R.~M., \& {Basu}, S. 2017, \aap, 605, A75

\bibitem[{{Donati} \& {Landstreet}(2009)}]{Donati2009}
{Donati}, J.~F. \& {Landstreet}, J.~D. 2009, \araa, 47, 333

\bibitem[{{Eggenberger} {et~al.}(2019){Eggenberger}, {Deheuvels}, {Miglio}, {Ekstr{\"o}m}, {Georgy}, {Meynet}, {Lagarde}, {Salmon}, {Buldgen}, {Montalb{\'a}n}, {Spada}, \& {Ballot}}]{Eggenberger2019I}
{Eggenberger}, P., {Deheuvels}, S., {Miglio}, A., {et~al.} 2019, \aap, 621, A66

\bibitem[{{Eggenberger} {et~al.}(2017){Eggenberger}, {Lagarde}, {Miglio}, {Montalb{\'a}n}, {Ekstr{\"o}m}, {Georgy}, {Meynet}, {Salmon}, {Ceillier}, {Garc{\'\i}a}, {Mathis}, {Deheuvels}, {Maeder}, {den Hartogh}, \& {Hirschi}}]{Eggenberger2017}
{Eggenberger}, P., {Lagarde}, N., {Miglio}, A., {et~al.} 2017, \aap, 599, A18

\bibitem[{{Eggenberger} {et~al.}(2012){Eggenberger}, {Montalb{\'a}n}, \& {Miglio}}]{Eggenberger2012}
{Eggenberger}, P., {Montalb{\'a}n}, J., \& {Miglio}, A. 2012, \aap, 544, L4

\bibitem[{{Ferrario} {et~al.}(2020){Ferrario}, {Wickramasinghe}, \& {Kawka}}]{Ferrario2020}
{Ferrario}, L., {Wickramasinghe}, D., \& {Kawka}, A. 2020, Advances in Space Research, 66, 1025

\bibitem[{{Foreman-Mackey} {et~al.}(2013){Foreman-Mackey}, {Hogg}, {Lang}, \& {Goodman}}]{Foreman-Mackey2013PASP}
{Foreman-Mackey}, D., {Hogg}, D.~W., {Lang}, D., \& {Goodman}, J. 2013, \pasp, 125, 306

\bibitem[{{Fuller} {et~al.}(2015){Fuller}, {Cantiello}, {Stello}, {Garcia}, \& {Bildsten}}]{Fuller2015}
{Fuller}, J., {Cantiello}, M., {Stello}, D., {Garcia}, R.~A., \& {Bildsten}, L. 2015, Science, 350, 423

\bibitem[{{Fuller} {et~al.}(2014){Fuller}, {Lecoanet}, {Cantiello}, \& {Brown}}]{Fuller2014}
{Fuller}, J., {Lecoanet}, D., {Cantiello}, M., \& {Brown}, B. 2014, \apj, 796, 17

\bibitem[{{Fuller} {et~al.}(2019){Fuller}, {Piro}, \& {Jermyn}}]{Fuller2019}
{Fuller}, J., {Piro}, A.~L., \& {Jermyn}, A.~S. 2019, \mnras, 485, 3661

\bibitem[{{Garc{\'\i}a} {et~al.}(2014){Garc{\'\i}a}, {Ceillier}, {Salabert}, {Mathur}, {van Saders}, {Pinsonneault}, {Ballot}, {Beck}, {Bloemen}, {Campante}, {Davies}, {do Nascimento}, {Mathis}, {Metcalfe}, {Nielsen}, {Su{\'a}rez}, {Chaplin}, {Jim{\'e}nez}, \& {Karoff}}]{garcia14}
{Garc{\'\i}a}, R.~A., {Ceillier}, T., {Salabert}, D., {et~al.} 2014, \aap, 572, A34

\bibitem[{{Gehan} {et~al.}(2018){Gehan}, {Mosser}, {Michel}, {Samadi}, \& {Kallinger}}]{Gehan2018}
{Gehan}, C., {Mosser}, B., {Michel}, E., {Samadi}, R., \& {Kallinger}, T. 2018, \aap, 616, A24

\bibitem[{{Gizon} \& {Solanki}(2003)}]{Gizon2003}
{Gizon}, L. \& {Solanki}, S.~K. 2003, \apj, 589, 1009

\bibitem[{{Gomes} \& {Lopes}(2020)}]{Gomes2020}
{Gomes}, P. \& {Lopes}, I. 2020, \mnras, 496, 620

\bibitem[{{Gough} \& {Thompson}(1990)}]{Gough1990}
{Gough}, D.~O. \& {Thompson}, M.~J. 1990, \mnras, 242, 25

\bibitem[{{Gouhier} {et~al.}(2022){Gouhier}, {Jouve}, \& {Ligni{\`e}res}}]{Gouhier2022aa}
{Gouhier}, B., {Jouve}, L., \& {Ligni{\`e}res}, F. 2022, \aap, {}

\bibitem[{{Goupil} {et~al.}(2013){Goupil}, {Mosser}, {Marques}, {Ouazzani}, {Belkacem}, {Lebreton}, \& {Samadi}}]{Goupil13}
{Goupil}, M.~J., {Mosser}, B., {Marques}, J.~P., {et~al.} 2013, \aap, 549, A75

\bibitem[{{Grosjean} {et~al.}(2014){Grosjean}, {Dupret}, {Belkacem}, {Montalban}, {Samadi}, \& {Mosser}}]{Grosjean14}
{Grosjean}, M., {Dupret}, M.~A., {Belkacem}, K., {et~al.} 2014, \aap, 572, A11

\bibitem[{{Hasan} {et~al.}(2005){Hasan}, {Zahn}, \& {Christensen-Dalsgaard}}]{Hasan2005}
{Hasan}, S.~S., {Zahn}, J.~P., \& {Christensen-Dalsgaard}, J. 2005, \aap, 444, L29

\bibitem[{{Hekker} \& {Christensen-Dalsgaard}(2017)}]{hekker17}
{Hekker}, S. \& {Christensen-Dalsgaard}, J. 2017, \aapr, 25, 1

\bibitem[{{Hermes} {et~al.}(2017){Hermes}, {G{\"a}nsicke}, {Kawaler}, {Greiss}, {Tremblay}, {Gentile Fusillo}, {Raddi}, {Fanale}, {Bell}, {Dennihy}, {Fuchs}, {Dunlap}, {Clemens}, {Montgomery}, {Winget}, {Chote}, {Marsh}, \& {Redfield}}]{hermes17}
{Hermes}, J.~J., {G{\"a}nsicke}, B.~T., {Kawaler}, S.~D., {et~al.} 2017, \apjs, 232, 23

\bibitem[{{Huber} {et~al.}(2009){Huber}, {Stello}, {Bedding}, {Chaplin}, {Arentoft}, {Quirion}, \& {Kjeldsen}}]{Huber2009}
{Huber}, D., {Stello}, D., {Bedding}, T.~R., {et~al.} 2009, Communications in Asteroseismology, 160, 74

\bibitem[{{Johnston}(2021)}]{Johnston2021}
{Johnston}, C. 2021, \aap, 655, A29

\bibitem[{{Jouve} {et~al.}(2015){Jouve}, {Gastine}, \& {Ligni{\`e}res}}]{Jouve15}
{Jouve}, L., {Gastine}, T., \& {Ligni{\`e}res}, F. 2015, \aap, 575, A106

\bibitem[{{Jouve} {et~al.}(2020){Jouve}, {Ligni{\`e}res}, \& {Gaurat}}]{Jouve20}
{Jouve}, L., {Ligni{\`e}res}, F., \& {Gaurat}, M. 2020, \aap, 641, A13

\bibitem[{{Kjeldsen} \& {Bedding}(1995)}]{Kjeldsen1995}
{Kjeldsen}, H. \& {Bedding}, T.~R. 1995, \aap, 293, 87

\bibitem[{{Kurtz} {et~al.}(2014){Kurtz}, {Saio}, {Takata}, {Shibahashi}, {Murphy}, \& {Sekii}}]{Kurtz2014}
{Kurtz}, D.~W., {Saio}, H., {Takata}, M., {et~al.} 2014, \mnras, 444, 102

\bibitem[{{Kuszlewicz} {et~al.}(2023){Kuszlewicz}, {Hon}, \& {Huber}}]{kuszlewicz23}
{Kuszlewicz}, J.~S., {Hon}, M., \& {Huber}, D. 2023, arXiv e-prints, arXiv:2307.06482

\bibitem[{{Landstreet}(1992)}]{Landstreet1992}
{Landstreet}, J.~D. 1992, \aapr, 4, 35

\bibitem[{{Li} {et~al.}(2022{\natexlab{a}}){Li}, {Deheuvels}, {Ballot}, \& {Ligni{\`e}res}}]{LiGang_2022_nature}
{Li}, G., {Deheuvels}, S., {Ballot}, J., \& {Ligni{\`e}res}, F. 2022{\natexlab{a}}, \nat, 610, 43

\bibitem[{{Li} {et~al.}(2020){Li}, {Van Reeth}, {Bedding}, {Murphy}, {Antoci}, {Ouazzani}, \& {Barbara}}]{LiGang2020}
{Li}, G., {Van Reeth}, T., {Bedding}, T.~R., {et~al.} 2020, \mnras, 491, 3586

\bibitem[{{Li} {et~al.}(2022{\natexlab{b}}){Li}, {Li}, {Bi}, {Bedding}, {Davies}, \& {Du}}]{Li2022keplerRG}
{Li}, T., {Li}, Y., {Bi}, S., {et~al.} 2022{\natexlab{b}}, \apj, 927, 167

\bibitem[{{Li} {et~al.}(2022{\natexlab{c}}){Li}, {Bedding}, {Murphy}, {Stello}, {Chen}, {Huber}, {Joyce}, {Marks}, {Zhang}, {Bi}, {Colman}, {Hayden}, {Hey}, {Li}, {Montet}, {Sharma}, \& {Wu}}]{li22_mass_transfer}
{Li}, Y., {Bedding}, T.~R., {Murphy}, S.~J., {et~al.} 2022{\natexlab{c}}, Nature Astronomy, 6, 673

\bibitem[{{Ligni{\`e}res} {et~al.}(2009){Ligni{\`e}res}, {Petit}, {B{\"o}hm}, \& {Auri{\`e}re}}]{Lignieres09}
{Ligni{\`e}res}, F., {Petit}, P., {B{\"o}hm}, T., \& {Auri{\`e}re}, M. 2009, \aap, 500, L41

\bibitem[{{Loi}(2020)}]{loi20}
{Loi}, S.~T. 2020, \mnras, 493, 5726

\bibitem[{{Loi}(2021)}]{Loi2021}
{Loi}, S.~T. 2021, \mnras, 504, 3711

\bibitem[{{Lomb}(1976)}]{Lomb1976}
{Lomb}, N.~R. 1976, \apss, 39, 447

\bibitem[{{Lund} {et~al.}(2017){Lund}, {Silva Aguirre}, {Davies}, {Chaplin}, {Christensen-Dalsgaard}, {Houdek}, {White}, {Bedding}, {Ball}, {Huber}, {Antia}, {Lebreton}, {Latham}, {Handberg}, {Verma}, {Basu}, {Casagrande}, {Justesen}, {Kjeldsen}, \& {Mosumgaard}}]{Lund2017}
{Lund}, M.~N., {Silva Aguirre}, V., {Davies}, G.~R., {et~al.} 2017, \apj, 835, 172

\bibitem[{{Marques} {et~al.}(2013){Marques}, {Goupil}, {Lebreton}, {Talon}, {Palacios}, {Belkacem}, {Ouazzani}, {Mosser}, {Moya}, {Morel}, {Pichon}, {Mathis}, {Zahn}, {Turck-Chi{\`e}ze}, \& {Nghiem}}]{Marques2013}
{Marques}, J.~P., {Goupil}, M.~J., {Lebreton}, Y., {et~al.} 2013, \aap, 549, A74

\bibitem[{{Mathis} \& {Bugnet}(2023)}]{Mathis2023A&A}
{Mathis}, S. \& {Bugnet}, L. 2023, \aap, 676, L9

\bibitem[{{Mathis} {et~al.}(2021){Mathis}, {Bugnet}, {Prat}, {Augustson}, {Mathur}, \& {Garcia}}]{Mathis2021}
{Mathis}, S., {Bugnet}, L., {Prat}, V., {et~al.} 2021, \aap, 647, A122

\bibitem[{{Mosser} {et~al.}(2011){Mosser}, {Barban}, {Montalb{\'a}n}, {Beck}, {Miglio}, {Belkacem}, {Goupil}, {Hekker}, {De Ridder}, {Dupret}, {Elsworth}, {Noels}, {Baudin}, {Michel}, {Samadi}, {Auvergne}, {Baglin}, \& {Catala}}]{mosser11b}
{Mosser}, B., {Barban}, C., {Montalb{\'a}n}, J., {et~al.} 2011, \aap, 532, A86

\bibitem[{{Mosser} {et~al.}(2017{\natexlab{a}}){Mosser}, {Belkacem}, {Pin{\c{c}}on}, {Takata}, {Vrard}, {Barban}, {Goupil}, {Kallinger}, \& {Samadi}}]{Mosser2017_dipole_modes}
{Mosser}, B., {Belkacem}, K., {Pin{\c{c}}on}, C., {et~al.} 2017{\natexlab{a}}, \aap, 598, A62

\bibitem[{{Mosser} {et~al.}(2012{\natexlab{a}}){Mosser}, {Elsworth}, {Hekker}, {Huber}, {Kallinger}, {Mathur}, {Belkacem}, {Goupil}, {Samadi}, {Barban}, {Bedding}, {Chaplin}, {Garc{\'{\i}}a}, {Stello}, {De Ridder}, {Middour}, {Morris}, \& {Quintana}}]{mosser12c}
{Mosser}, B., {Elsworth}, Y., {Hekker}, S., {et~al.} 2012{\natexlab{a}}, \aap, 537, A30

\bibitem[{{Mosser} {et~al.}(2018){Mosser}, {Gehan}, {Belkacem}, {Samadi}, {Michel}, \& {Goupil}}]{mosser18}
{Mosser}, B., {Gehan}, C., {Belkacem}, K., {et~al.} 2018, \aap, 618, A109

\bibitem[{{Mosser} {et~al.}(2012{\natexlab{b}}){Mosser}, {Goupil}, {Belkacem}, {Marques}, {Beck}, {Bloemen}, {De Ridder}, {Barban}, {Deheuvels}, {Elsworth}, {Hekker}, {Kallinger}, {Ouazzani}, {Pinsonneault}, {Samadi}, {Stello}, {Garc{\'\i}a}, {Klaus}, {Li}, {Mathur}, \& {Morris}}]{Mosser2012}
{Mosser}, B., {Goupil}, M.~J., {Belkacem}, K., {et~al.} 2012{\natexlab{b}}, \aap, 548, A10

\bibitem[{{Mosser} {et~al.}(2017{\natexlab{b}}){Mosser}, {Pin{\c{c}}on}, {Belkacem}, {Takata}, \& {Vrard}}]{Mosser2017_q}
{Mosser}, B., {Pin{\c{c}}on}, C., {Belkacem}, K., {Takata}, M., \& {Vrard}, M. 2017{\natexlab{b}}, \aap, 600, A1

\bibitem[{{Mosser} {et~al.}(2015{\natexlab{a}}){Mosser}, {Vrard}, {Belkacem}, {Deheuvels}, \& {Goupil}}]{Mosser2015}
{Mosser}, B., {Vrard}, M., {Belkacem}, K., {Deheuvels}, S., \& {Goupil}, M.~J. 2015{\natexlab{a}}, \aap, 584, A50

\bibitem[{{Mosser} {et~al.}(2015{\natexlab{b}}){Mosser}, {Vrard}, {Belkacem}, {Deheuvels}, \& {Goupil}}]{mosser15}
{Mosser}, B., {Vrard}, M., {Belkacem}, K., {Deheuvels}, S., \& {Goupil}, M.~J. 2015{\natexlab{b}}, \aap, 584, A50

\bibitem[{{Ouazzani} {et~al.}(2019){Ouazzani}, {Marques}, {Goupil}, {Christophe}, {Antoci}, {Salmon}, \& {Ballot}}]{Ouazzani2019}
{Ouazzani}, R.~M., {Marques}, J.~P., {Goupil}, M.~J., {et~al.} 2019, \aap, 626, A121

\bibitem[{{Pin{\c{c}}on} {et~al.}(2016){Pin{\c{c}}on}, {Belkacem}, \& {Goupil}}]{Pincon2016}
{Pin{\c{c}}on}, C., {Belkacem}, K., \& {Goupil}, M.~J. 2016, \aap, 588, A122

\bibitem[{{R{\"u}diger} {et~al.}(2015){R{\"u}diger}, {Gellert}, {Spada}, \& {Tereshin}}]{Rudiger2015}
{R{\"u}diger}, G., {Gellert}, M., {Spada}, F., \& {Tereshin}, I. 2015, \aap, 573, A80

\bibitem[{{Rui} \& {Fuller}(2021)}]{Rui2021MNRAS}
{Rui}, N.~Z. \& {Fuller}, J. 2021, \mnras, 508, 1618

\bibitem[{{Rui} \& {Fuller}(2023)}]{Rui2023MNRAS}
{Rui}, N.~Z. \& {Fuller}, J. 2023, \mnras, 523, 582

\bibitem[{{Scargle}(1982)}]{Scargle1982}
{Scargle}, J.~D. 1982, \apj, 263, 835

\bibitem[{{Shibahashi}(1979)}]{shibahashi79}
{Shibahashi}, H. 1979, \pasj, 31, 87

\bibitem[{{Spada} {et~al.}(2016){Spada}, {Gellert}, {Arlt}, \& {Deheuvels}}]{Spada2016}
{Spada}, F., {Gellert}, M., {Arlt}, R., \& {Deheuvels}, S. 2016, \aap, 589, A23

\bibitem[{{Spruit}(1999)}]{Spruit99}
{Spruit}, H.~C. 1999, \aap, 349, 189

\bibitem[{{Stello} {et~al.}(2016){Stello}, {Cantiello}, {Fuller}, {Huber}, {Garc{\'\i}a}, {Bedding}, {Bildsten}, \& {Silva Aguirre}}]{Stello2016}
{Stello}, D., {Cantiello}, M., {Fuller}, J., {et~al.} 2016, \nat, 529, 364

\bibitem[{{Takata}(2016)}]{takata16}
{Takata}, M. 2016, \pasj, 68, 109

\bibitem[{{Triana} {et~al.}(2017){Triana}, {Corsaro}, {De Ridder}, {Bonanno}, {P{\'e}rez Hern{\'a}ndez}, \& {Garc{\'\i}a}}]{Triana2017}
{Triana}, S.~A., {Corsaro}, E., {De Ridder}, J., {et~al.} 2017, \aap, 602, A62

\bibitem[{{Unno} {et~al.}(1989){Unno}, {Osaki}, {Ando}, {Saio}, \& {Shibahashi}}]{Unno1989}
{Unno}, W., {Osaki}, Y., {Ando}, H., {Saio}, H., \& {Shibahashi}, H. 1989, {Nonradial oscillations of stars} (University of Tokyo Press, Tokyo)

\bibitem[{{Van Reeth} {et~al.}(2016){Van Reeth}, {Tkachenko}, \& {Aerts}}]{VanReeth2016}
{Van Reeth}, T., {Tkachenko}, A., \& {Aerts}, C. 2016, \aap, 593, A120

\bibitem[{{Vrard} {et~al.}(2016){Vrard}, {Mosser}, \& {Samadi}}]{Vrard2016}
{Vrard}, M., {Mosser}, B., \& {Samadi}, R. 2016, \aap, 588, A87

\bibitem[{{Wade} {et~al.}(2012){Wade}, {Grunhut}, \& {MiMeS Collaboration}}]{Wade2012}
{Wade}, G.~A., {Grunhut}, J.~H., \& {MiMeS Collaboration}. 2012, in Astronomical Society of the Pacific Conference Series, Vol. 464, Circumstellar Dynamics at High Resolution, ed. A.~C. {Carciofi} \& T.~{Rivinius}, 405

\bibitem[{{Yu} {et~al.}(2018){Yu}, {Huber}, {Bedding}, {Stello}, {Hon}, {Murphy}, \& {Khanna}}]{Yu2018}
{Yu}, J., {Huber}, D., {Bedding}, T.~R., {et~al.} 2018, \apjs, 236, 42

\bibitem[{{Zahn}(1992)}]{Zahn1992}
{Zahn}, J.~P. 1992, \aap, 265, 115

\end{thebibliography}

%The normal \'{e}chelle diagram of KIC 5696081. The frequencies are folded by the large separation $\Delta\nu = 17.9010\,\mu\mathrm{Hz}$. The modes which are used in the fitting procedure are marked by the colour symbols: green `+' for $m=1$, purple `$\bullet$' for m=0, and red `-' for m=-1 modes. The best-fitting frequencies are marked by the black `+', `$\bullet$', `-' symbols at the top of each panel.

\begin{appendix}

\section{\'{E}chelle diagram of KIC\,9467102}

We show the normal \'{e}chelle diagram of KIC\,9467102 as an example of serious splitting overlap discussed in section~\ref{subsubsec:splitting_identification}.

\begin{figure*}
    \centering
    \includegraphics[width=0.8\linewidth]{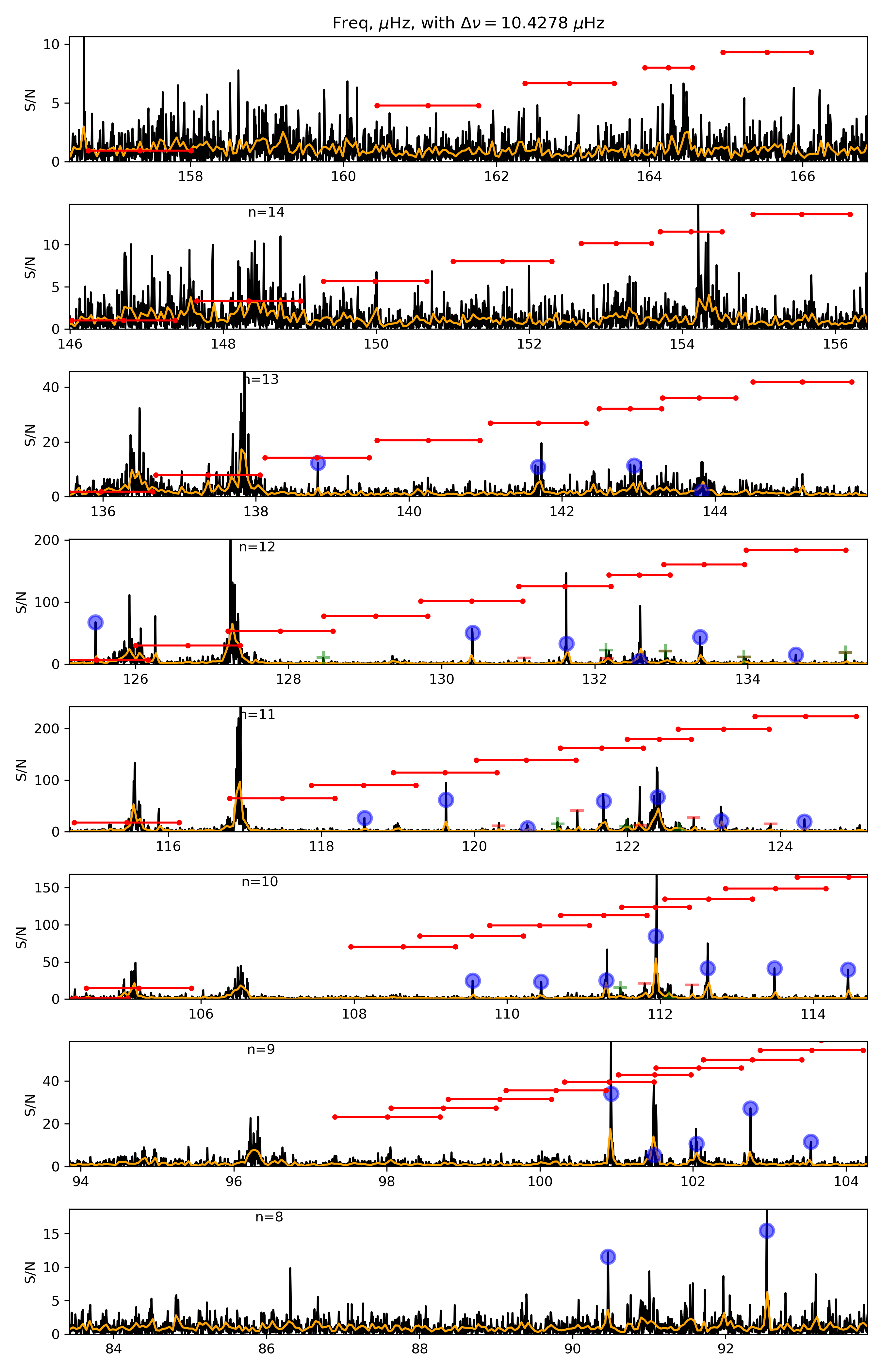}
    \caption{Normal \'{e}chelle diagram for KIC\,9467102. The frequencies are folded by the large separation $\Delta\nu = 10.4278\,\mu\mathrm{Hz}$. The modes that are used in the fitting procedure are marked by the colour symbols: green `+' for $m=1$, purple `$\bullet$' for m=0, and red `-' for m=-1 modes. The best-fitting results are shown by the horizontal red lines, which connect the components in each triplet. Note that the overlap is serious. }\label{apdxfig:echelle_diagram_9467102}
\end{figure*}

\section{All the diagrams} \label{appendixsection:all_figures}
We show the related diagrams of all the \starnumber stars. For each star, there are four diagrams shown by the following order: 
\begin{itemize}
    \item Normal \'{e}chelle diagram.
    \item Stretched \'{e}chelle diagram.
    \item Splitting asymmetry as a function of frequency.
    \item Corner diagram of the MCMC fit result.
\end{itemize}
%\begin{figure*}
%    \centering
%    \includegraphics[width = 0.8\linewidth]{figures/KIC005696081_magnetism_echelle.png}
%    \caption{The normal \'{e}chelle diagram of KIC 5696081. The frequencies are folded by the large separation $\Delta\nu = 16.4420\,\mu\mathrm{Hz}$. The modes which are used in the fitting procedure are marked by the colour symbols: green `+' for $m=1$, purple `$\bullet$' for m=0, and red `-' for m=-1 modes. The best-fitting frequencies are marked by the black `+', `$\bullet$', `-' symbols at the top of each panel. }
 %   \label{fig:echelle_diagram_5696081}
%\end{figure*}

%\iffalse

\begin{figure*}
\centering
\includegraphics[width = 0.8\linewidth]{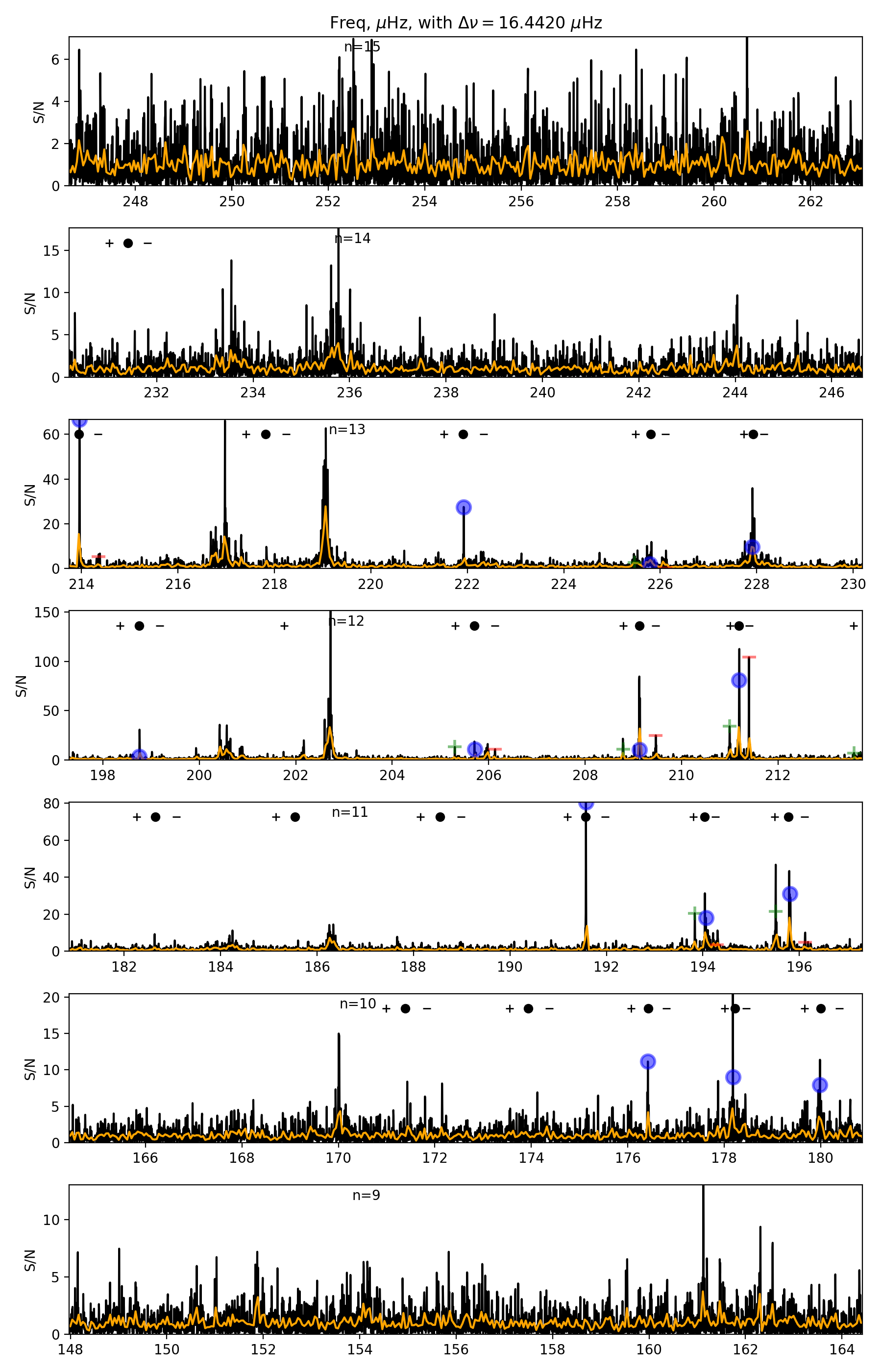}
\caption{Normal \'{e}chelle diagram of KIC\,4458118. The frequencies are folded by the large separation $\Delta\nu = 16.4420\,\mu\mathrm{Hz}$. The modes which are used in the fitting procedure are marked by the colour symbols: green `+' for $m=1$, purple `$\bullet$' for m=0, and red `-' for m=-1 modes. The best-fitting frequencies are marked by the black `+', `$\bullet$', `-' symbols at the top of each panel.}
\label{apdxfig:echelle_diagram_4458118}
\end{figure*}

\begin{figure*}
\centering
\includegraphics[width = 0.8\linewidth]{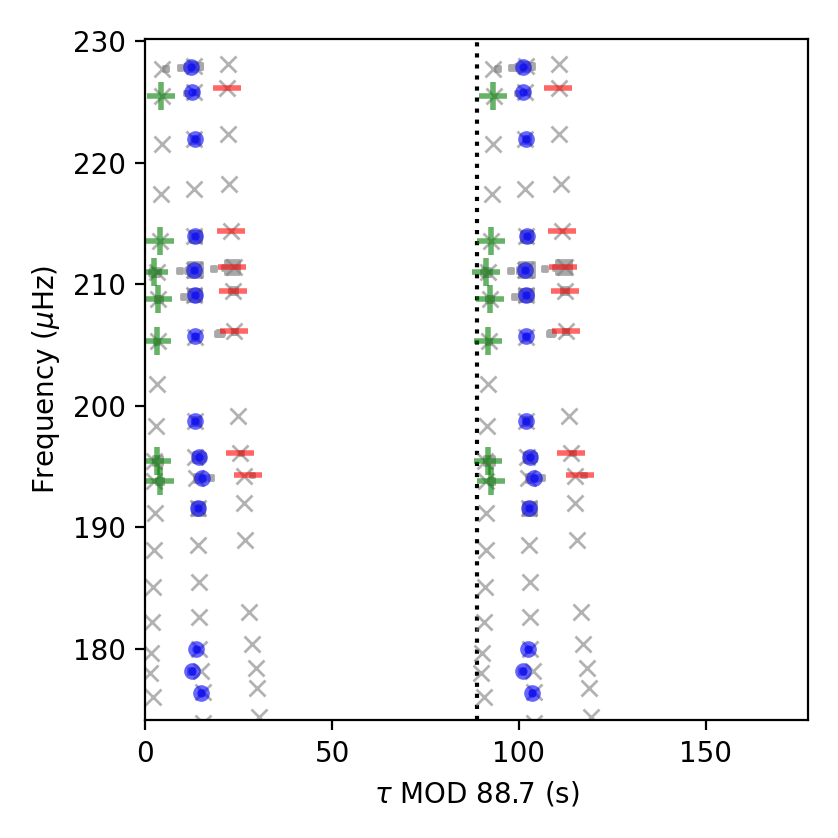}
\caption{Stretched \'{e}chelle diagram of KIC\,4458118. Symbols are explained in Fig.~\ref{fig:stretched_KIC5792889}. }
\label{apdxfig:echelle_diagram_stretched_4458118}
\end{figure*}

\begin{figure*}
\centering
\includegraphics[width = 0.6\linewidth]{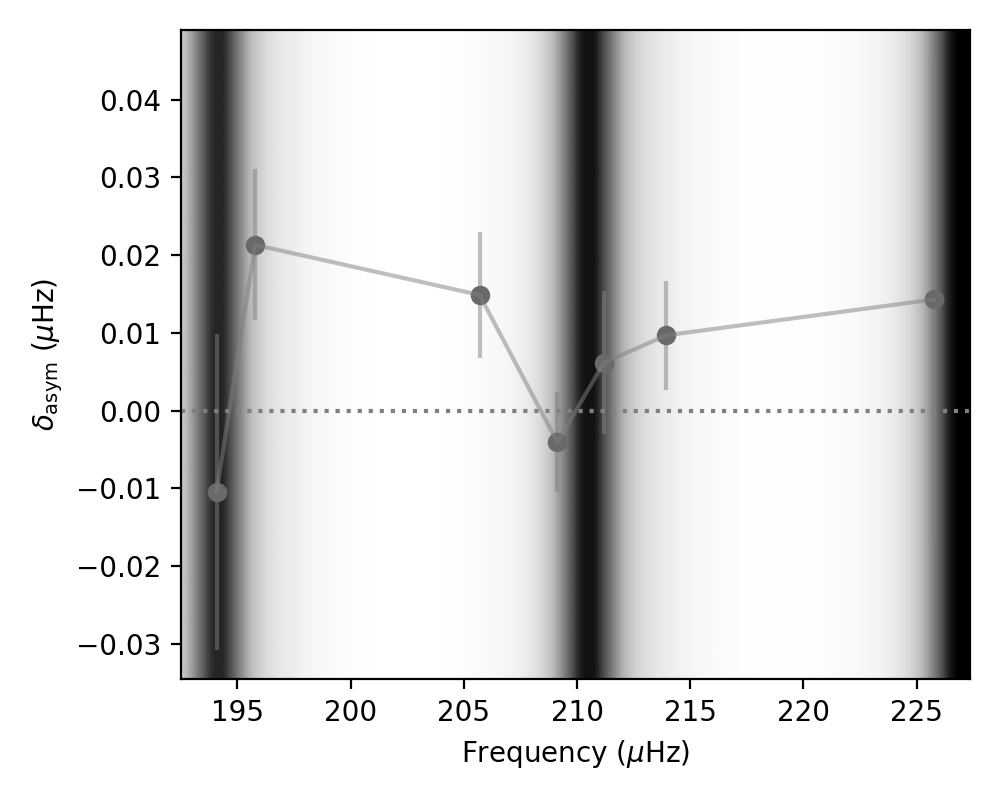}
\caption{Splitting asymmetries as a function of frequency of KIC\,4458118. Symbols are explained in Fig.~\ref{fig:KIC5696081_asymmtry_vs_freq}.}
\label{apdxfig:asymmetry_4458118}
\end{figure*}

\begin{figure*}
\centering
\includegraphics[width = 0.8\linewidth]{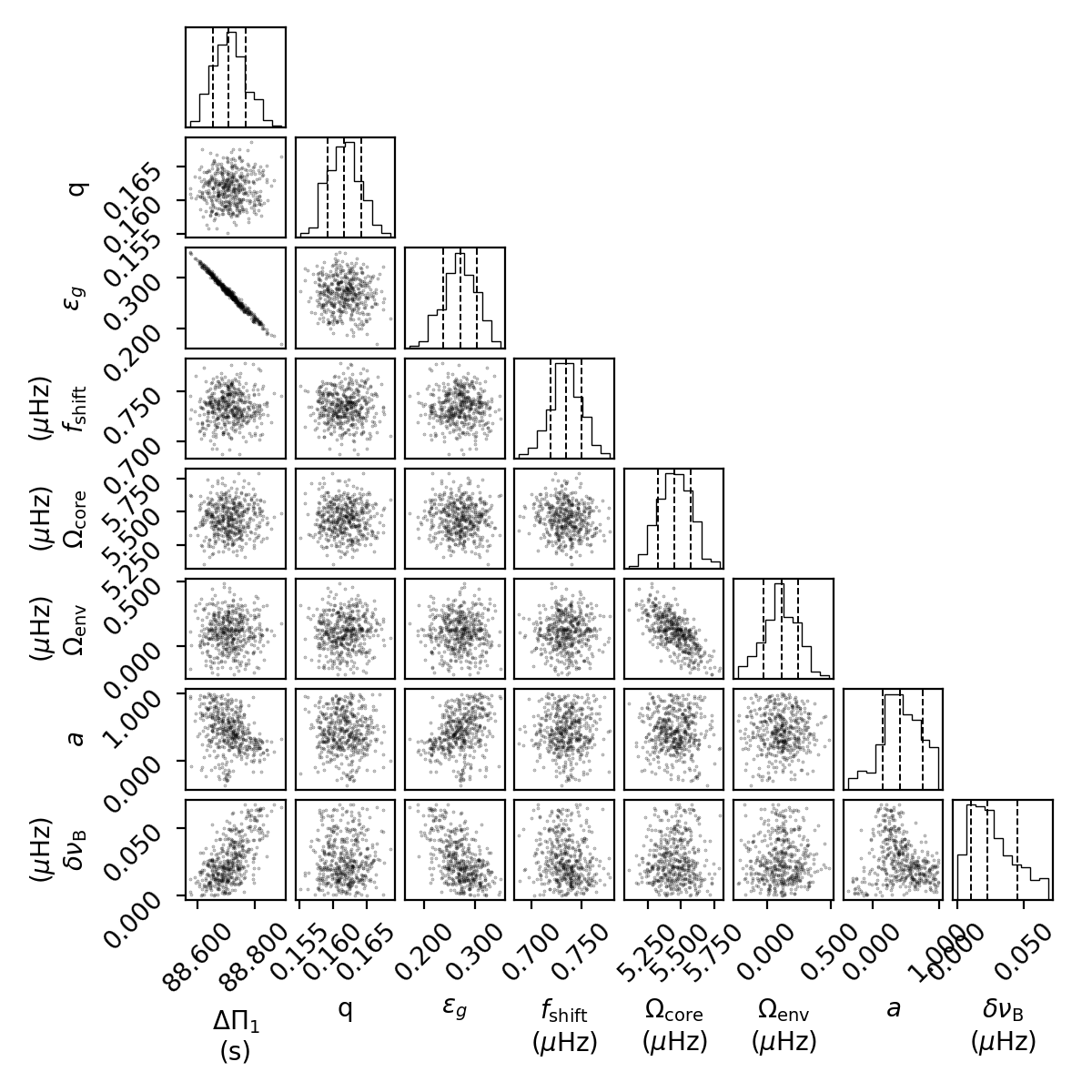}
\caption{Corner diagram of the MCMC fitting result of KIC\,4458118.}
\label{apdxfig:corner_4458118}
\end{figure*}

\begin{figure*}
\centering
\includegraphics[width = 0.8\linewidth]{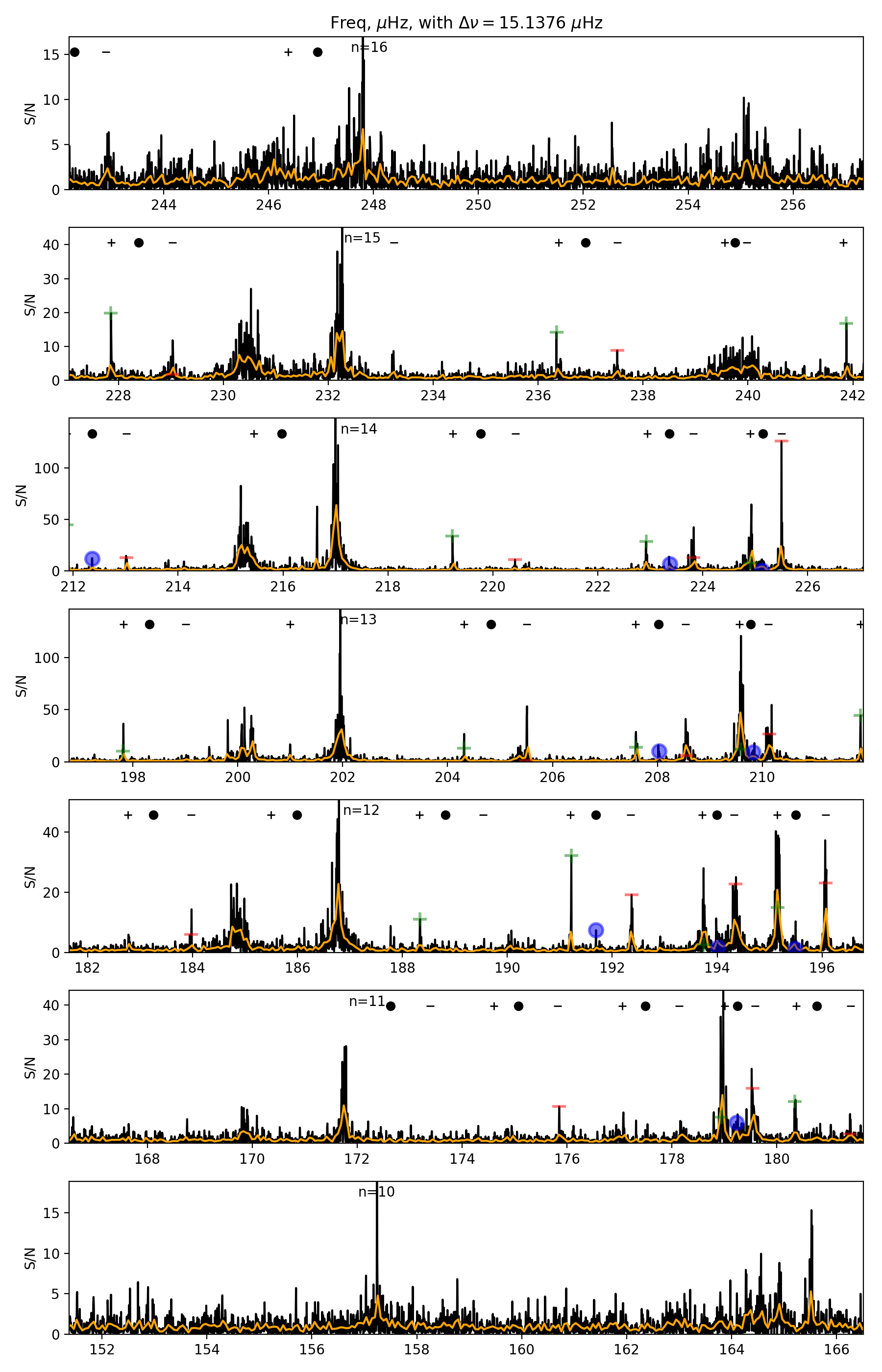}
\caption{Normal \'{e}chelle diagram of KIC\,5196300. See Fig.~\ref{apdxfig:echelle_diagram_4458118} for the explanations of the symbols.}
\label{apdxfig:echelle_diagram_5196300}
\end{figure*}

\begin{figure*}
\centering
\includegraphics[width = 0.6\linewidth]{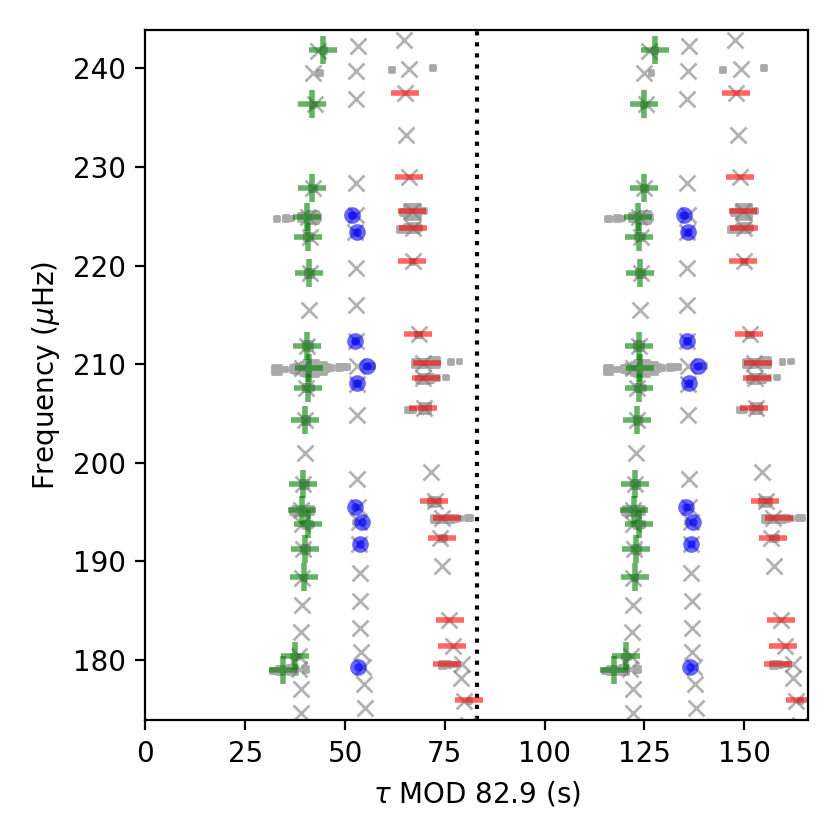}
\caption{Stretched \'{e}chelle diagram of KIC\,5196300. Symbols are explained in Fig.~\ref{fig:stretched_KIC5792889}. }
\label{apdxfig:echelle_diagram_stretched_5196300}
\end{figure*}

\begin{figure*}
\centering
\includegraphics[width = 0.8\linewidth]{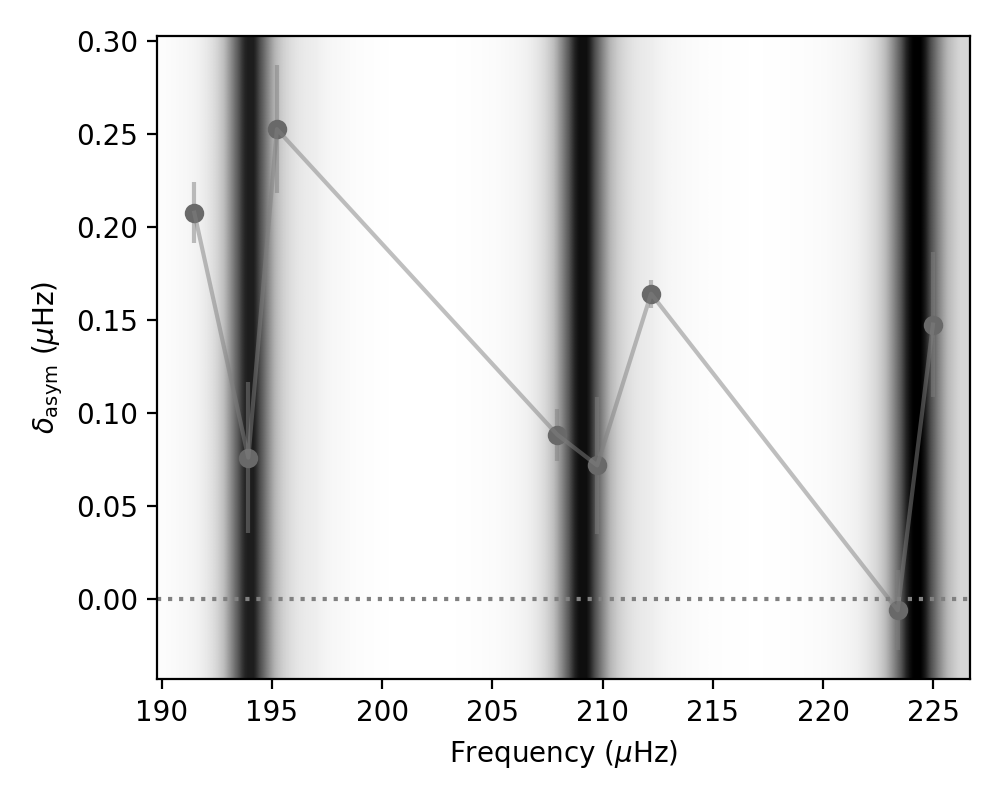}
\caption{Splitting asymmetries as a function of frequency of KIC\,5196300. Symbols are explained in Fig.~\ref{fig:KIC5696081_asymmtry_vs_freq}.}
\label{apdxfig:asymmetry_5196300}
\end{figure*}

\begin{figure*}
\centering
\includegraphics[width = 0.8\linewidth]{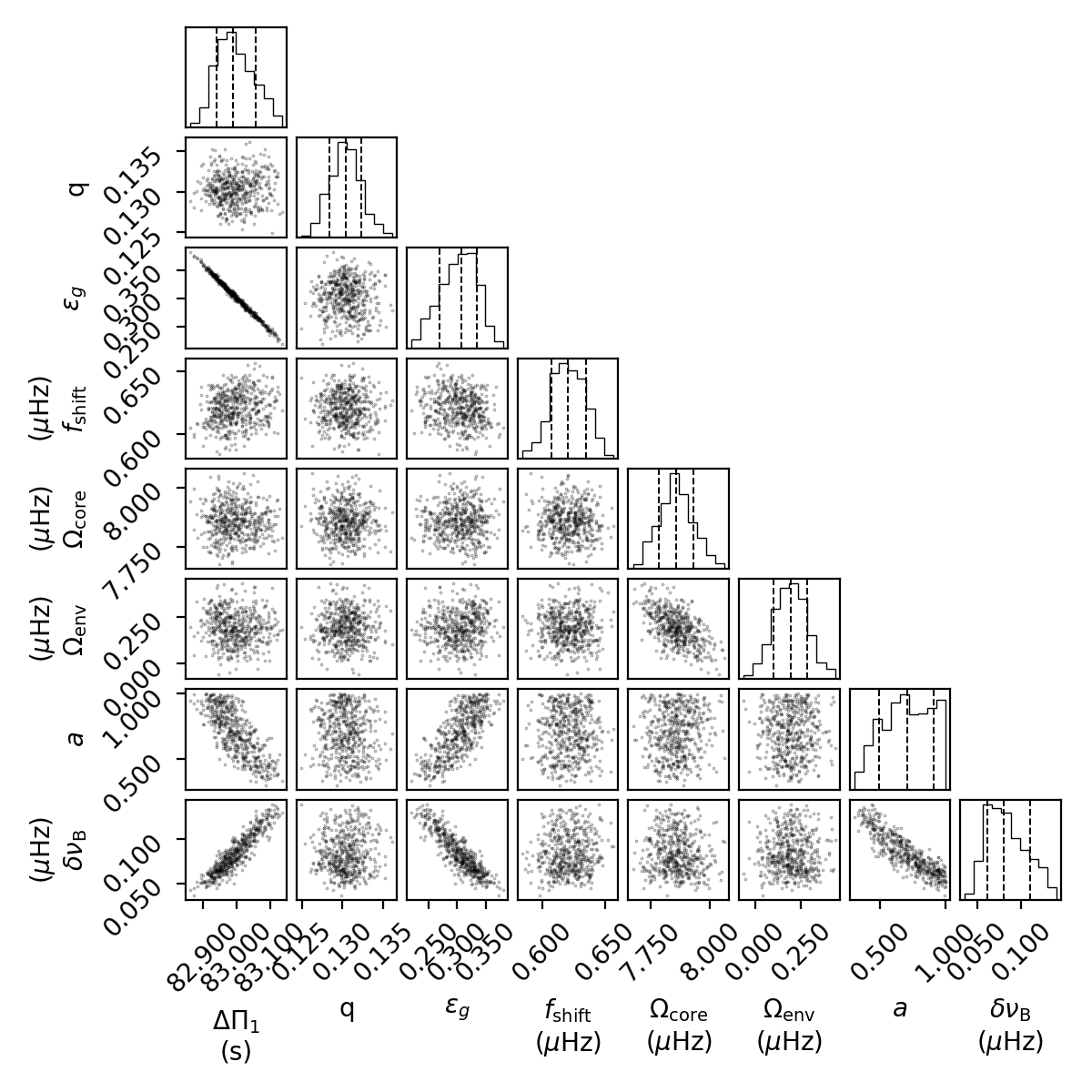}
\caption{Corner diagram of the MCMC fitting result of KIC\,5196300.}
\label{apdxfig:corner_5196300}
\end{figure*}

\begin{figure*}
\centering
\includegraphics[width = 0.8\linewidth]{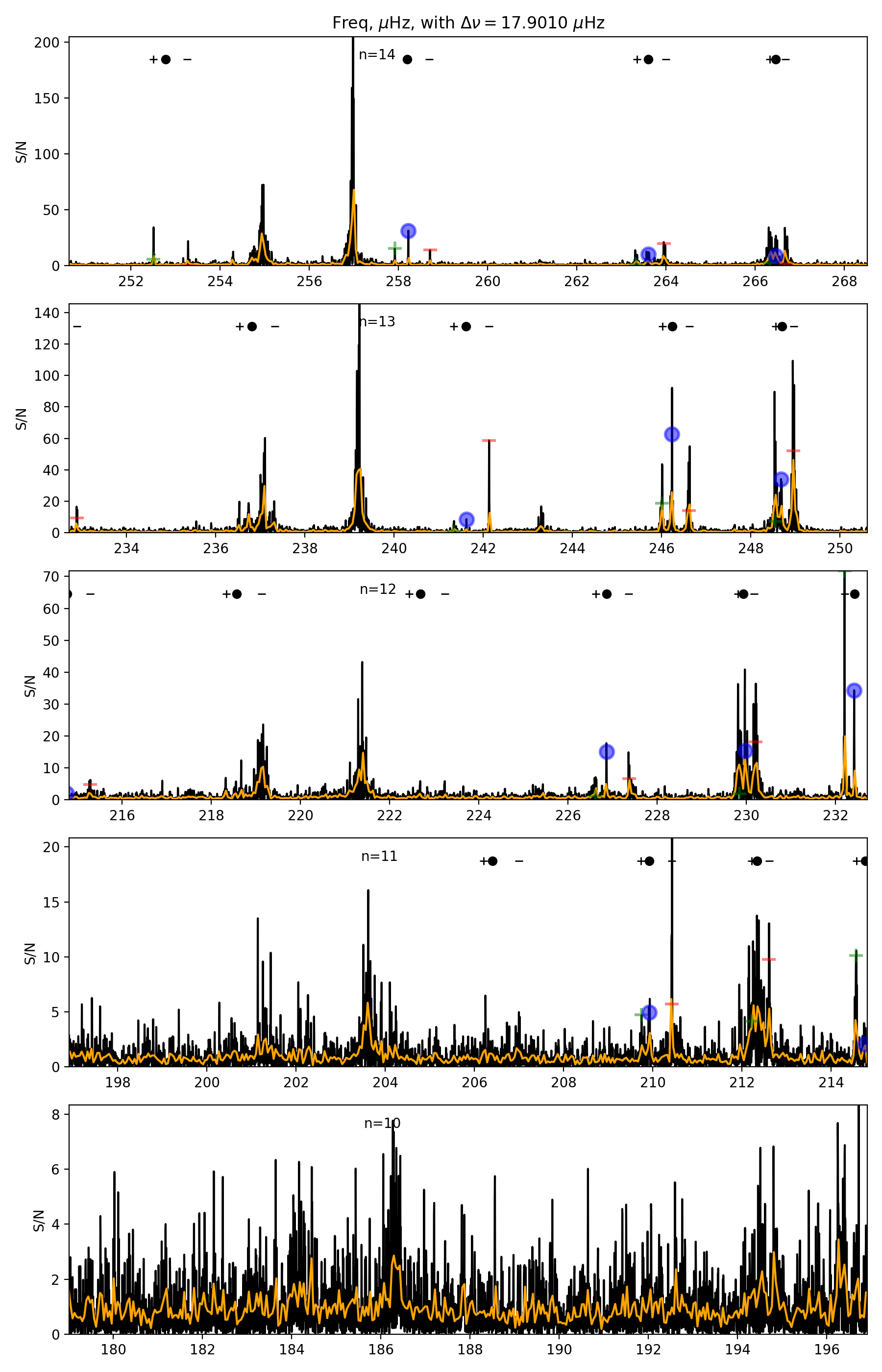}
\caption{Normal \'{e}chelle diagram of KIC\,5696081. See Fig.~\ref{apdxfig:echelle_diagram_4458118} for the explanations of the symbols.}
\label{apdxfig:echelle_diagram_5696081}
\end{figure*}

\begin{figure*}
\centering
\includegraphics[width = 0.6\linewidth]{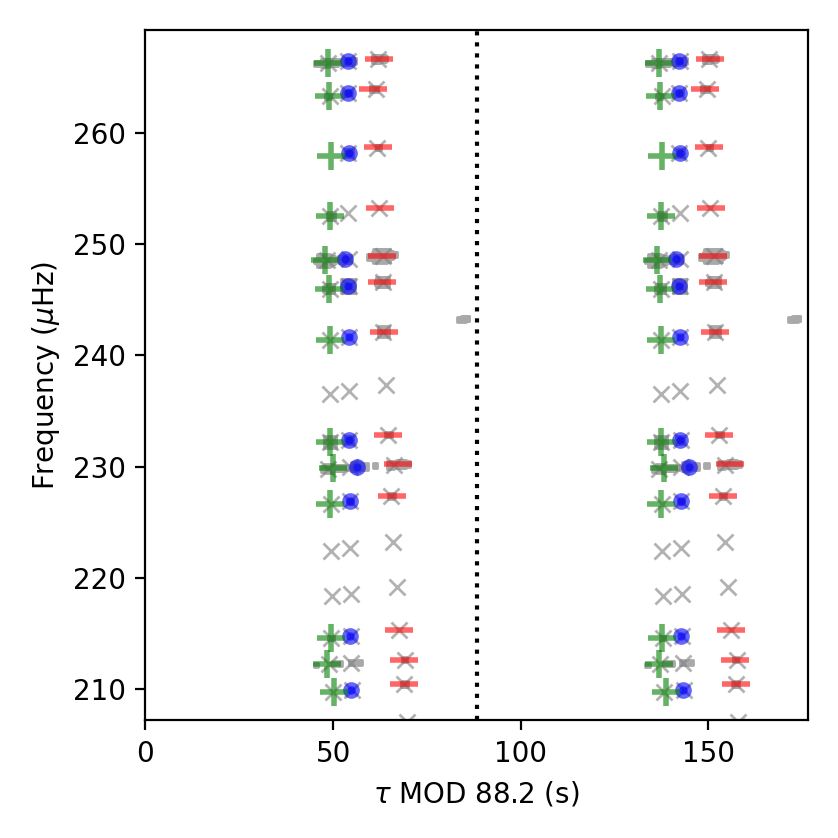}
\caption{Stretched \'{e}chelle diagram of KIC\,5696081. Symbols are explained in Fig.~\ref{fig:stretched_KIC5792889}. }
\label{apdxfig:echelle_diagram_stretched_5696081}
\end{figure*}

\begin{figure*}
\centering
\includegraphics[width = 0.8\linewidth]{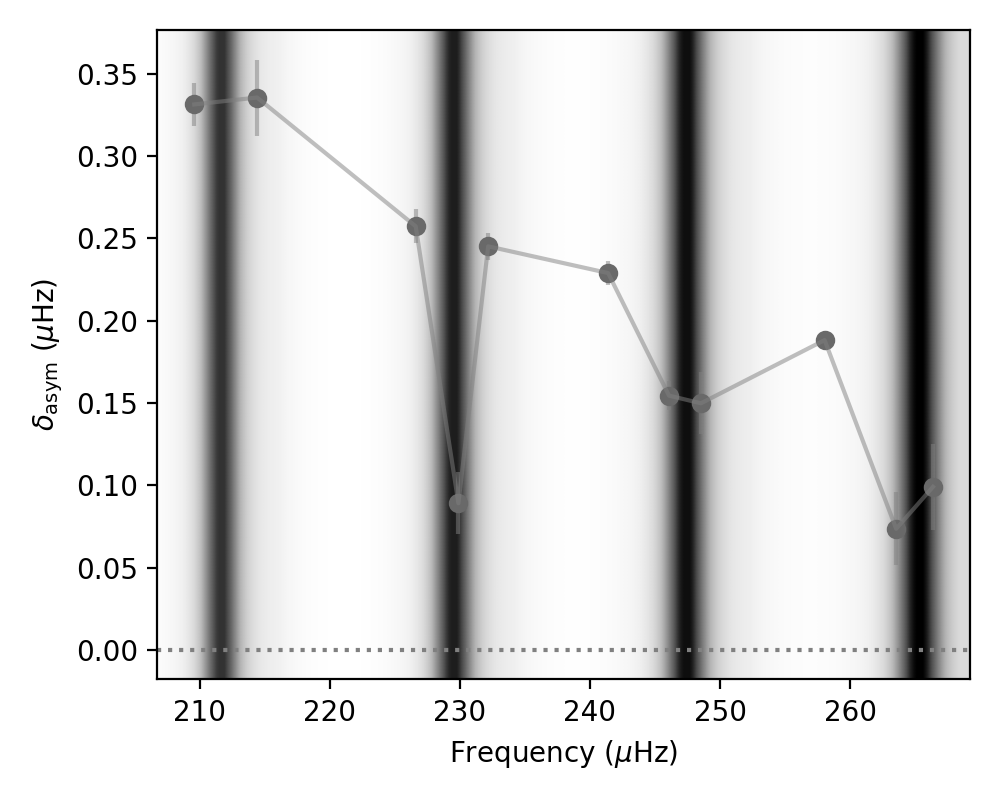}
\caption{Splitting asymmetries as a function of frequency of KIC\,5696081. Symbols are explained in Fig.~\ref{fig:KIC5696081_asymmtry_vs_freq}.}
\label{apdxfig:asymmetry_5696081}
\end{figure*}

\begin{figure*}
\centering
\includegraphics[width = 0.8\linewidth]{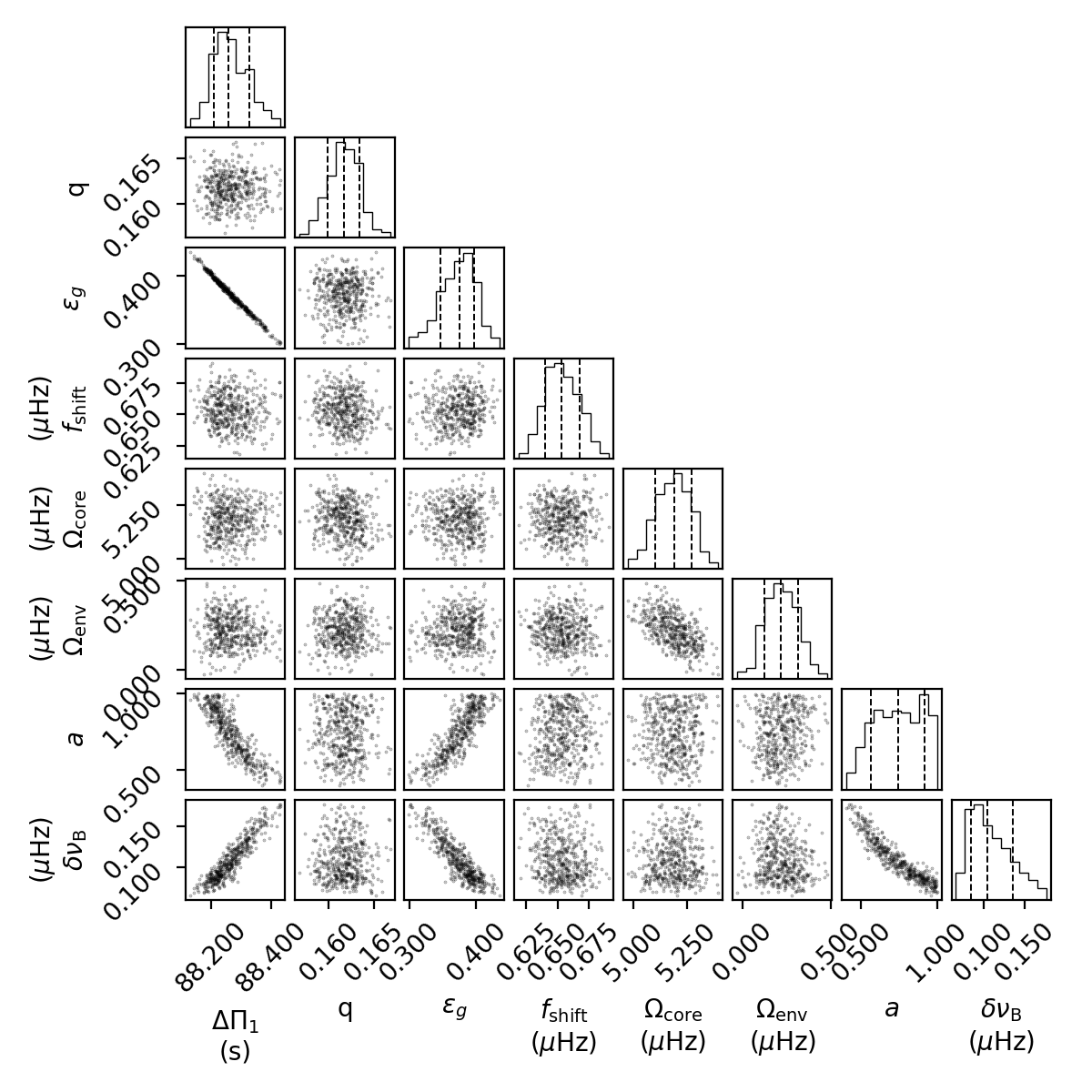}
\caption{Corner diagram of the MCMC fitting result of KIC\,5696081.}
\label{apdxfig:corner_5696081}
\end{figure*}

\begin{figure*}
\centering
\includegraphics[width = 0.8\linewidth]{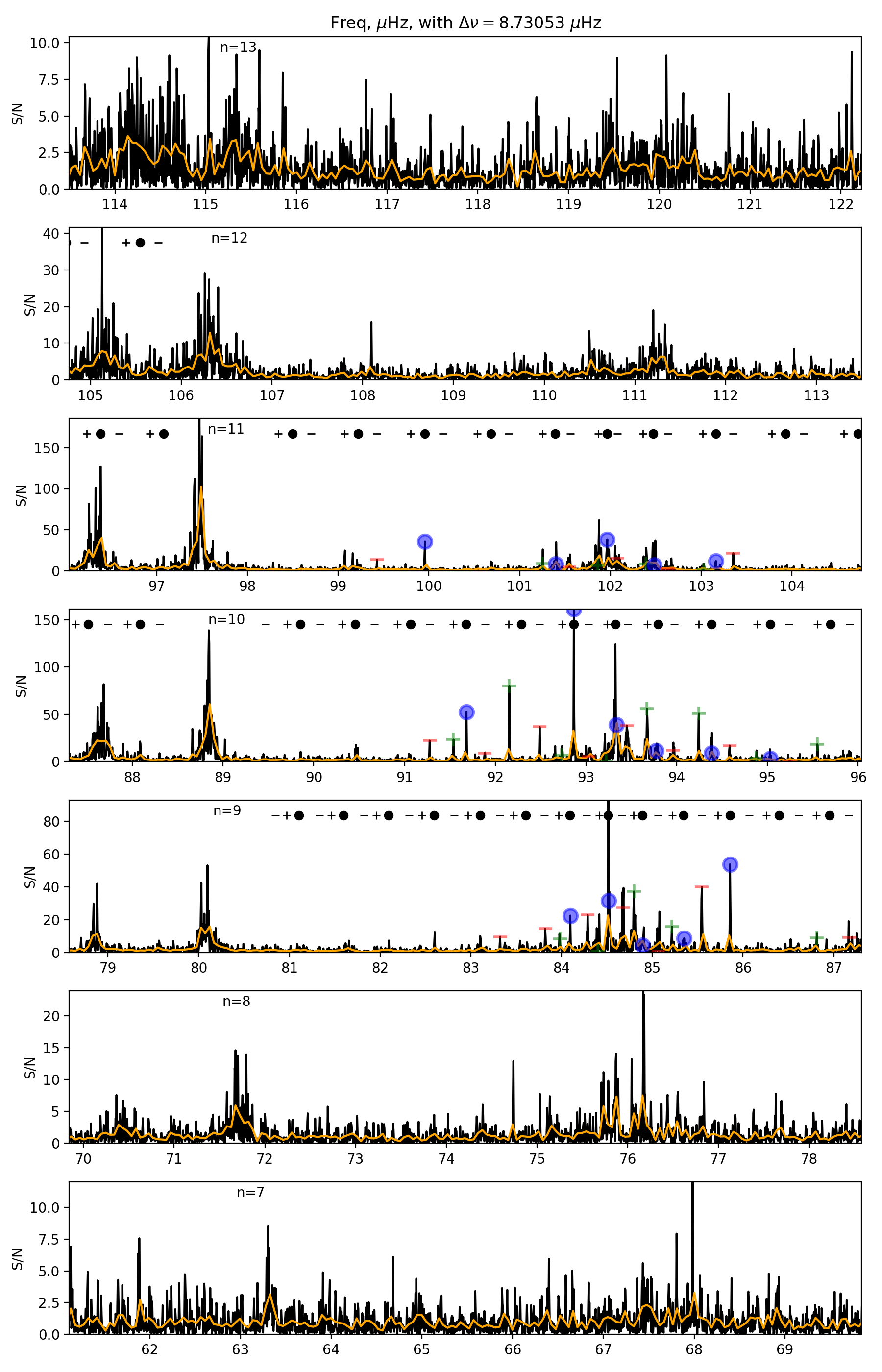}
\caption{Normal \'{e}chelle diagram of KIC\,6936091. See Fig.~\ref{apdxfig:echelle_diagram_4458118} for the explanations of the symbols.}
\label{apdxfig:echelle_diagram_6936091}
\end{figure*}

\begin{figure*}
\centering
\includegraphics[width = 0.6\linewidth]{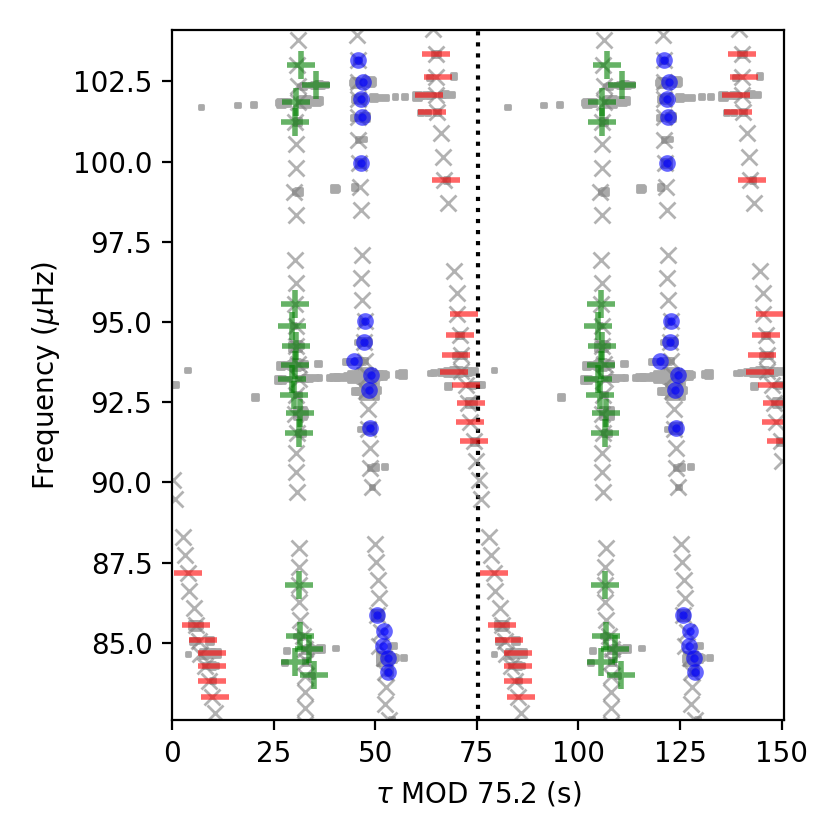}
\caption{Stretched \'{e}chelle diagram of KIC\,6936091. Symbols are explained in Fig.~\ref{fig:stretched_KIC5792889}. }
\label{apdxfig:echelle_diagram_stretched_6936091}
\end{figure*}

\begin{figure*}
\centering
\includegraphics[width = 0.8\linewidth]{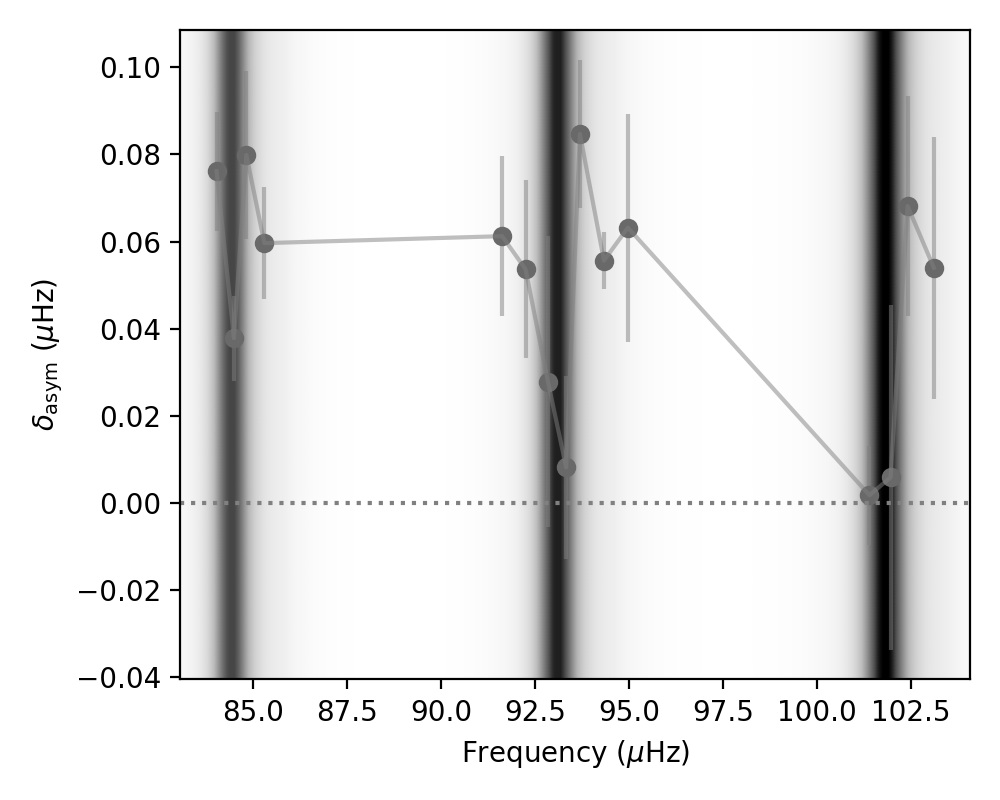}
\caption{Splitting asymmetries as a function of frequency of KIC\,6936091. Symbols are explained in Fig.~\ref{fig:KIC5696081_asymmtry_vs_freq}.}
\label{apdxfig:asymmetry_6936091}
\end{figure*}

\begin{figure*}
\centering
\includegraphics[width = 0.8\linewidth]{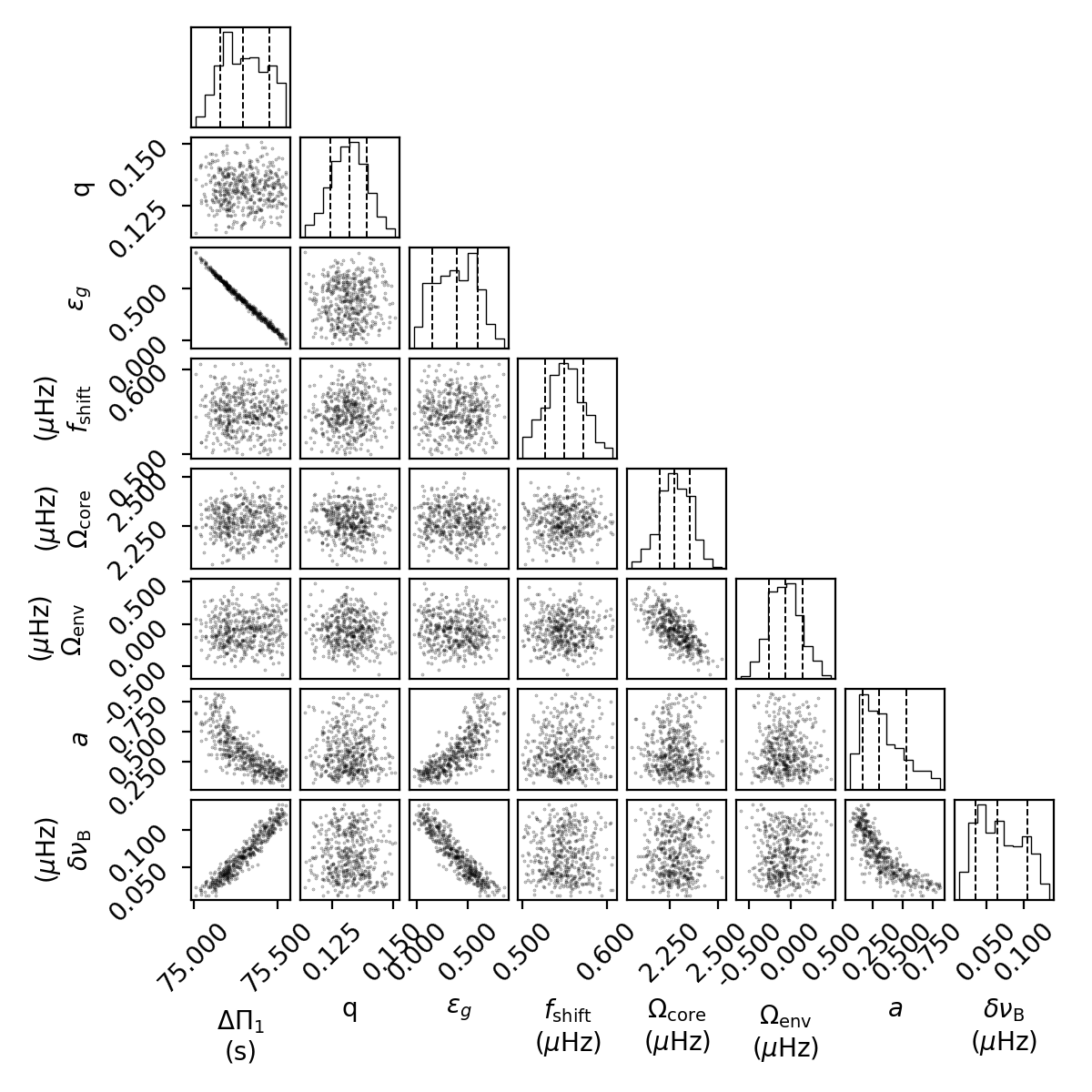}
\caption{Corner diagram of the MCMC fitting result of KIC\,6936091.}
\label{apdxfig:corner_6936091}
\end{figure*}

\begin{figure*}
\centering
\includegraphics[width = 0.8\linewidth]{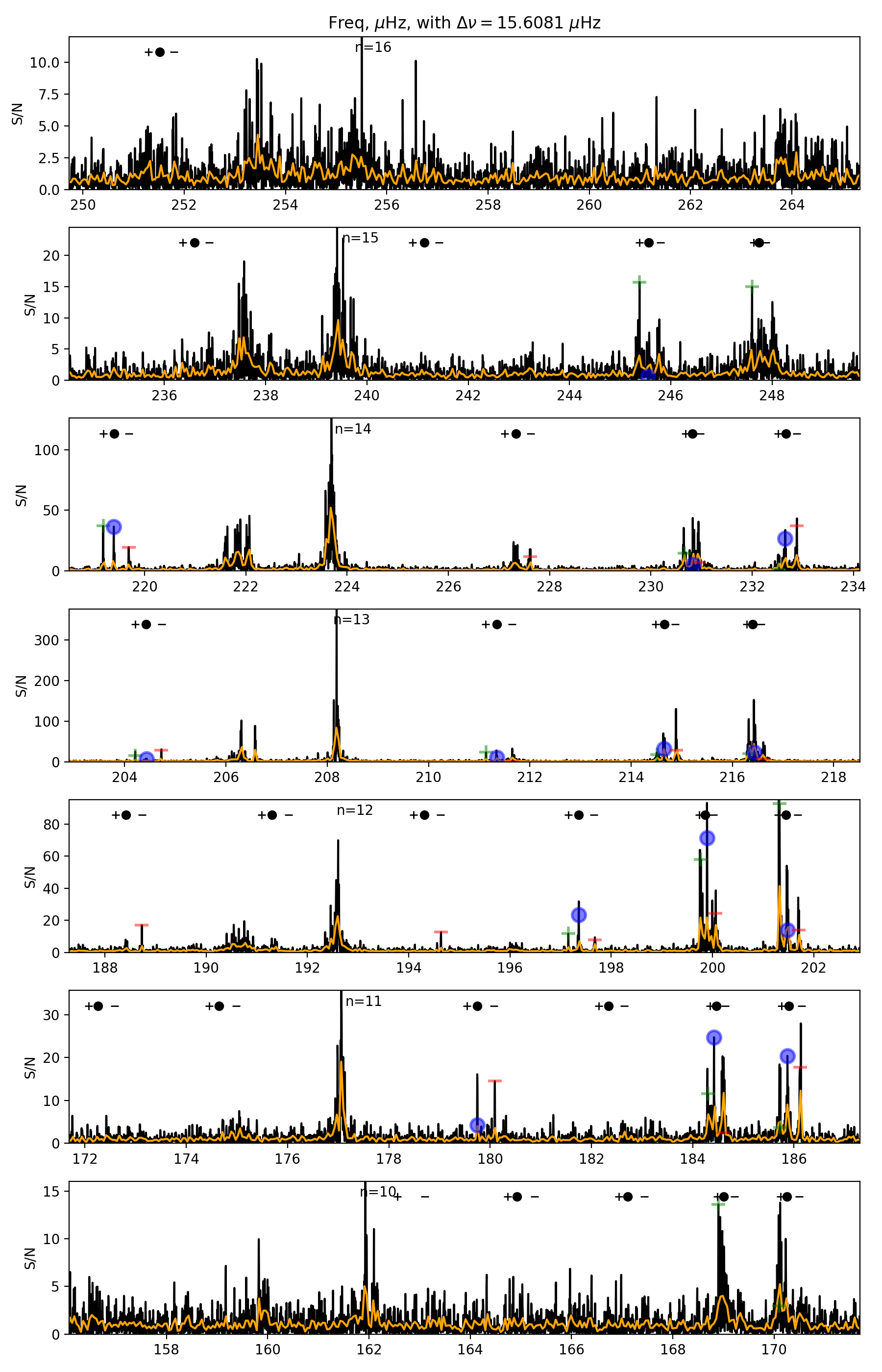}
\caption{Normal \'{e}chelle diagram of KIC\,7009365. See Fig.~\ref{apdxfig:echelle_diagram_4458118} for the explanations of the symbols.}
\label{apdxfig:echelle_diagram_7009365}
\end{figure*}

\begin{figure*}
\centering
\includegraphics[width = 0.6\linewidth]{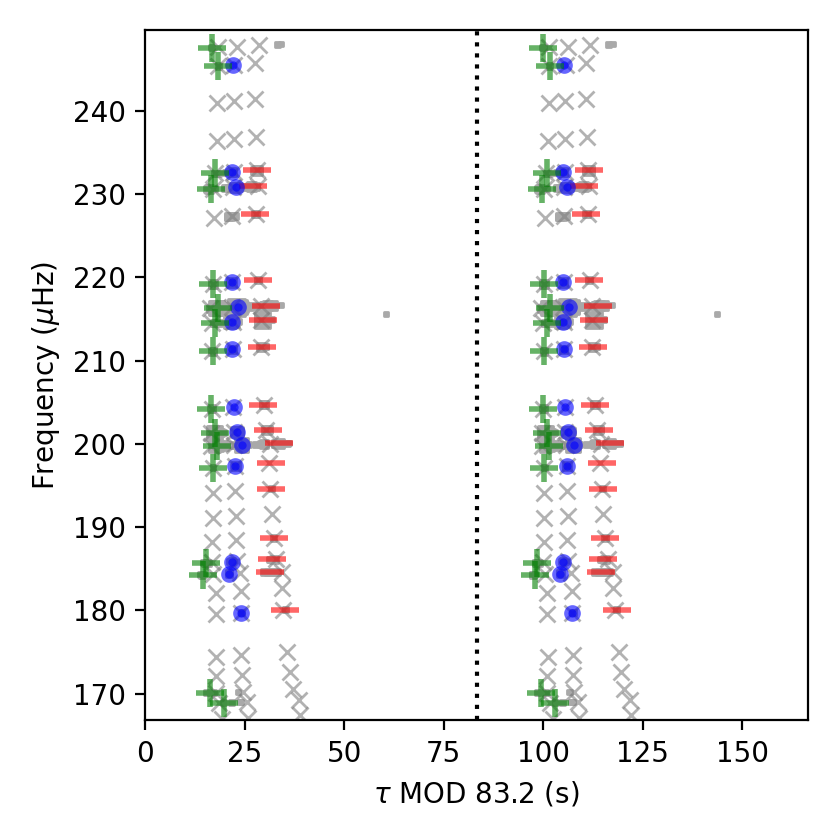}
\caption{Stretched \'{e}chelle diagram of KIC\,7009365. Symbols are explained in Fig.~\ref{fig:stretched_KIC5792889}. }
\label{apdxfig:echelle_diagram_stretched_7009365}
\end{figure*}

\begin{figure*}
\centering
\includegraphics[width = 0.8\linewidth]{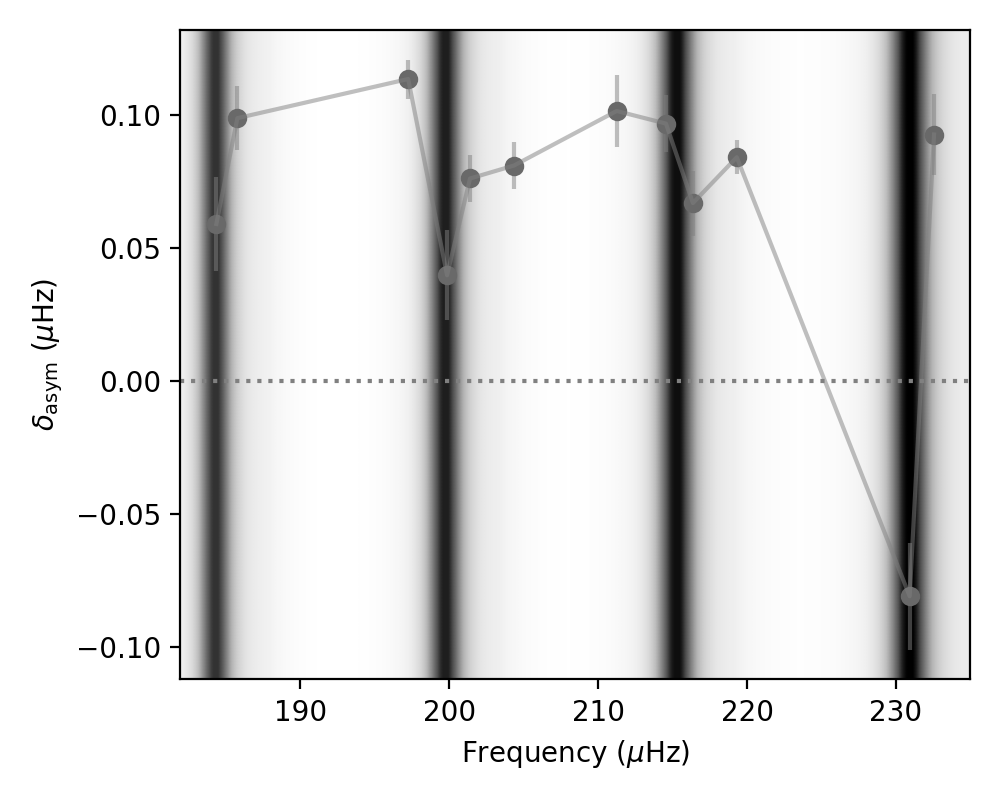}
\caption{Splitting asymmetries as a function of frequency of KIC\,7009365. Symbols are explained in Fig.~\ref{fig:KIC5696081_asymmtry_vs_freq}.}
\label{apdxfig:asymmetry_7009365}
\end{figure*}

\begin{figure*}
\centering
\includegraphics[width = 0.8\linewidth]{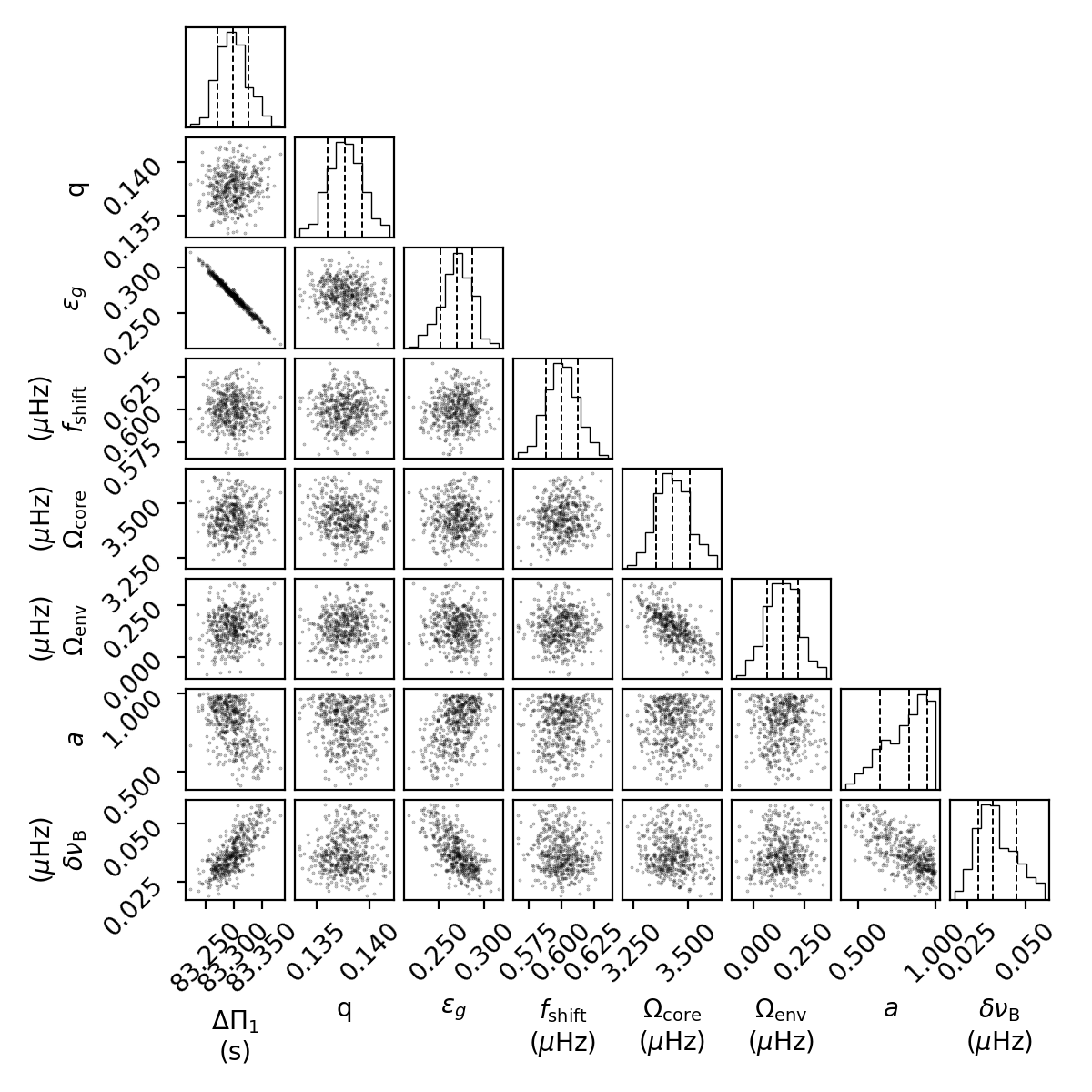}
\caption{Corner diagram of the MCMC fitting result of KIC\,7009365.}
\label{apdxfig:corner_7009365}
\end{figure*}

\begin{figure*}
\centering
\includegraphics[width = 0.8\linewidth]{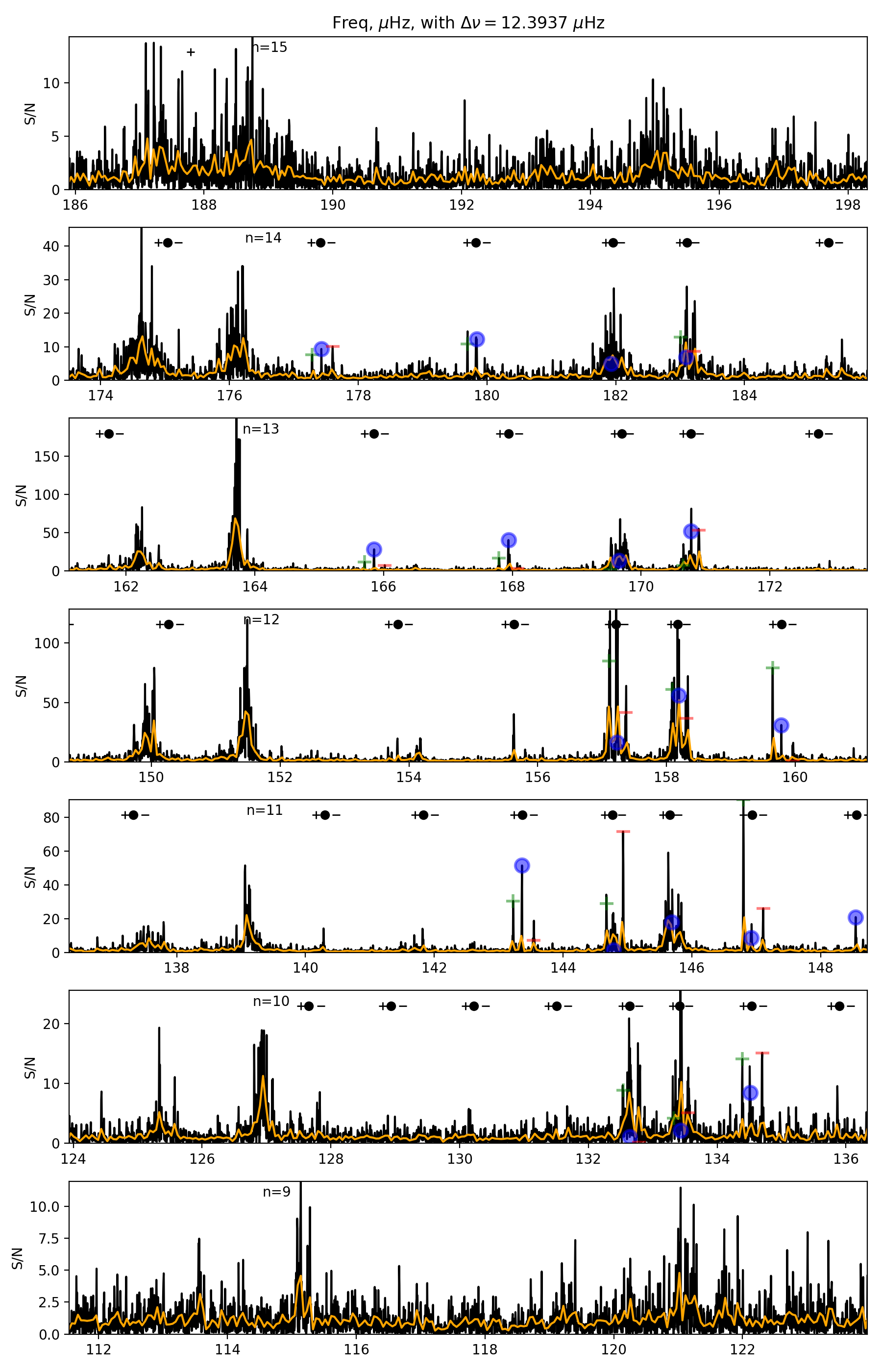}
\caption{Normal \'{e}chelle diagram of KIC\,7518143. See Fig.~\ref{apdxfig:echelle_diagram_4458118} for the explanations of the symbols.}
\label{apdxfig:echelle_diagram_7518143}
\end{figure*}

\begin{figure*}
\centering
\includegraphics[width = 0.6\linewidth]{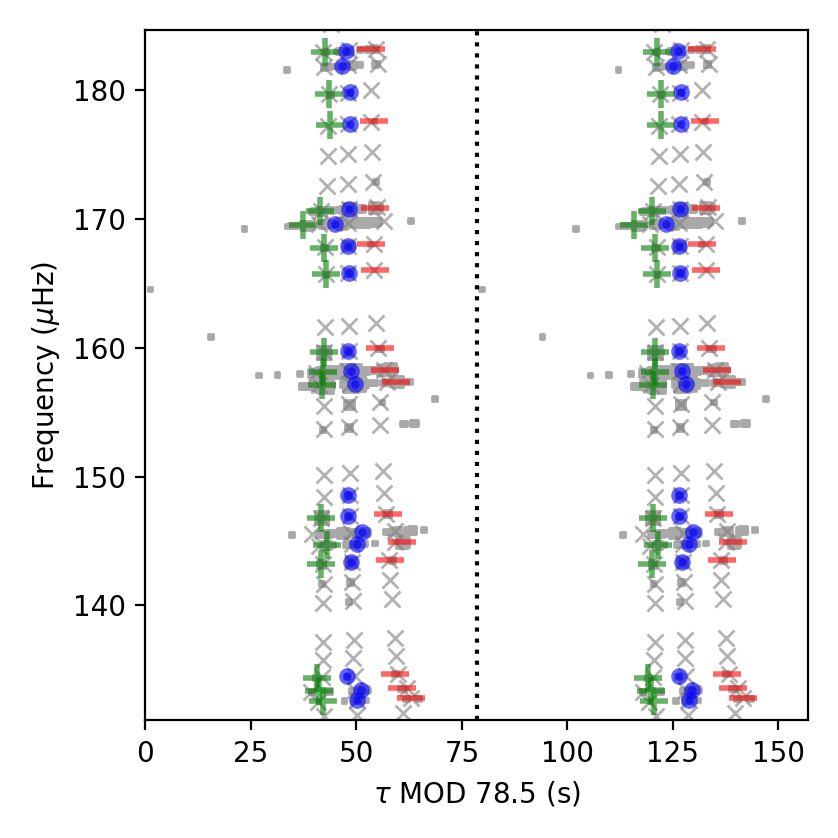}
\caption{Stretched \'{e}chelle diagram of KIC\,7518143. Symbols are explained in Fig.~\ref{fig:stretched_KIC5792889}. }
\label{apdxfig:echelle_diagram_stretched_7518143}
\end{figure*}

\begin{figure*}
\centering
\includegraphics[width = 0.8\linewidth]{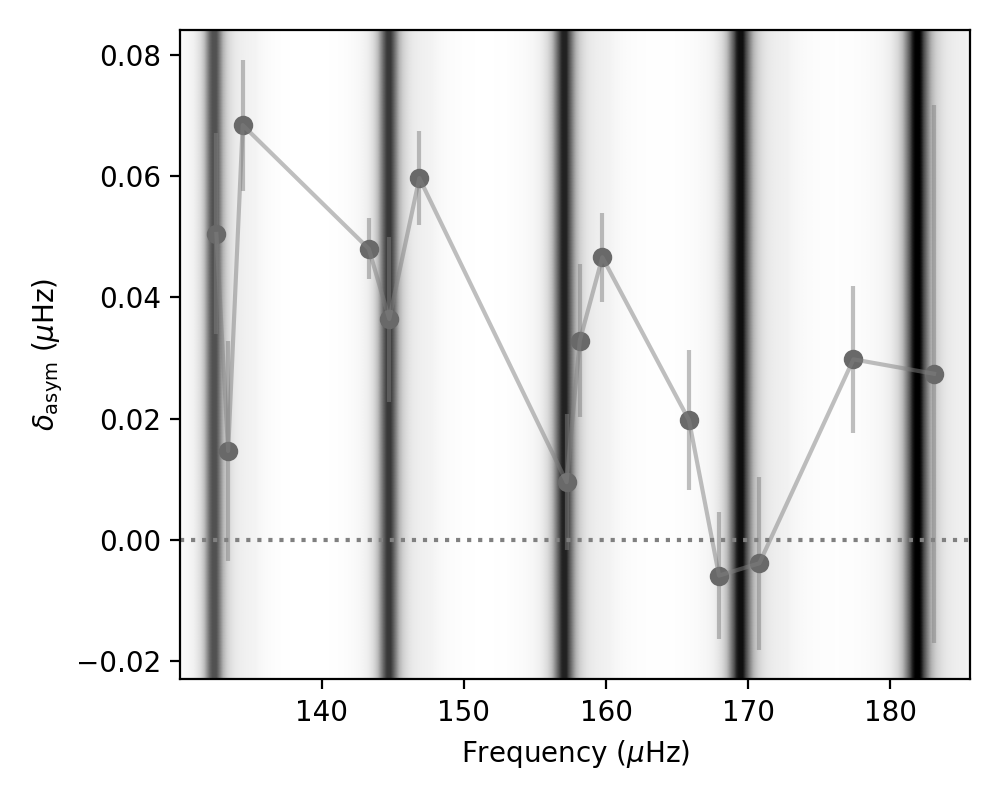}
\caption{Splitting asymmetries as a function of frequency of KIC\,7518143. Symbols are explained in Fig.~\ref{fig:KIC5696081_asymmtry_vs_freq}.}
\label{apdxfig:asymmetry_7518143}
\end{figure*}

\begin{figure*}
\centering
\includegraphics[width = 0.8\linewidth]{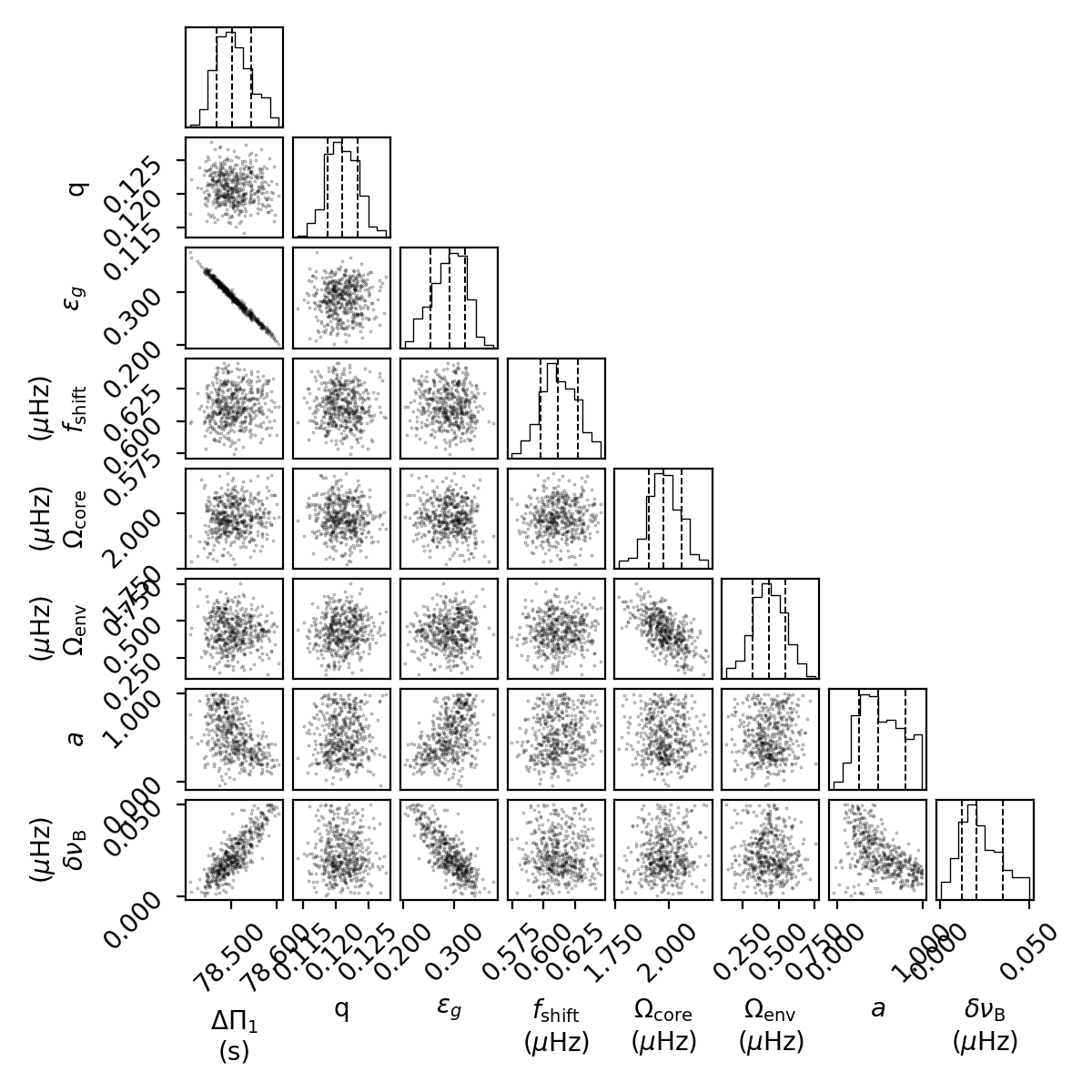}
\caption{Corner diagram of the MCMC fitting result of KIC\,7518143.}
\label{apdxfig:corner_7518143}
\end{figure*}

\begin{figure*}
\centering
\includegraphics[width = 0.8\linewidth]{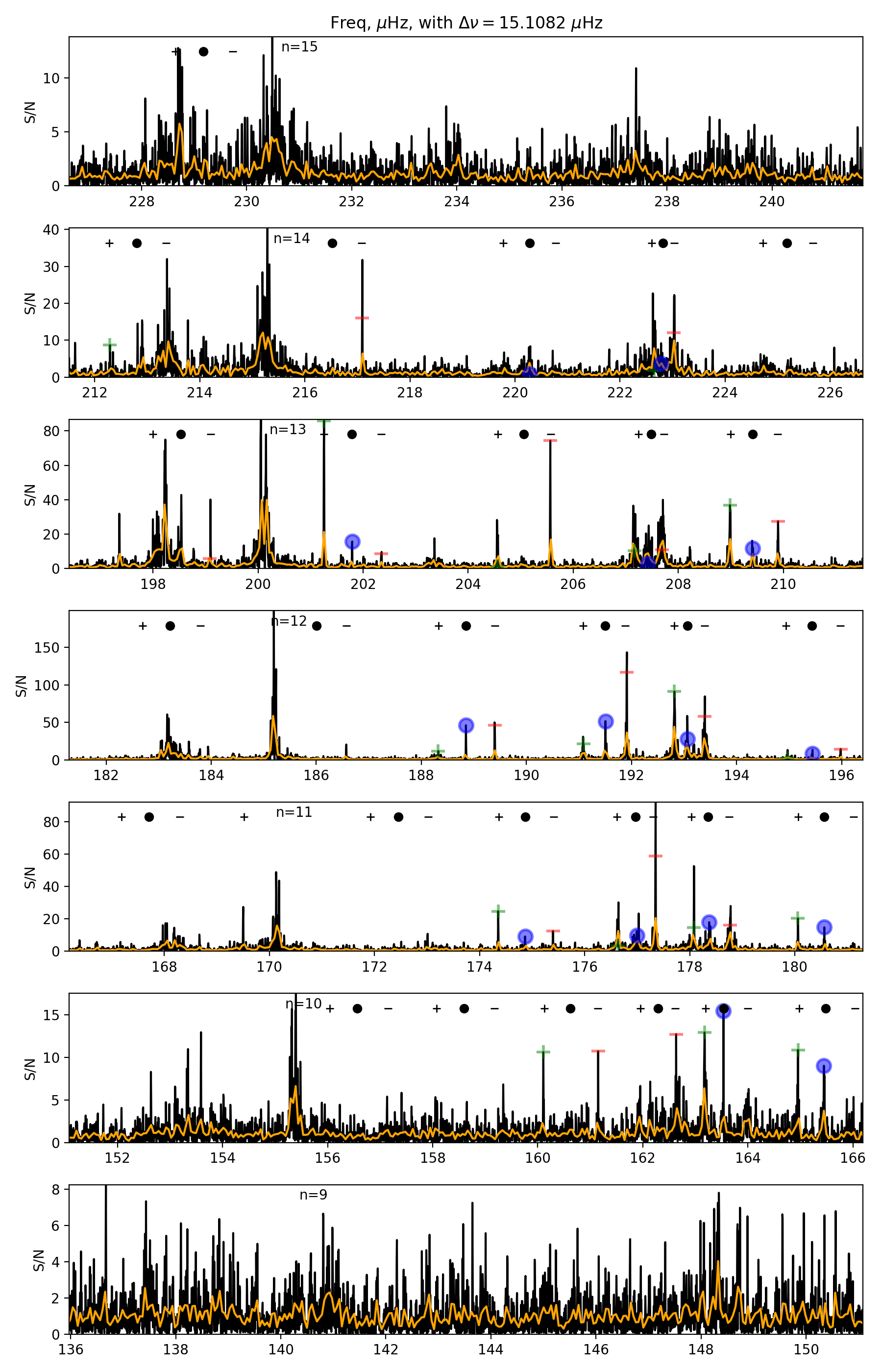}
\caption{Normal \'{e}chelle diagram of KIC\,8540034. See Fig.~\ref{apdxfig:echelle_diagram_4458118} for the explanations of the symbols.}
\label{apdxfig:echelle_diagram_8540034}
\end{figure*}

\begin{figure*}
\centering
\includegraphics[width = 0.6\linewidth]{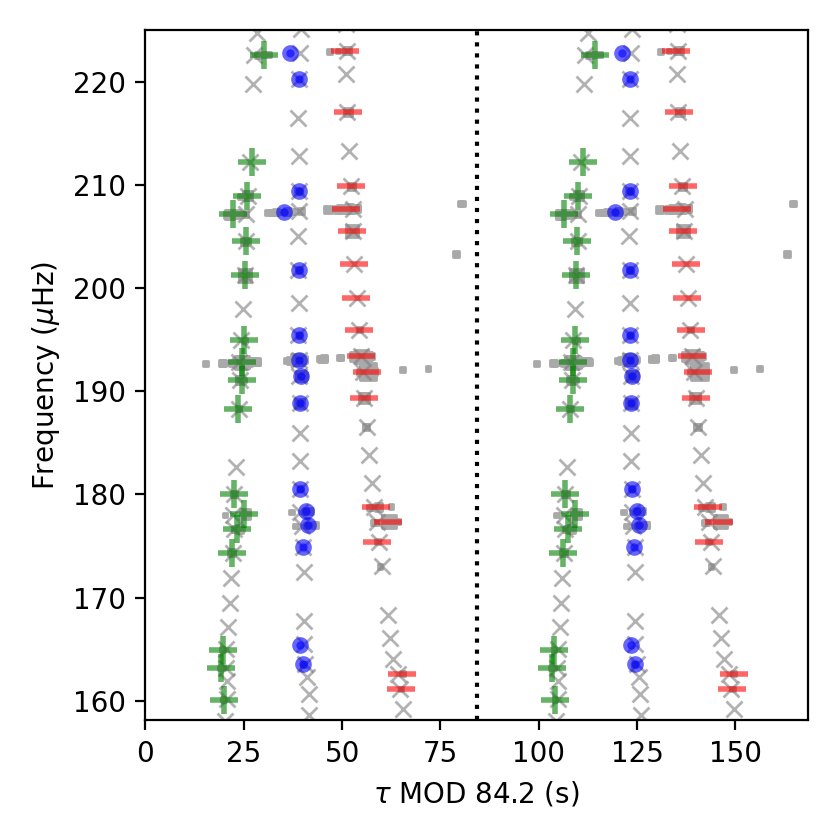}
\caption{Stretched \'{e}chelle diagram of KIC\,8540034. Symbols are explained in Fig.~\ref{fig:stretched_KIC5792889}. }
\label{apdxfig:echelle_diagram_stretched_8540034}
\end{figure*}

\begin{figure*}
\centering
\includegraphics[width = 0.8\linewidth]{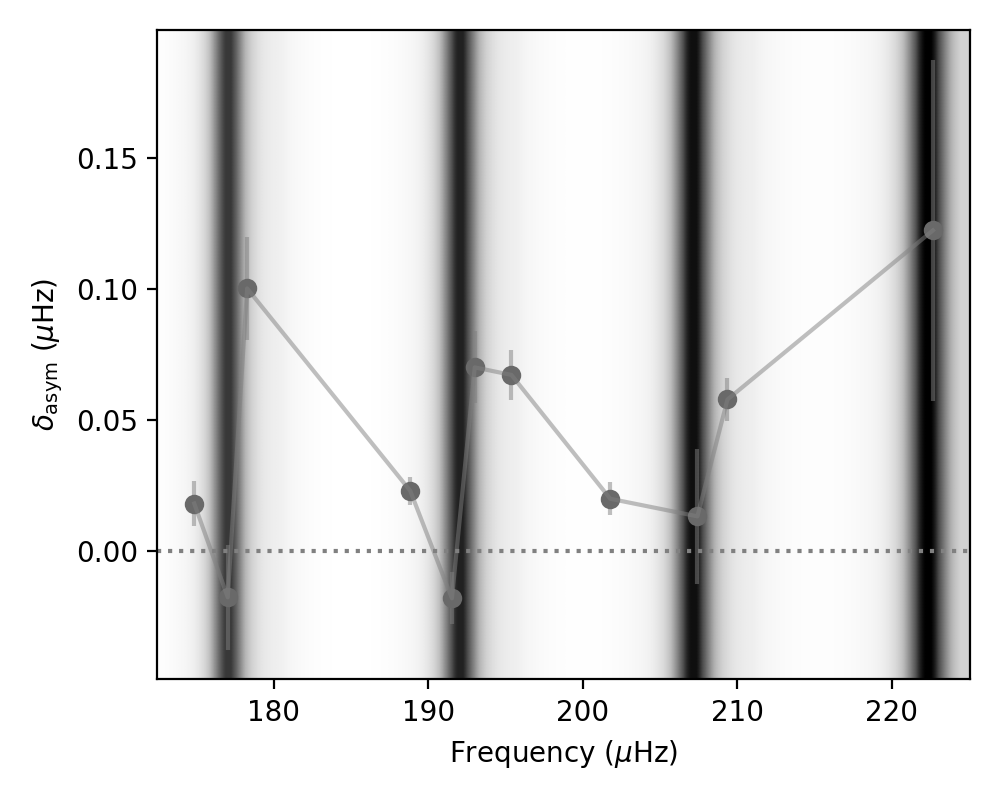}
\caption{Splitting asymmetries as a function of frequency of KIC\,8540034. Symbols are explained in Fig.~\ref{fig:KIC5696081_asymmtry_vs_freq}.}
\label{apdxfig:asymmetry_8540034}
\end{figure*}

\begin{figure*}
\centering
\includegraphics[width = 0.8\linewidth]{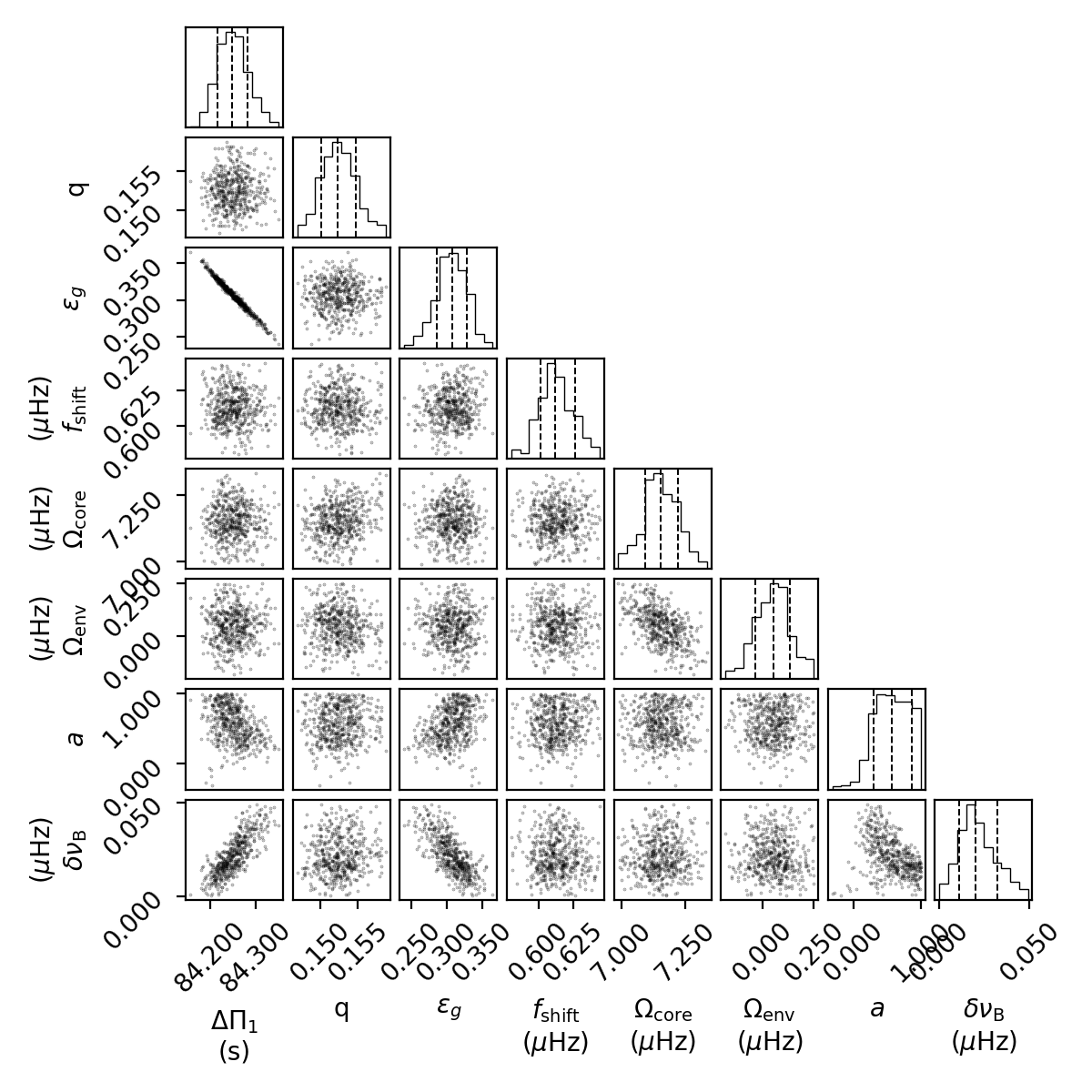}
\caption{Corner diagram of the MCMC fitting result of KIC\,8540034.}
\label{apdxfig:corner_8540034}
\end{figure*}

\begin{figure*}
\centering
\includegraphics[width = 0.8\linewidth]{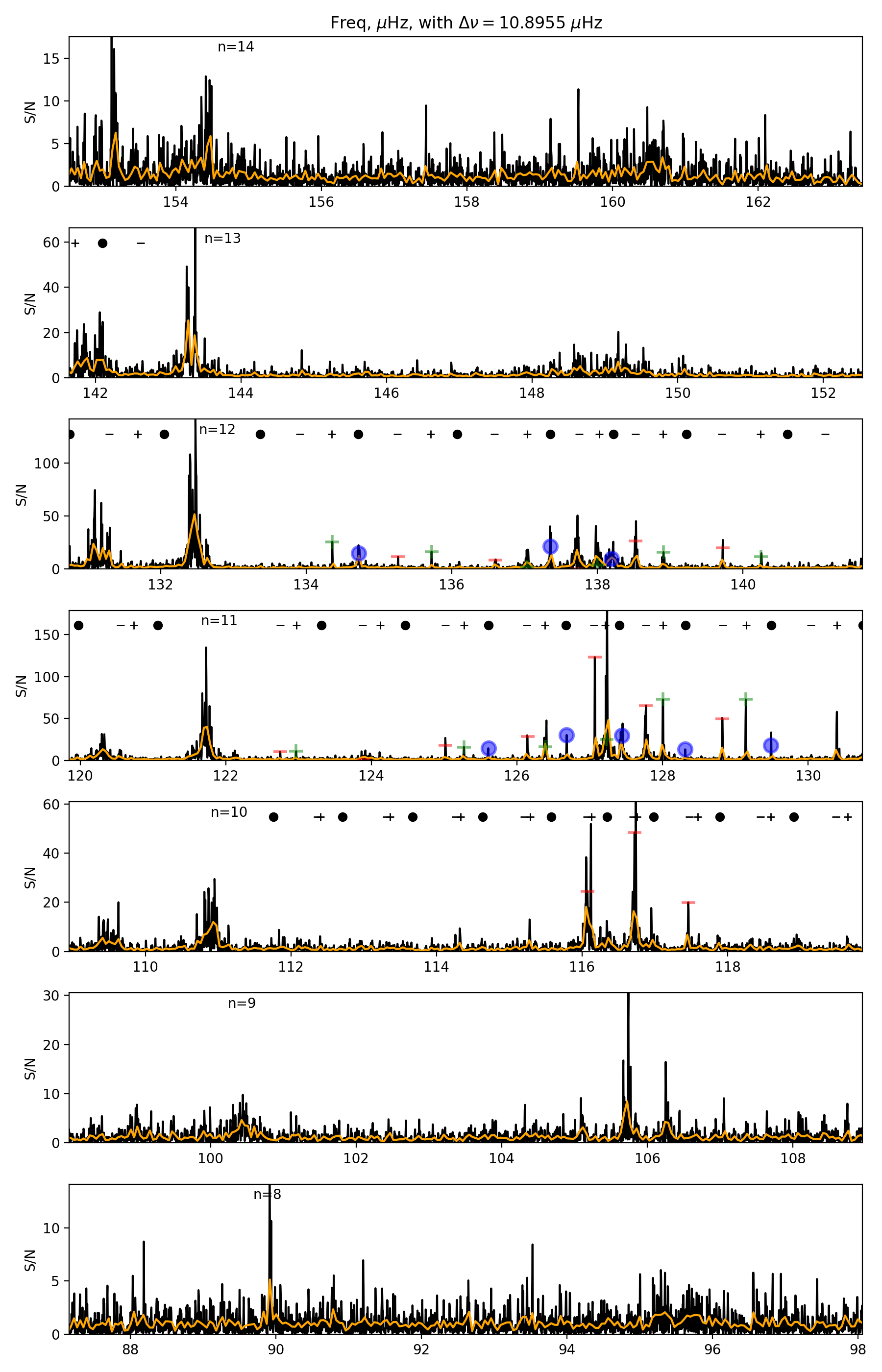}
\caption{Normal \'{e}chelle diagram of KIC\,8619145. See Fig.~\ref{apdxfig:echelle_diagram_4458118} for the explanations of the symbols.}
\label{apdxfig:echelle_diagram_8619145}
\end{figure*}

\begin{figure*}
\centering
\includegraphics[width = 0.6\linewidth]{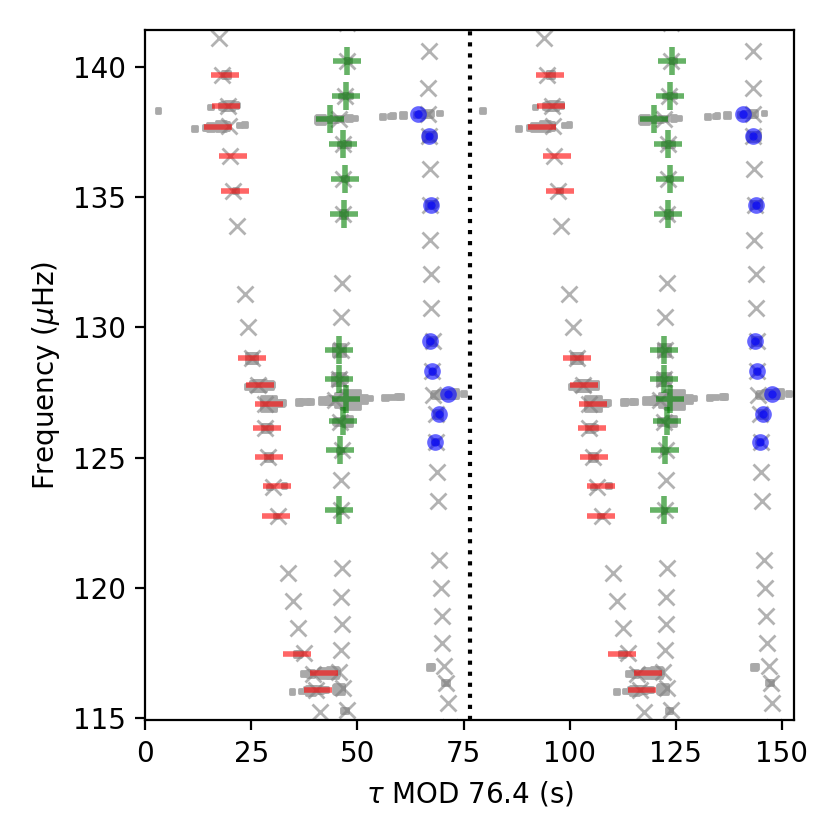}
\caption{Stretched \'{e}chelle diagram of KIC\,8619145. Symbols are explained in Fig.~\ref{fig:stretched_KIC5792889}. }
\label{apdxfig:echelle_diagram_stretched_8619145}
\end{figure*}

\begin{figure*}
\centering
\includegraphics[width = 0.8\linewidth]{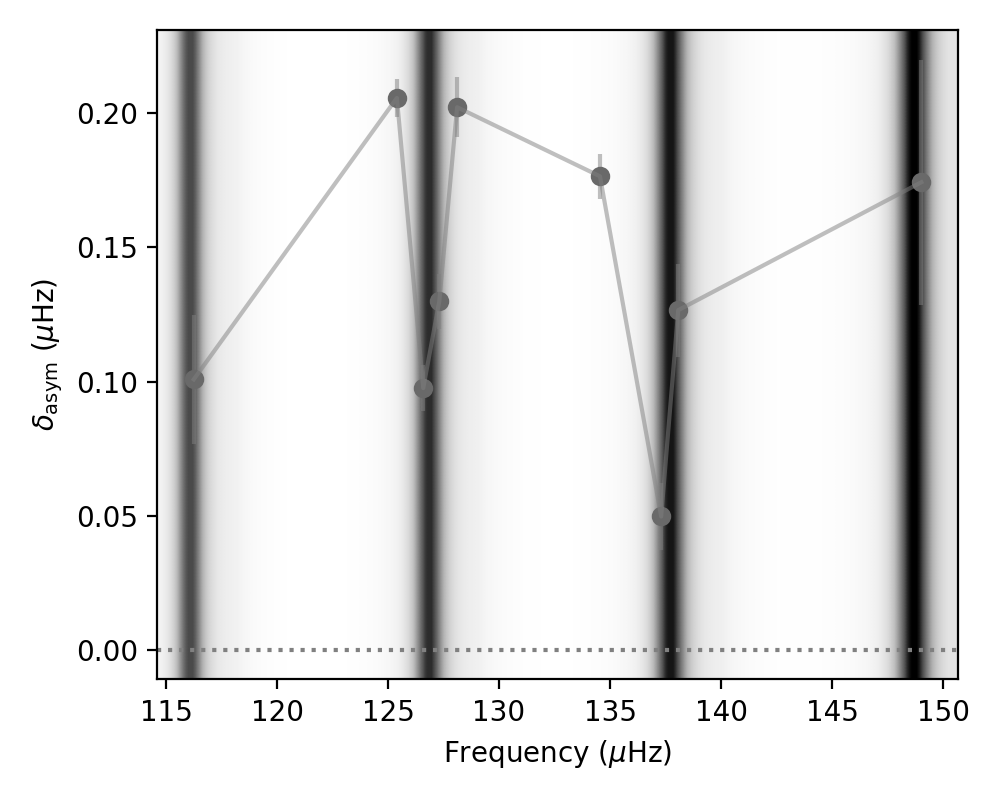}
\caption{Splitting asymmetries as a function of frequency of KIC\,8619145. Symbols are explained in Fig.~\ref{fig:KIC5696081_asymmtry_vs_freq}.}
\label{apdxfig:asymmetry_8619145}
\end{figure*}

\begin{figure*}
\centering
\includegraphics[width = 0.8\linewidth]{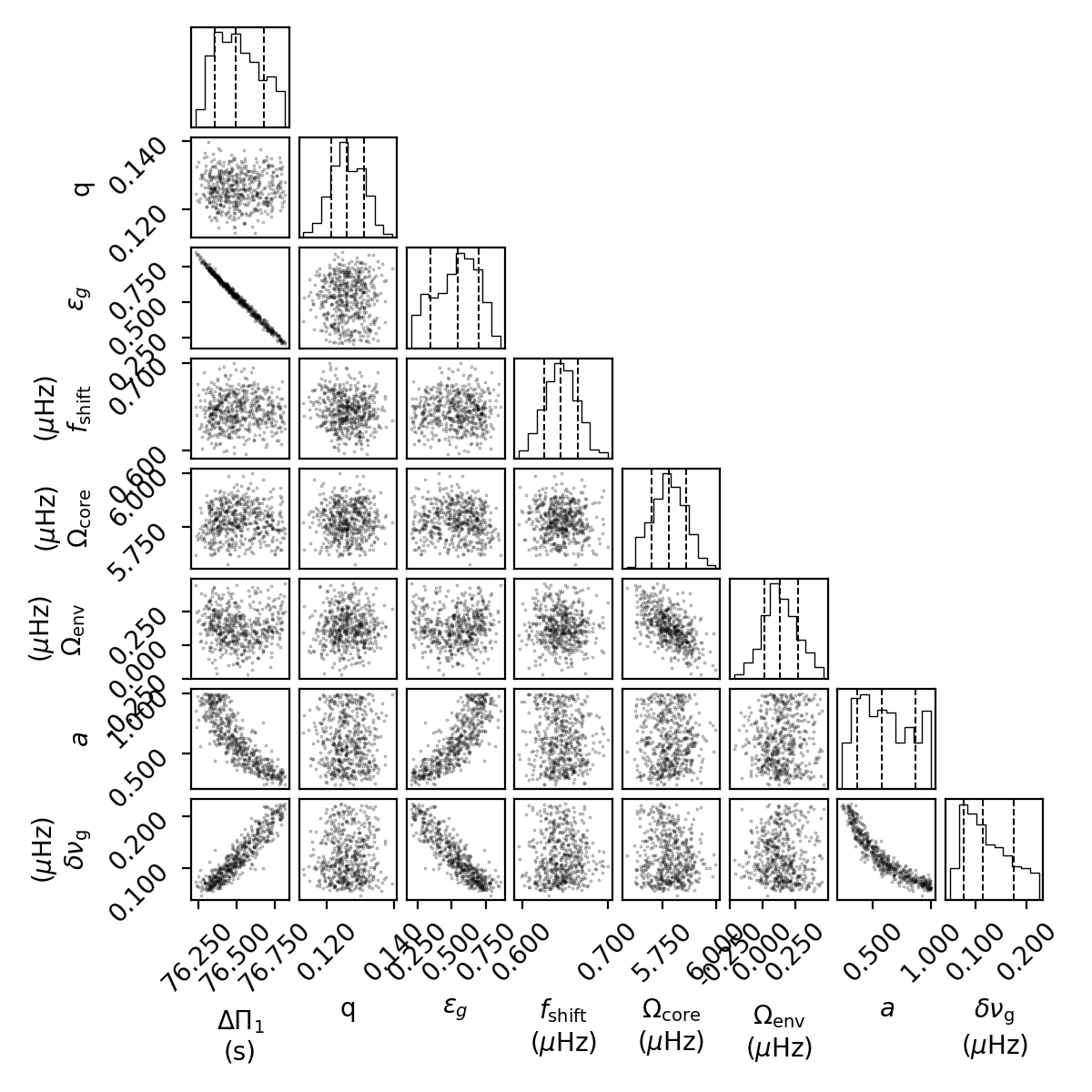}
\caption{Corner diagram of the MCMC fitting result of KIC\,8619145.}
\label{apdxfig:corner_8619145}
\end{figure*}

\begin{figure*}
\centering
\includegraphics[width = 0.8\linewidth]{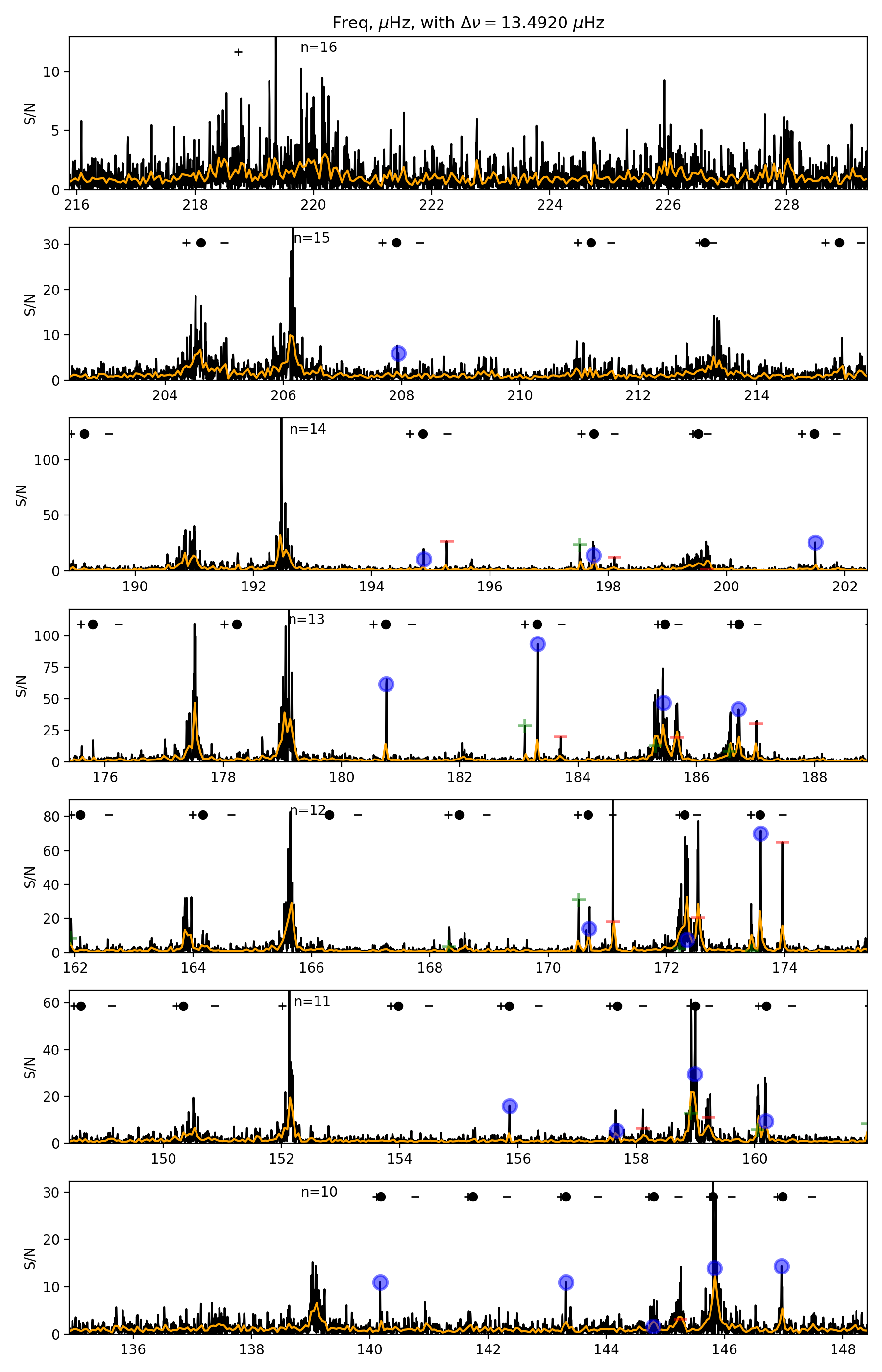}
\caption{Normal \'{e}chelle diagram of KIC\,8684542. See Fig.~\ref{apdxfig:echelle_diagram_4458118} for the explanations of the symbols.}
\label{apdxfig:echelle_diagram_8684542}
\end{figure*}

\begin{figure*}
\centering
\includegraphics[width = 0.6\linewidth]{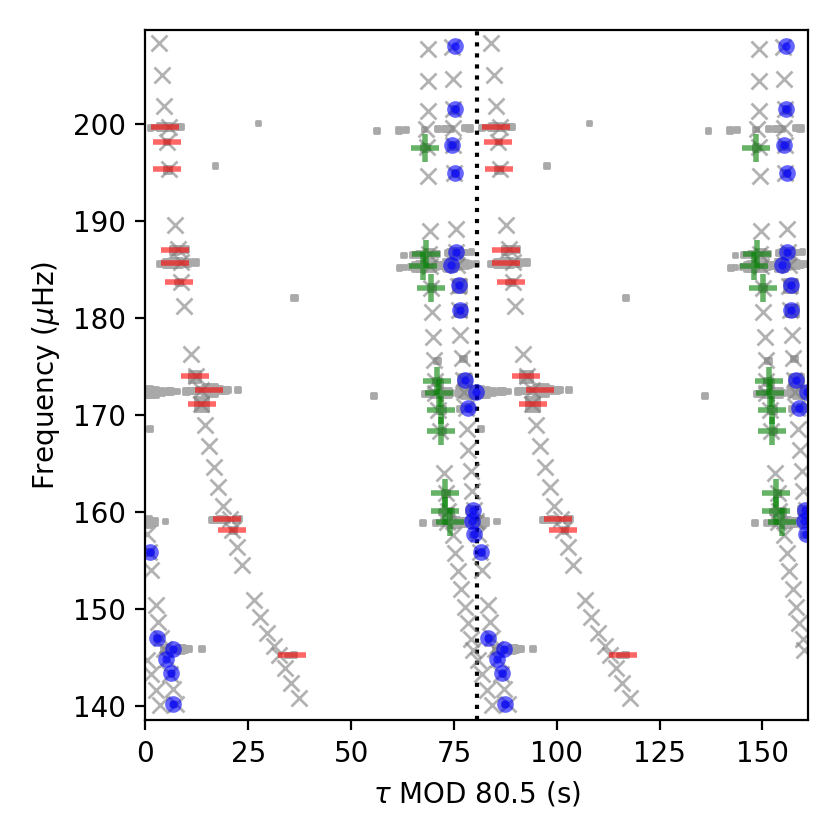}
\caption{Stretched \'{e}chelle diagram of KIC\,8684542. Symbols are explained in Fig.~\ref{fig:stretched_KIC5792889}. }
\label{apdxfig:echelle_diagram_stretched_8684542}
\end{figure*}

\begin{figure*}
\centering
\includegraphics[width = 0.8\linewidth]{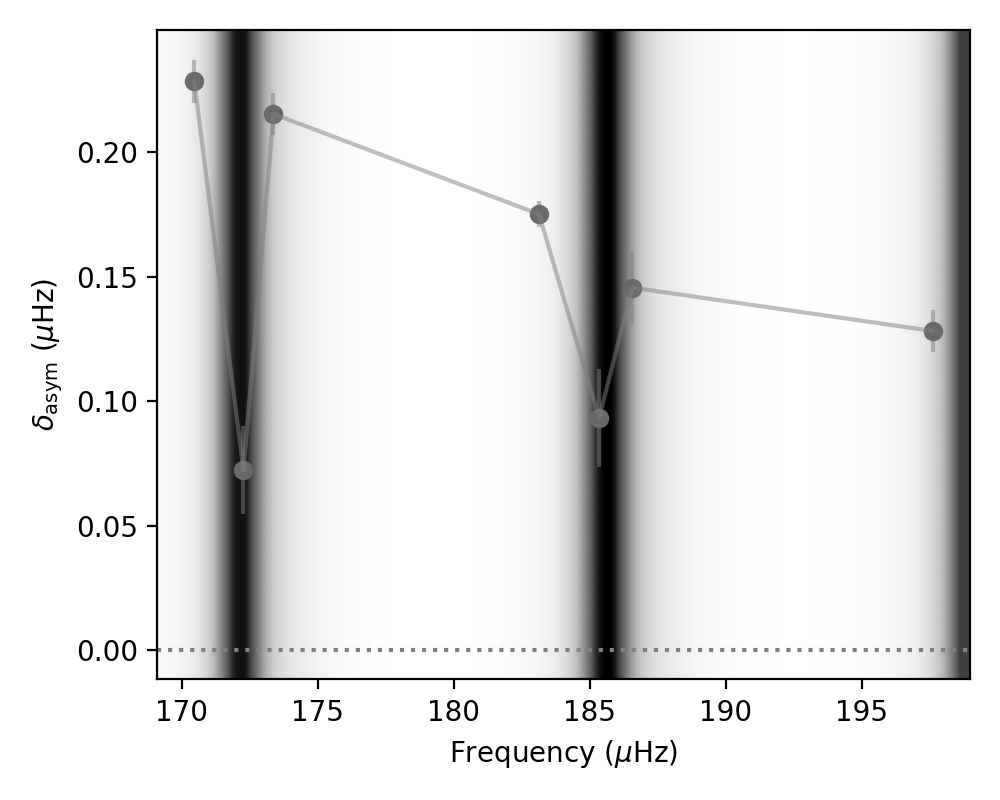}
\caption{Splitting asymmetries as a function of frequency of KIC\,8684542. Symbols are explained in Fig.~\ref{fig:KIC5696081_asymmtry_vs_freq}.}
\label{apdxfig:asymmetry_8684542}
\end{figure*}

\begin{figure*}
\centering
\includegraphics[width = 0.8\linewidth]{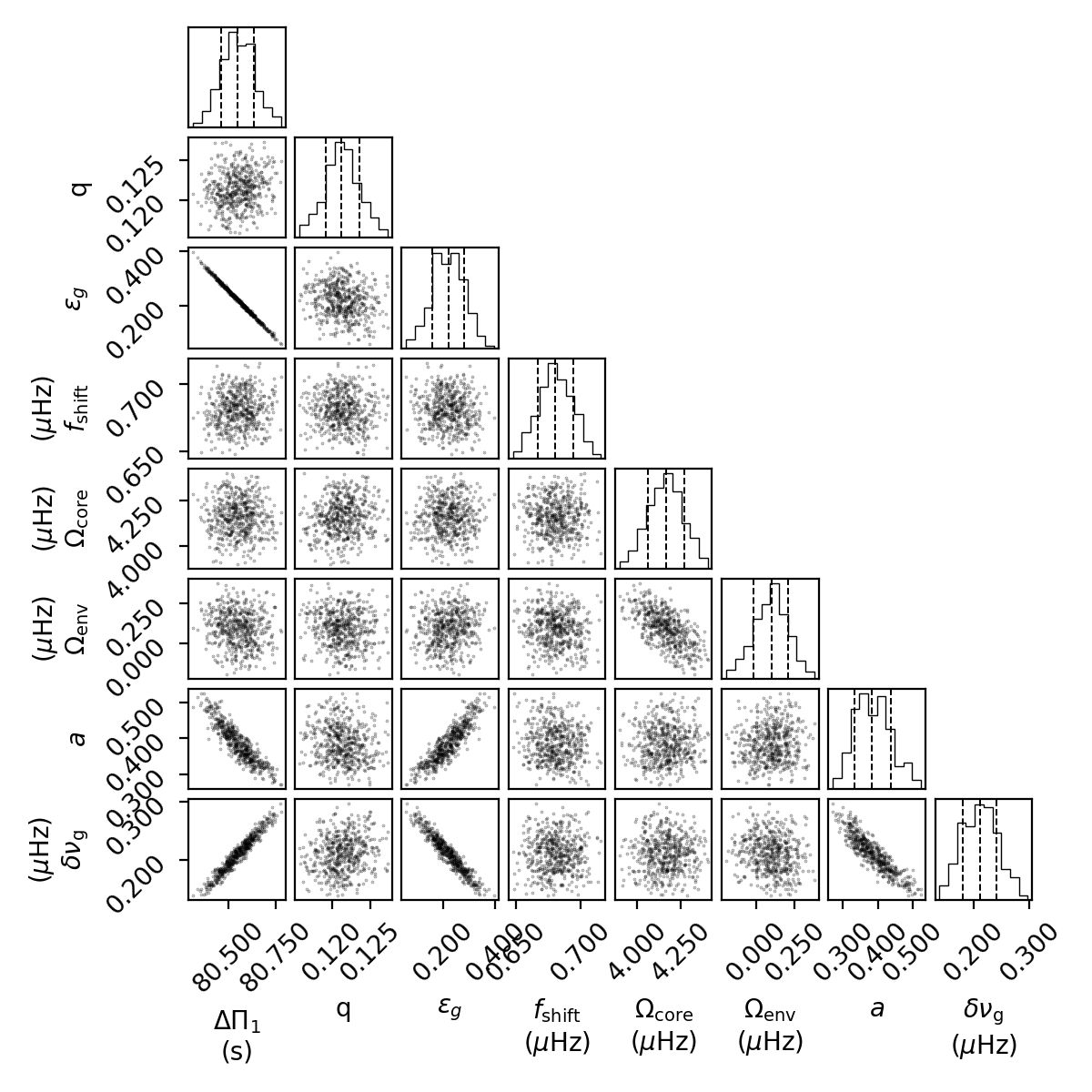}
\caption{Corner diagram of the MCMC fitting result of KIC\,8684542.}
\label{apdxfig:corner_8684542}
\end{figure*}

\begin{figure*}
\centering
\includegraphics[width = 0.8\linewidth]{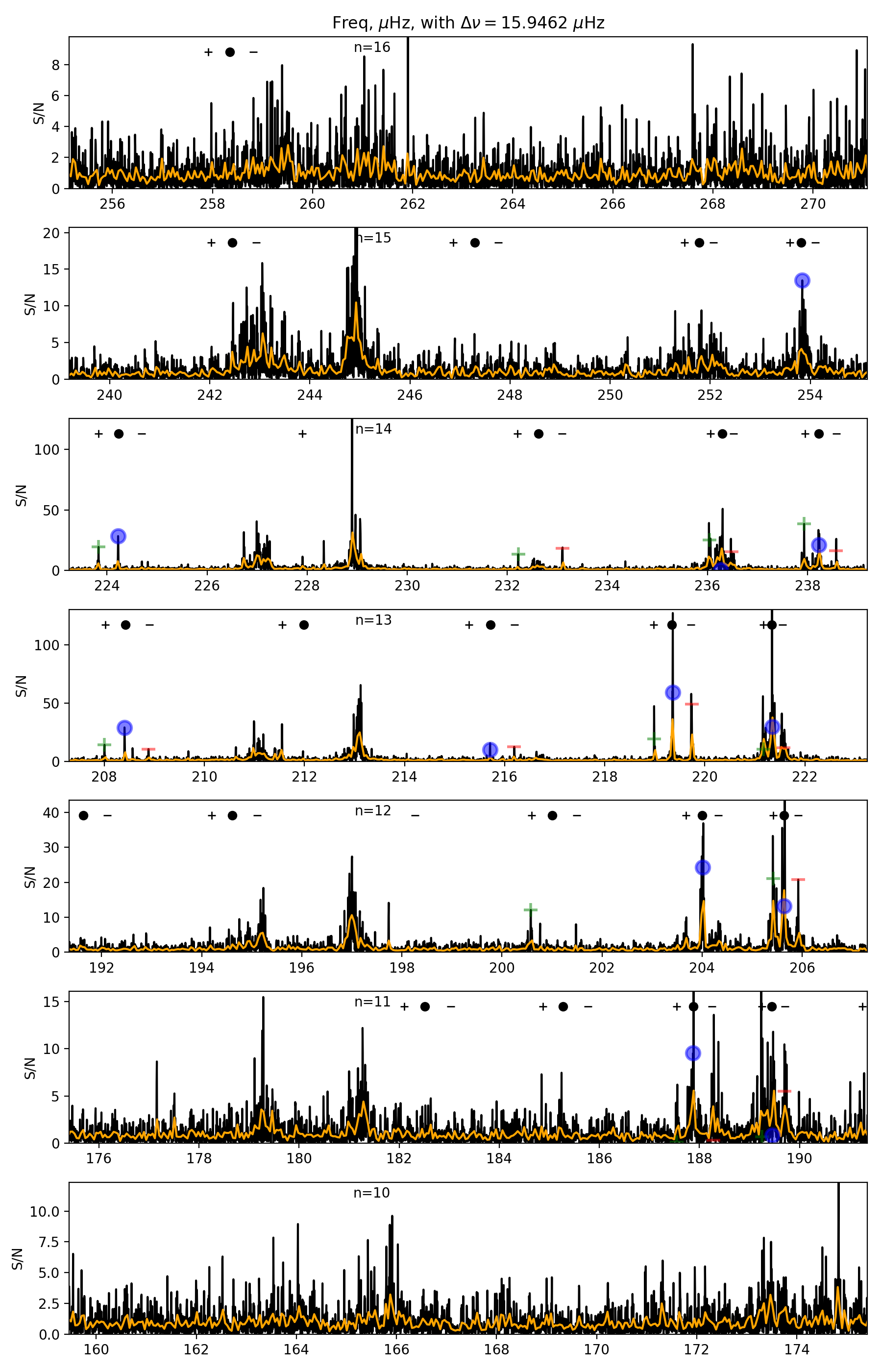}
\caption{Normal \'{e}chelle diagram of KIC\,9202471. See Fig.~\ref{apdxfig:echelle_diagram_4458118} for the explanations of the symbols.}
\label{apdxfig:echelle_diagram_9202471}
\end{figure*}

\begin{figure*}
\centering
\includegraphics[width = 0.6\linewidth]{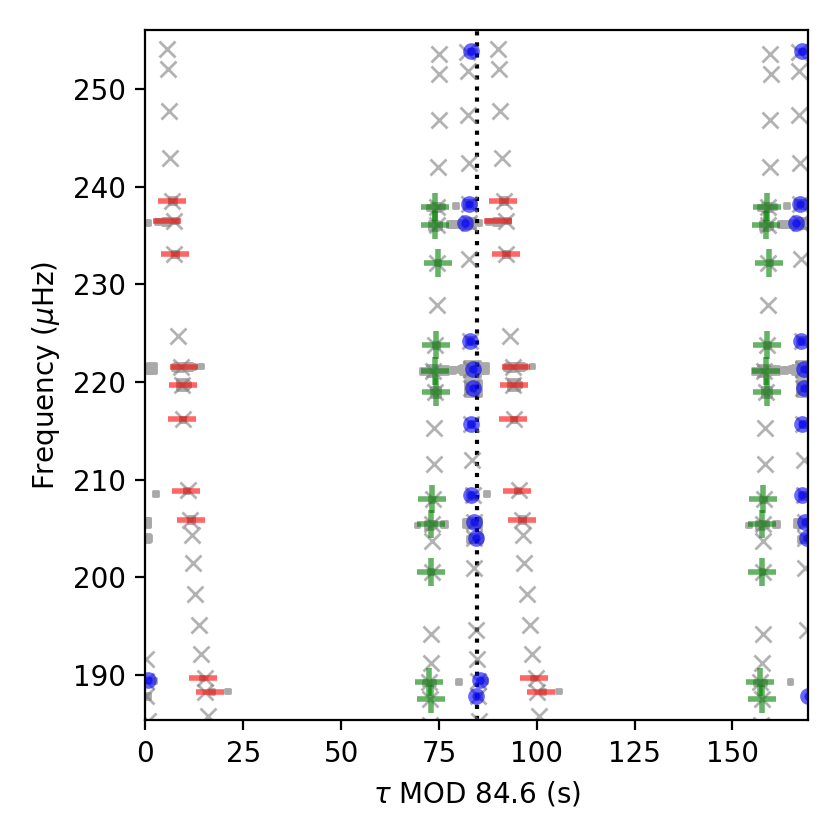}
\caption{Stretched \'{e}chelle diagram of KIC\,9202471. Symbols are explained in Fig.~\ref{fig:stretched_KIC5792889}. }
\label{apdxfig:echelle_diagram_stretched_9202471}
\end{figure*}

\begin{figure*}
\centering
\includegraphics[width = 0.8\linewidth]{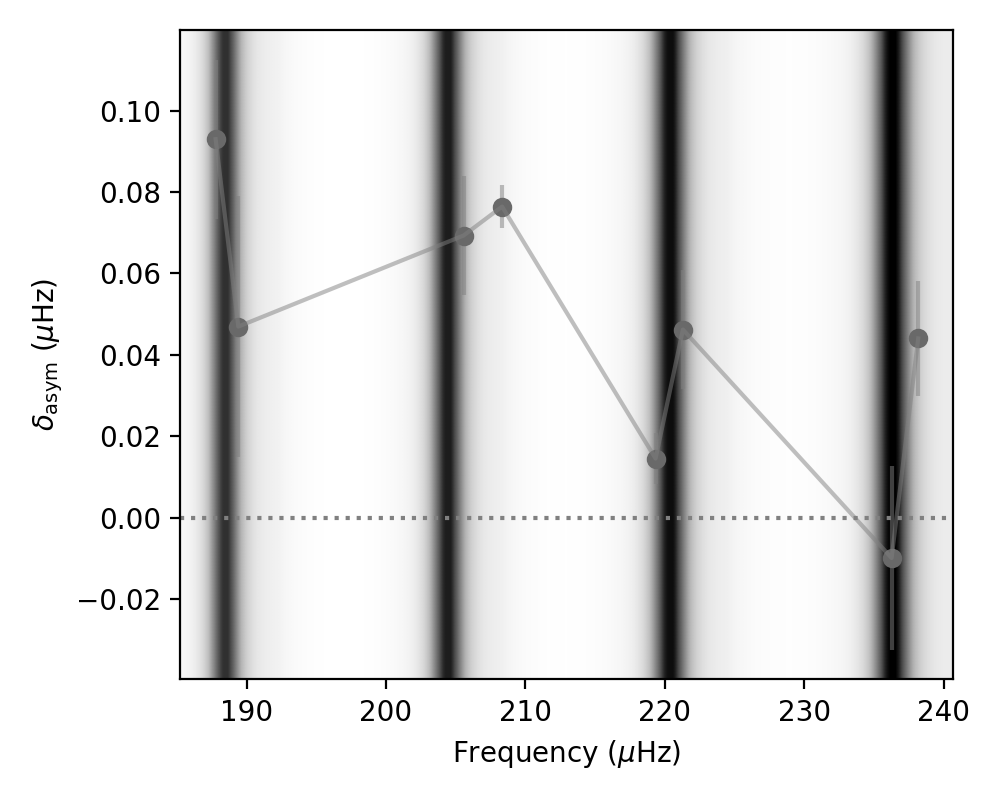}
\caption{Splitting asymmetries as a function of frequency of KIC\,9202471. Symbols are explained in Fig.~\ref{fig:KIC5696081_asymmtry_vs_freq}.}
\label{apdxfig:asymmetry_9202471}
\end{figure*}

\begin{figure*}
\centering
\includegraphics[width = 0.8\linewidth]{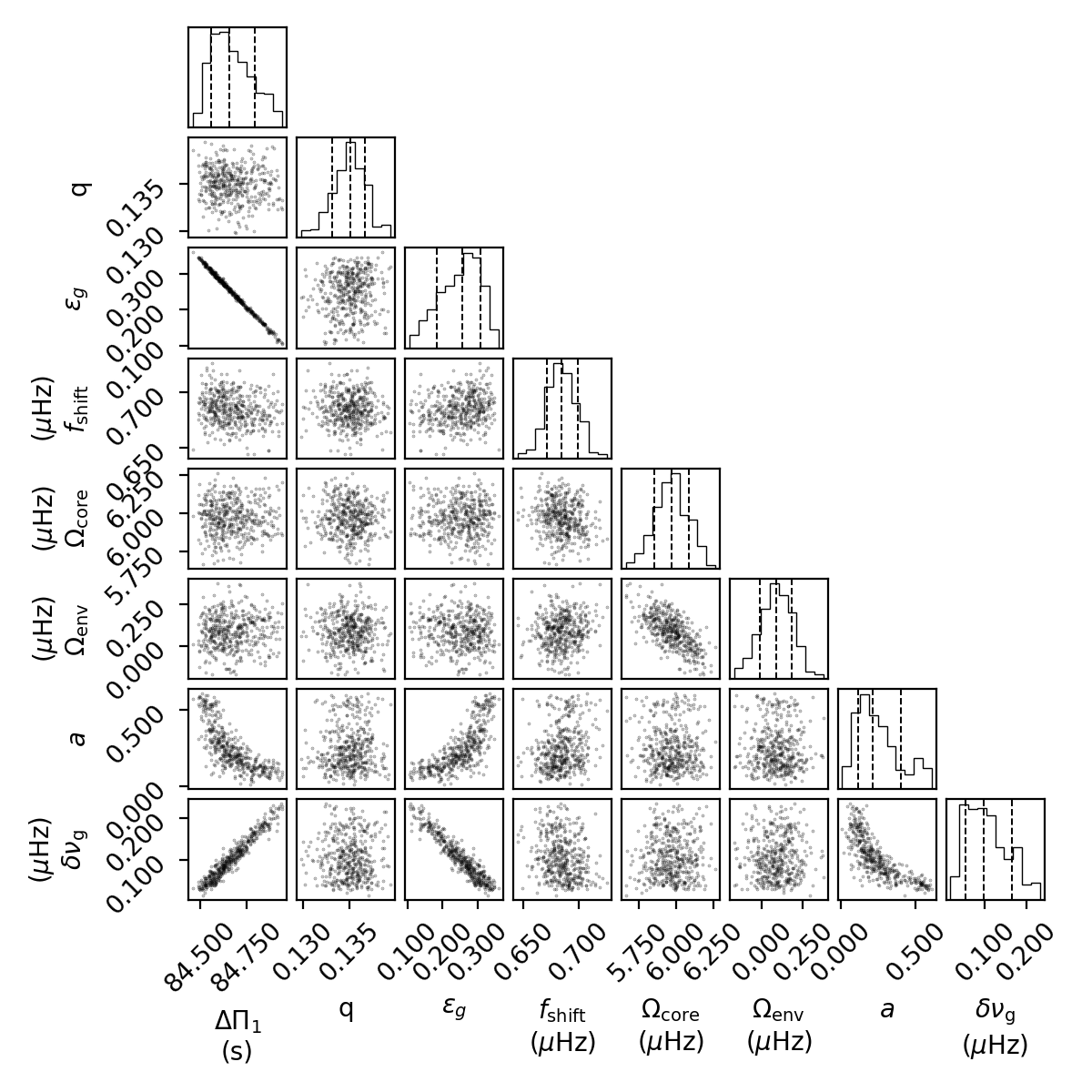}
\caption{Corner diagram of the MCMC fitting result of KIC\,9202471.}
\label{apdxfig:corner_9202471}
\end{figure*}

\begin{figure*}
\centering
\includegraphics[width = 0.8\linewidth]{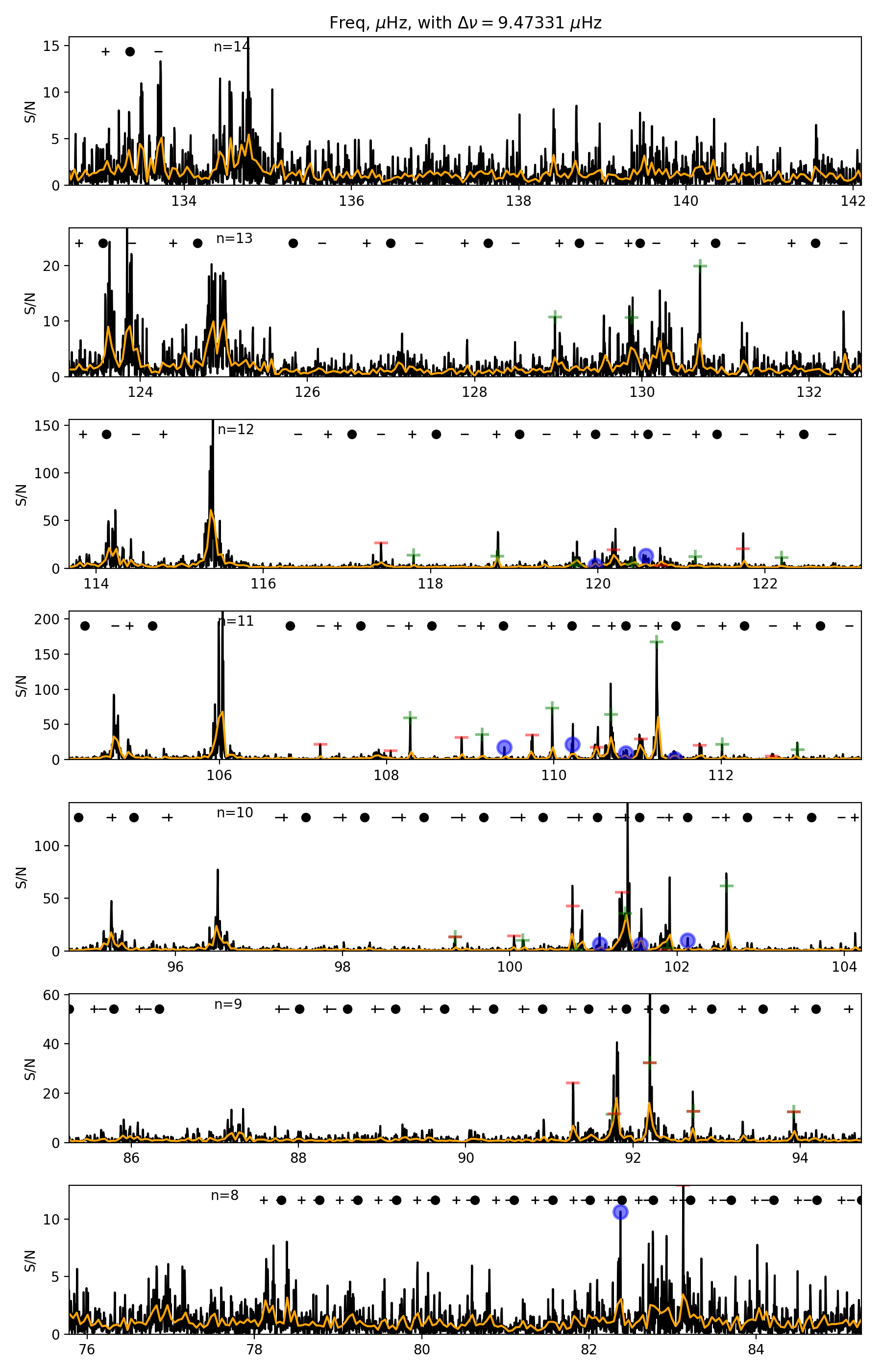}
\caption{Normal \'{e}chelle diagram of KIC\,9589420. See Fig.~\ref{apdxfig:echelle_diagram_4458118} for the explanations of the symbols.}
\label{apdxfig:echelle_diagram_9589420}
\end{figure*}

\begin{figure*}
\centering
\includegraphics[width = 0.6\linewidth]{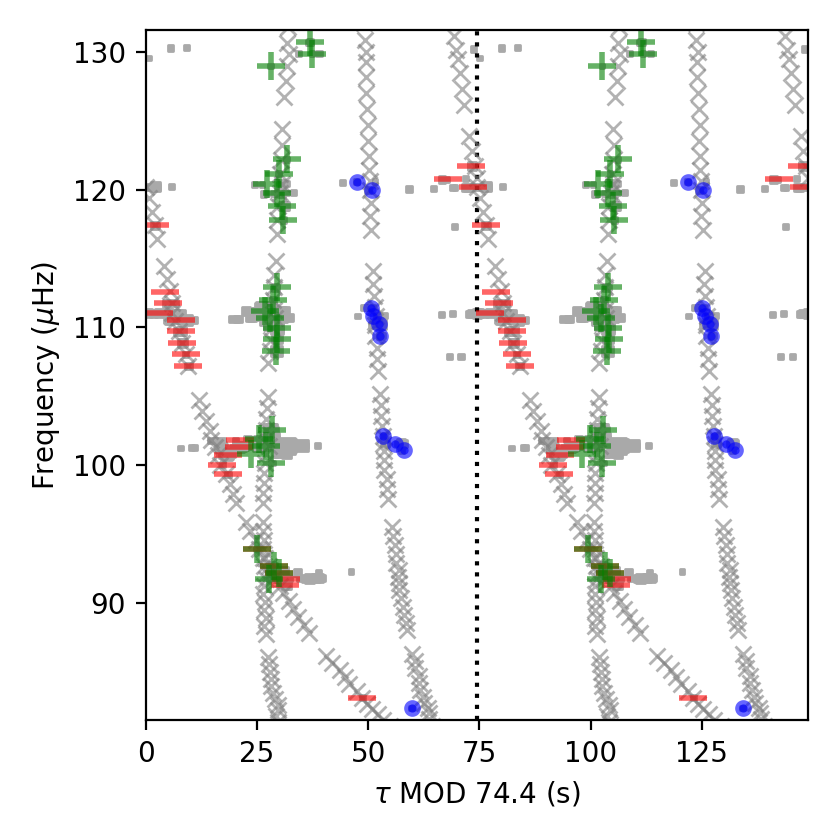}
\caption{Stretched \'{e}chelle diagram of KIC\,9589420. Symbols are explained in Fig.~\ref{fig:stretched_KIC5792889}. }
\label{apdxfig:echelle_diagram_stretched_9589420}
\end{figure*}

\begin{figure*}
\centering
\includegraphics[width = 0.8\linewidth]{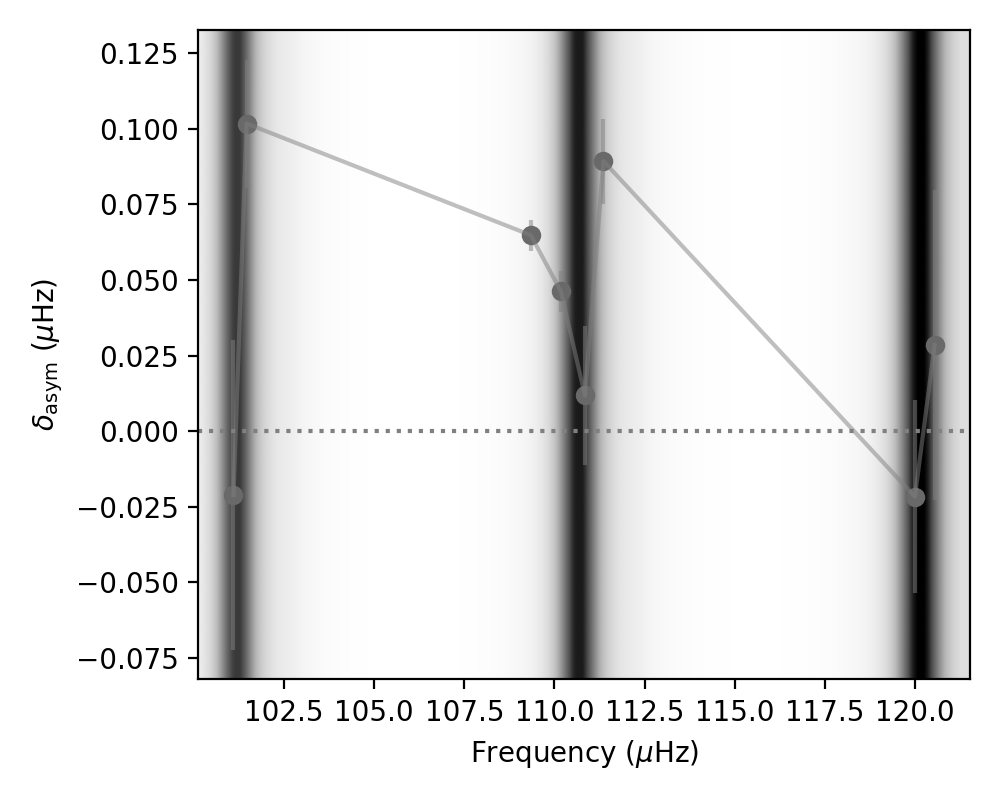}
\caption{Splitting asymmetries as a function of frequency of KIC\,9589420. Symbols are explained in Fig.~\ref{fig:KIC5696081_asymmtry_vs_freq}.}
\label{apdxfig:asymmetry_9589420}
\end{figure*}

\begin{figure*}
\centering
\includegraphics[width = 0.8\linewidth]{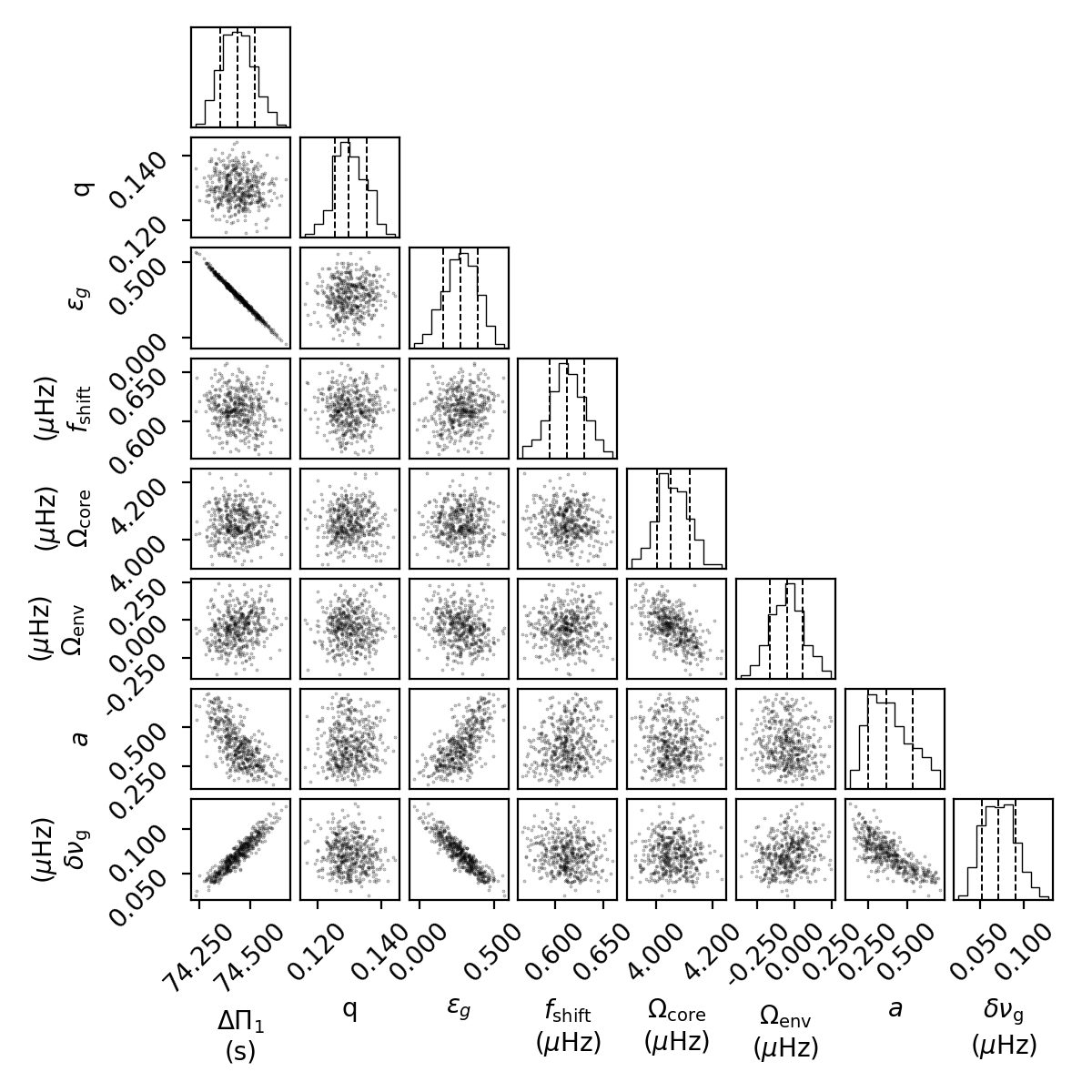}
\caption{Corner diagram of the MCMC fitting result of KIC\,9589420.}
\label{apdxfig:corner_9589420}
\end{figure*}

\begin{figure*}
\centering
\includegraphics[width = 0.8\linewidth]{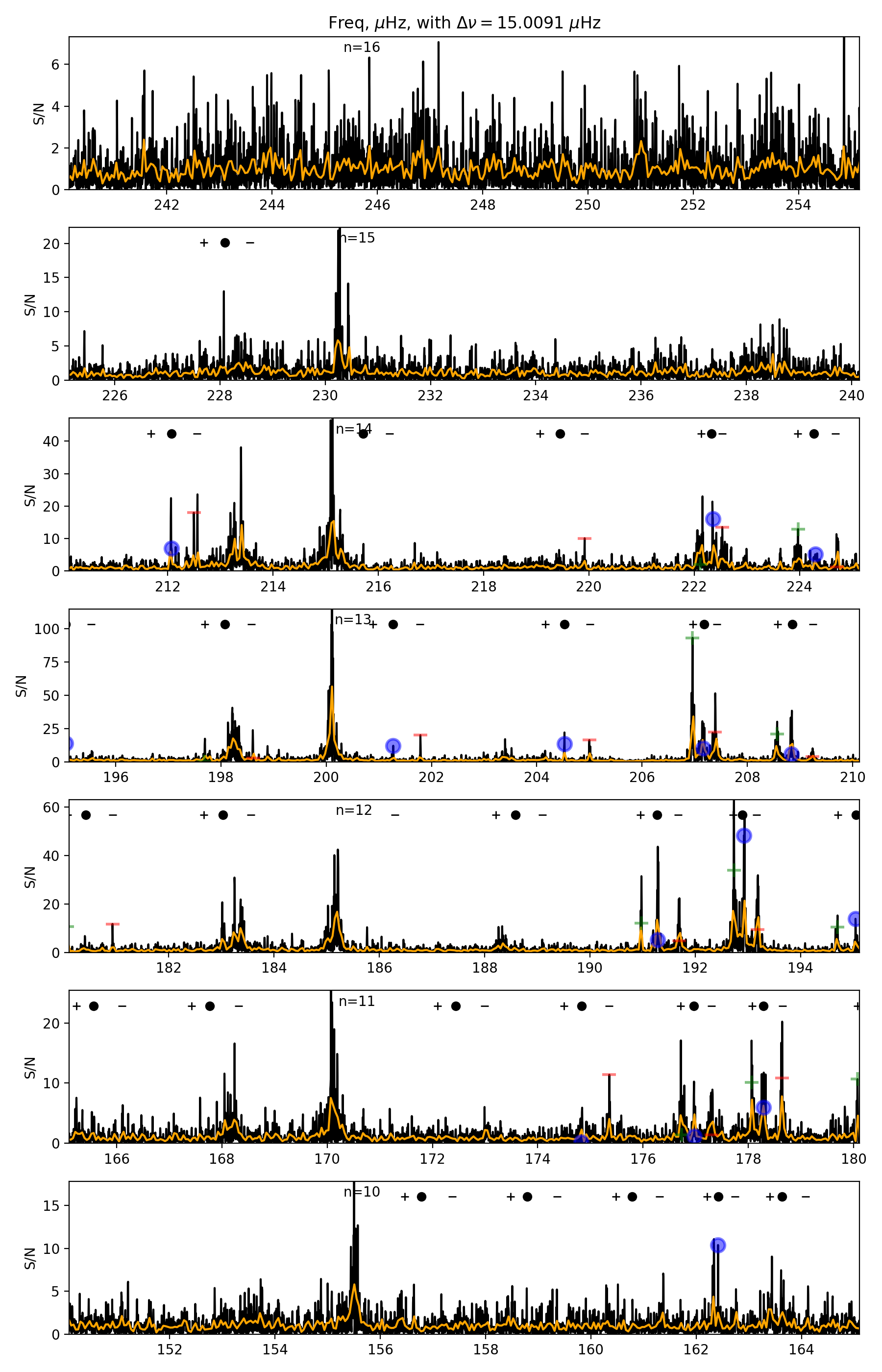}
\caption{Normal \'{e}chelle diagram of KIC\,10801792. See Fig.~\ref{apdxfig:echelle_diagram_4458118} for the explanations of the symbols.}
\label{apdxfig:echelle_diagram_10801792}
\end{figure*}

\begin{figure*}
\centering
\includegraphics[width = 0.6\linewidth]{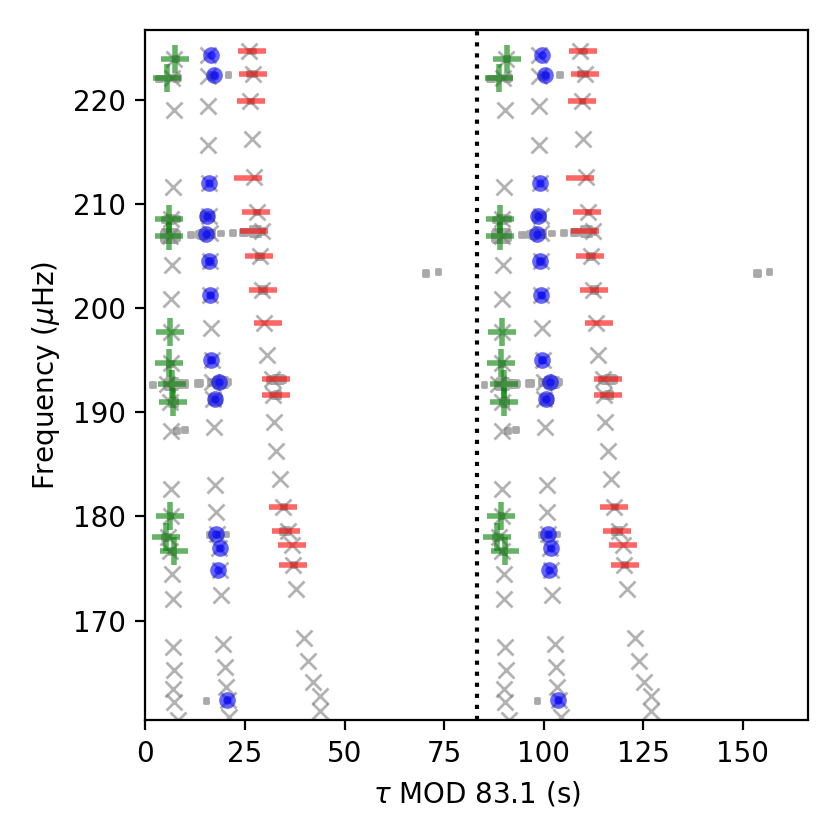}
\caption{Stretched \'{e}chelle diagram of KIC\,10801792. Symbols are explained in Fig.~\ref{fig:stretched_KIC5792889}. }
\label{apdxfig:echelle_diagram_stretched_10801792}
\end{figure*}

\begin{figure*}
\centering
\includegraphics[width = 0.8\linewidth]{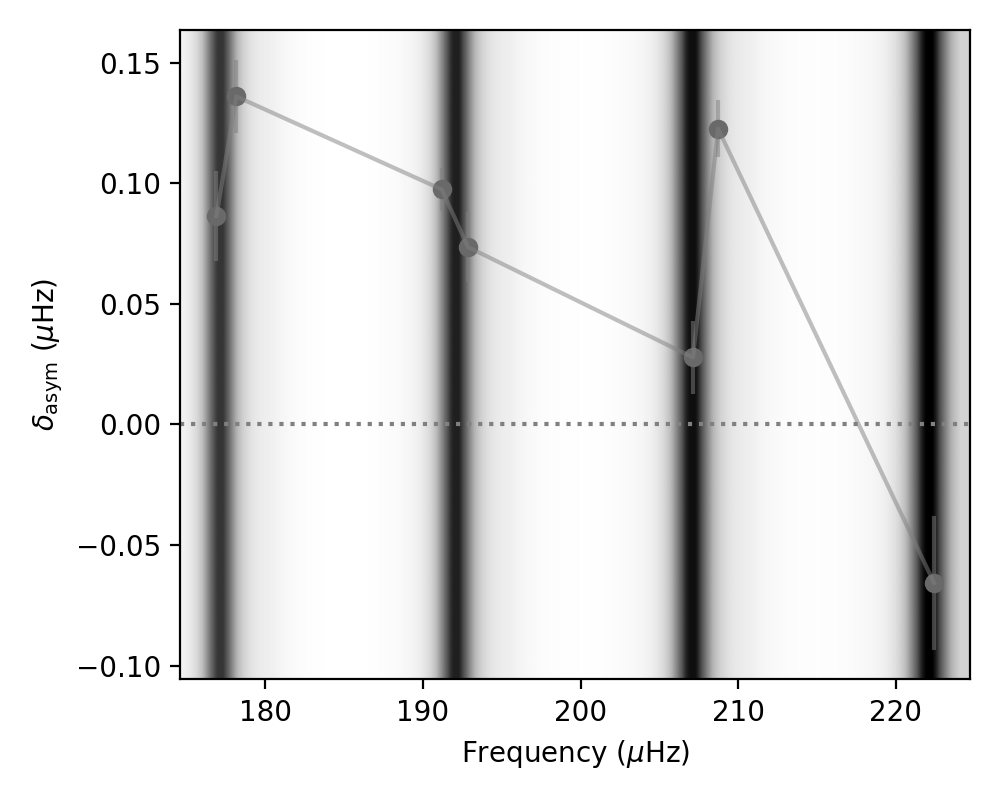}
\caption{Splitting asymmetries as a function of frequency of KIC\,10801792. Symbols are explained in Fig.~\ref{fig:KIC5696081_asymmtry_vs_freq}.}
\label{apdxfig:asymmetry_10801792}
\end{figure*}

\begin{figure*}
\centering
\includegraphics[width = 0.8\linewidth]{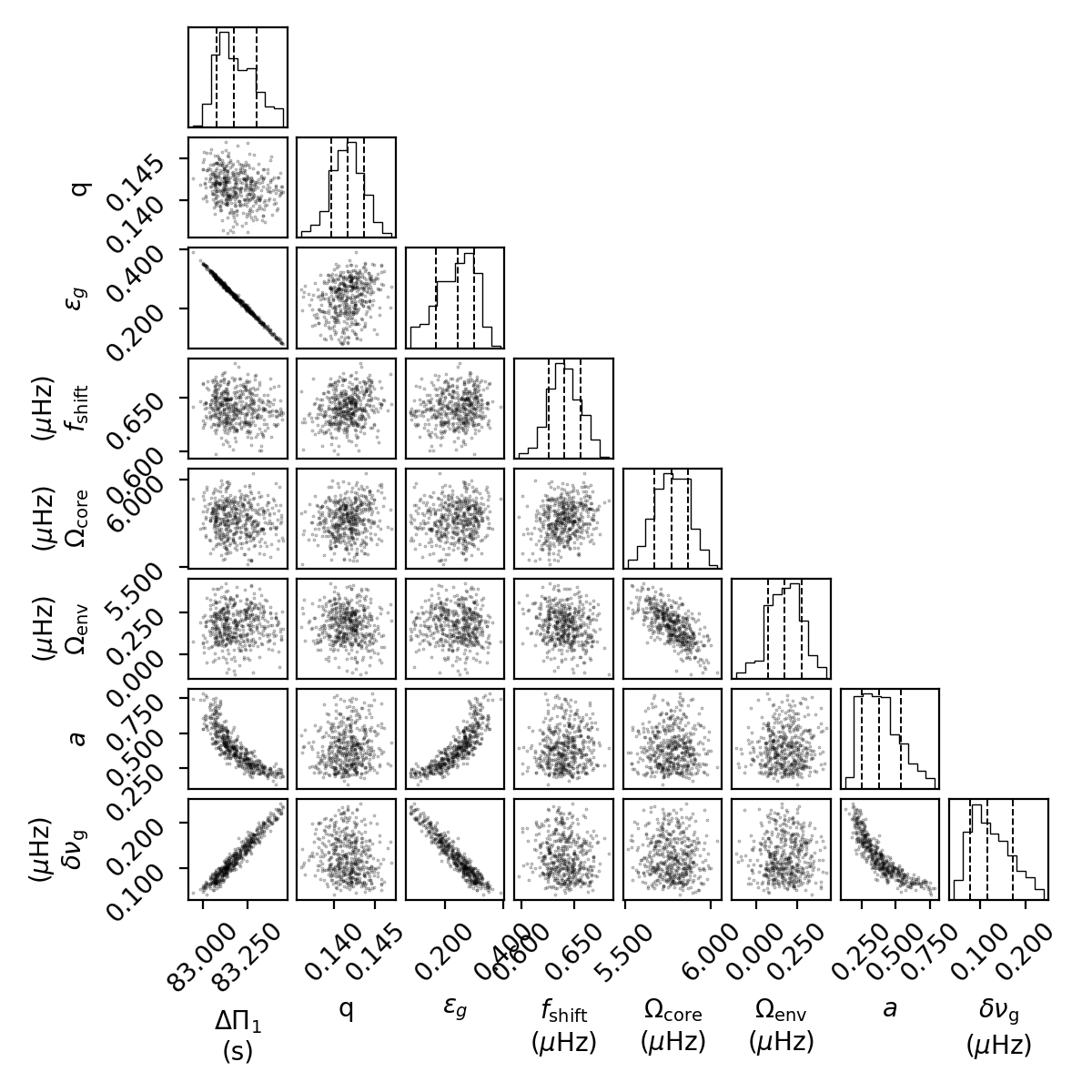}
\caption{Corner diagram of the MCMC fitting result of KIC\,10801792.}
\label{apdxfig:corner_10801792}
\end{figure*}

\begin{figure*}
\centering
\includegraphics[width = 0.8\linewidth]{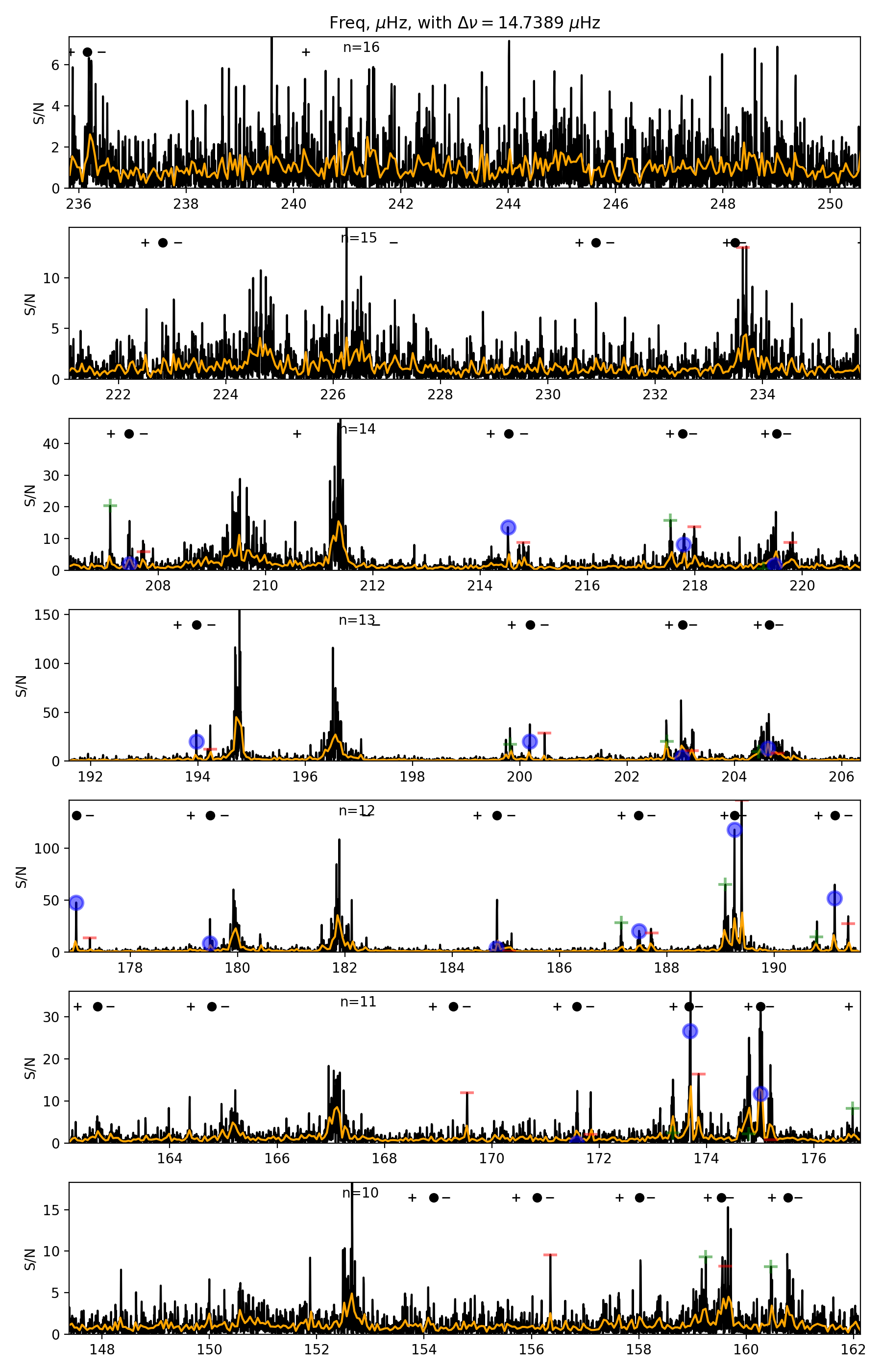}
\caption{Normal \'{e}chelle diagram of KIC\,11515377. See Fig.~\ref{apdxfig:echelle_diagram_4458118} for the explanations of the symbols.}
\label{apdxfig:echelle_diagram_11515377}
\end{figure*}

\begin{figure*}
\centering
\includegraphics[width = 0.6\linewidth]{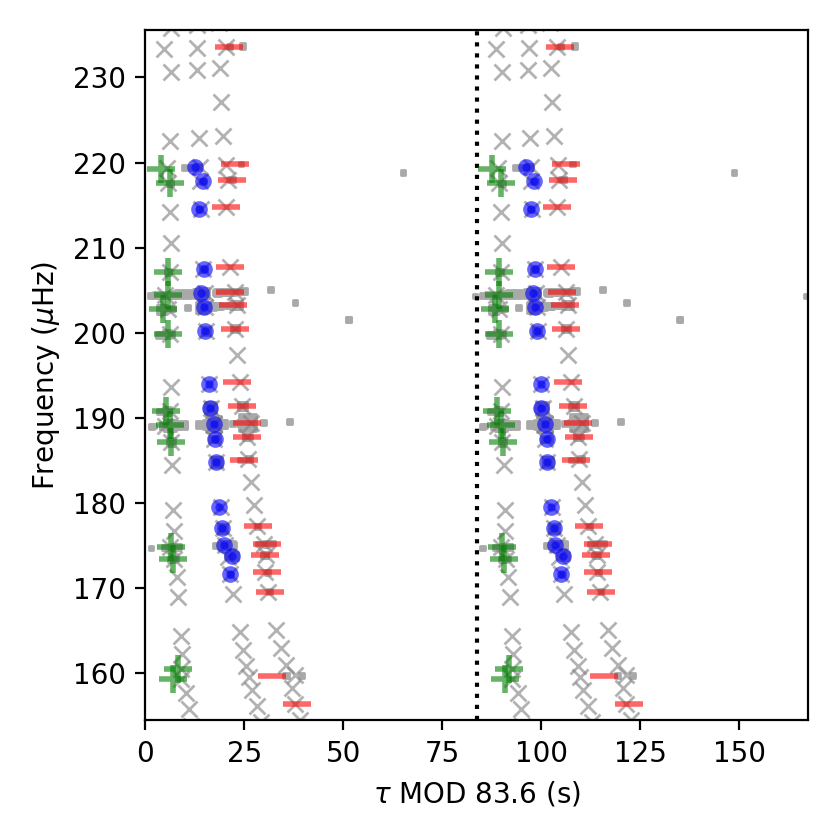}
\caption{Stretched \'{e}chelle diagram of KIC\,11515377. Symbols are explained in Fig.~\ref{fig:stretched_KIC5792889}. }
\label{apdxfig:echelle_diagram_stretched_11515377}
\end{figure*}

\begin{figure*}
\centering
\includegraphics[width = 0.8\linewidth]{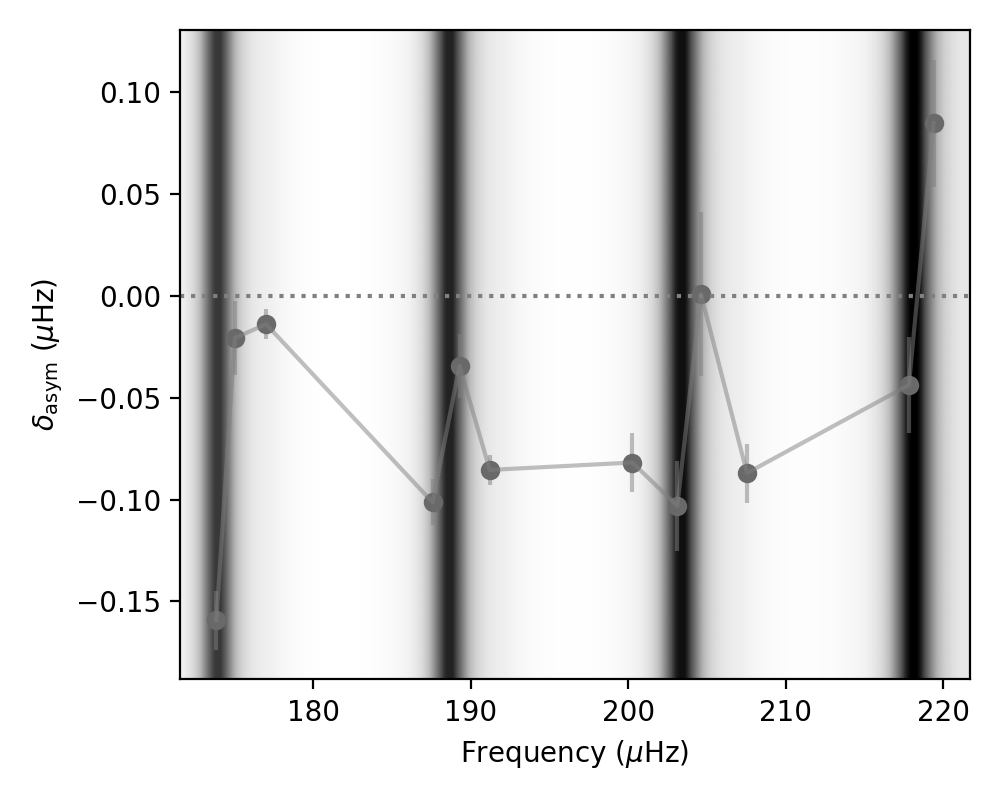}
\caption{Splitting asymmetries as a function of frequency of KIC\,11515377. Symbols are explained in Fig.~\ref{fig:KIC5696081_asymmtry_vs_freq}.}
\label{apdxfig:asymmetry_11515377}
\end{figure*}

\begin{figure*}
\centering
\includegraphics[width = 0.8\linewidth]{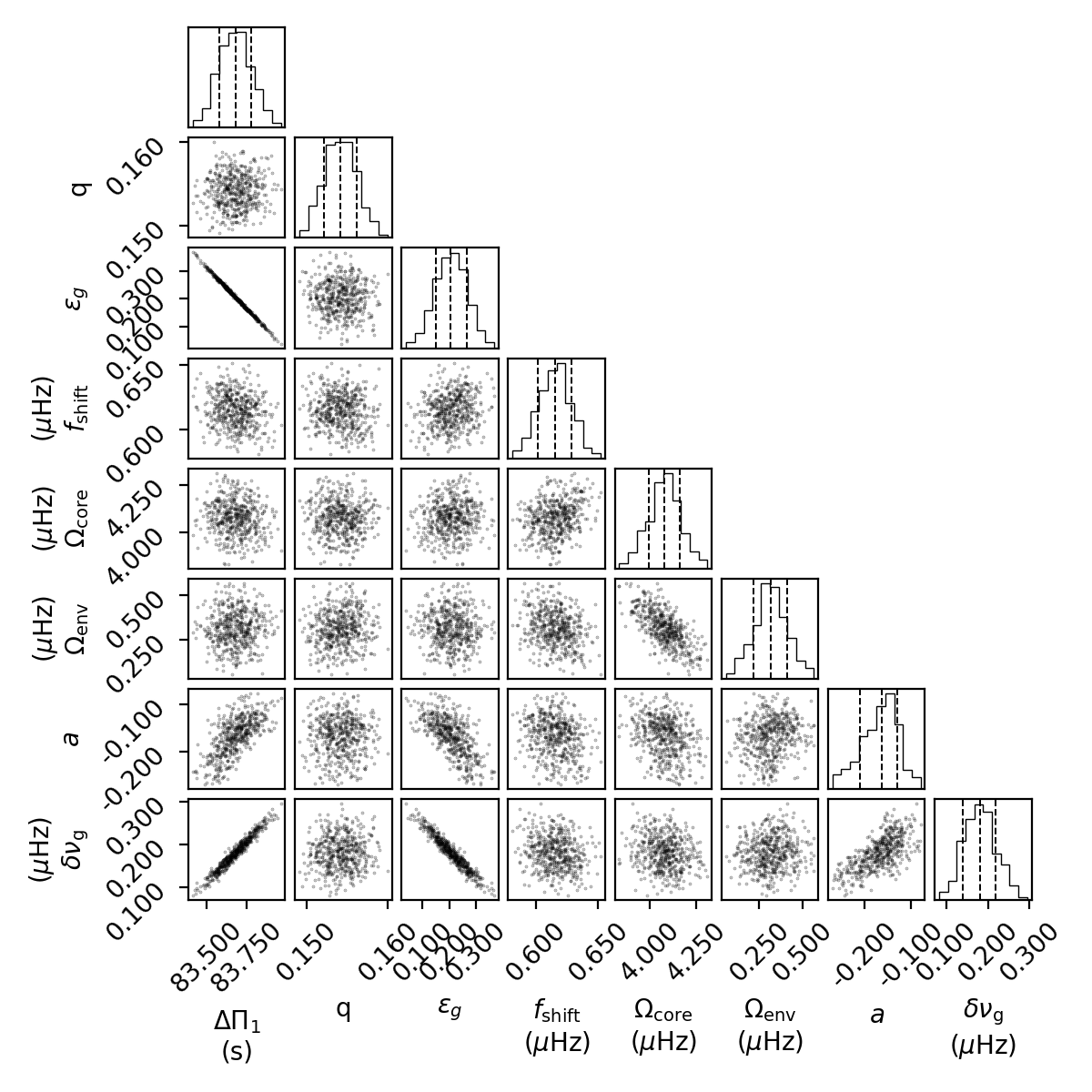}
\caption{Corner diagram of the MCMC fitting result of KIC\,11515377.}
\label{apdxfig:corner_11515377}
\end{figure*}

%\fi

\section{Other parameters} \label{appendixsec:other_paramters}
\subsection{$\Delta \nu$ vs $\Delta \Pi_1$}
\begin{figure}
    \centering
    \includegraphics[width=\linewidth]{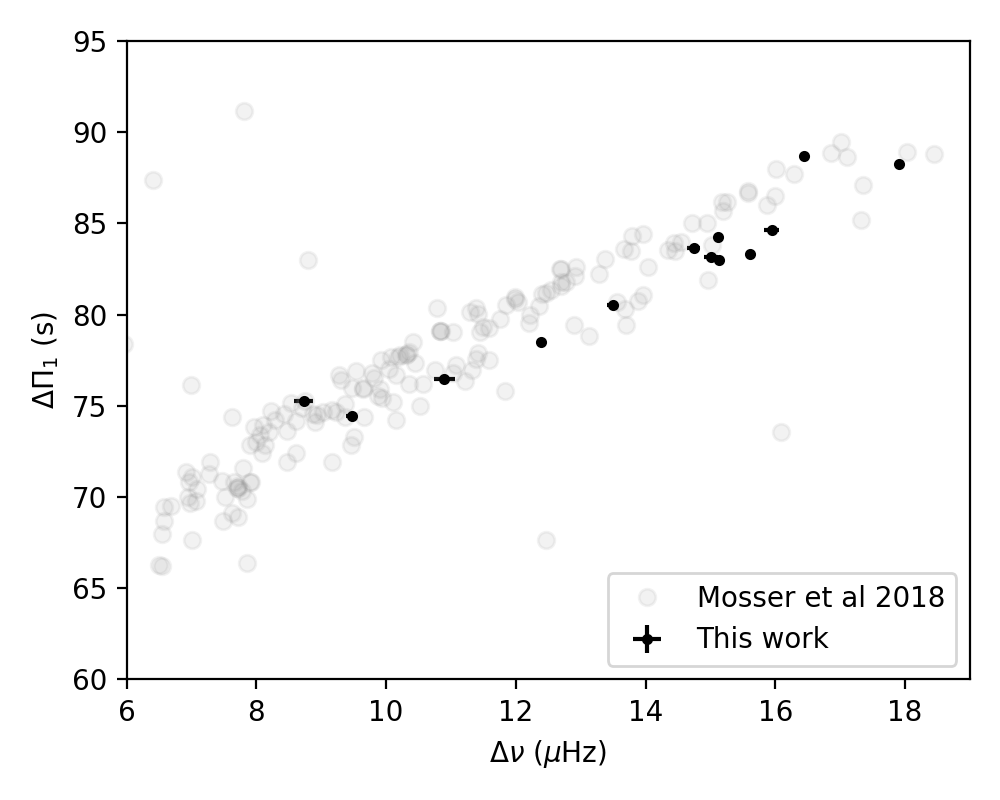}
    \caption{The correlation between $\Delta \nu$ and $\Delta\Pi_1$. The grey circles are given by \cite{mosser18}. }
    \label{fig:delta_nu_vs_delta_pi}
\end{figure}

Figure~\ref{fig:delta_nu_vs_delta_pi} displays the correlation between the measured period spacing $\Delta \Pi_1$ and the frequency separation $\Delta \nu$. The frequency separations are distributed between $\sim\!\!8$ to $\sim\!\!18\,\mathrm{\mu Hz}$ and the period spacings range from $\sim\!\!75$ to $\sim\!\!90\,\mathrm{s}$. All the stars are located at the well-defined $\Delta \Pi_1$--$\Delta \nu$ degenerate ridge, as shown by the grey dots by \cite{mosser18}, which implies that these stars did not undergo any mass transfer or binary merger \citep[e.g.][]{Deheuvels2022A&A, Rui2021MNRAS}. We find that our stars tend to locate close to the lower boundary of the $\Delta \Pi_1$--$\Delta \nu$ ridge, implying that they have higher masses \citep{Deheuvels2022A&A}. 

\subsection{$q$, $\varepsilon_\mathrm{g}$, and $f_\mathrm{shift}$} 

\begin{figure}
    \centering
    \includegraphics[width=\linewidth]{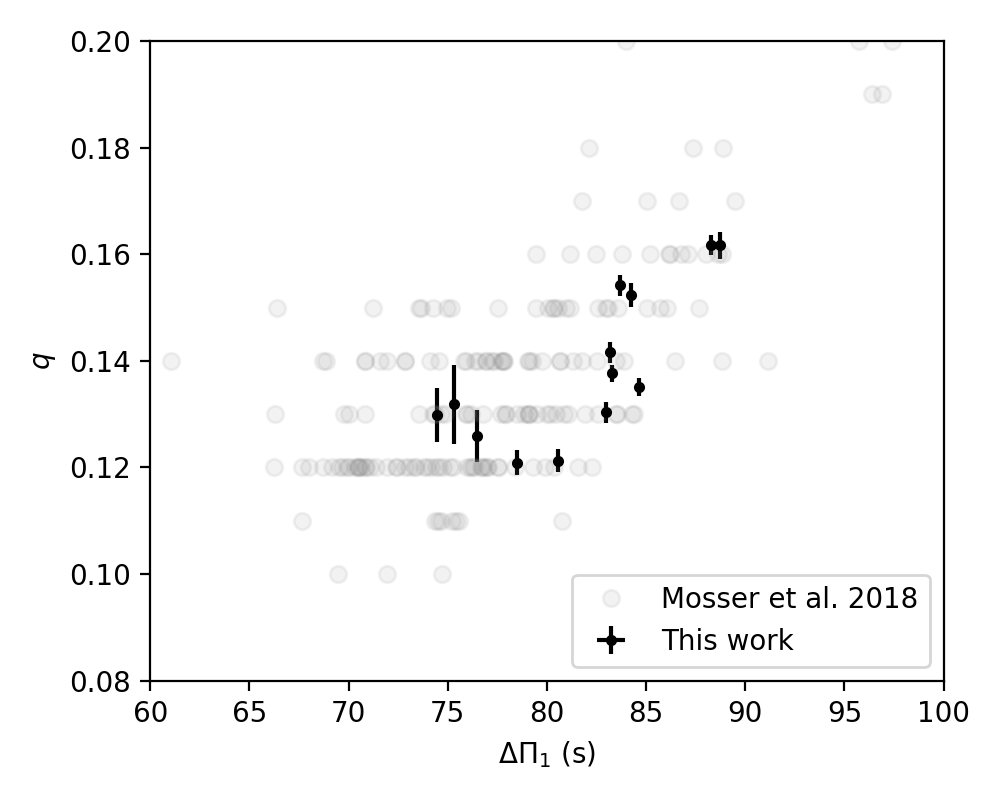}
    \caption{The coupling factor $q$ vs $\Delta\Pi_1$. The grey circles are reported by \cite{mosser18}.}
    \label{fig:coupling_factor}
\end{figure}

We show the values of the coupling factor $q$ in Fig.~\ref{fig:coupling_factor}. The $q$ values of our stars are in the typical range of red giant stars, and we also reproduce the increasing trend between $q$ and $\Delta \Pi_1$ \citep[e.g.][]{mosser18}.

\begin{figure}
    \centering
    \includegraphics[width=\linewidth]{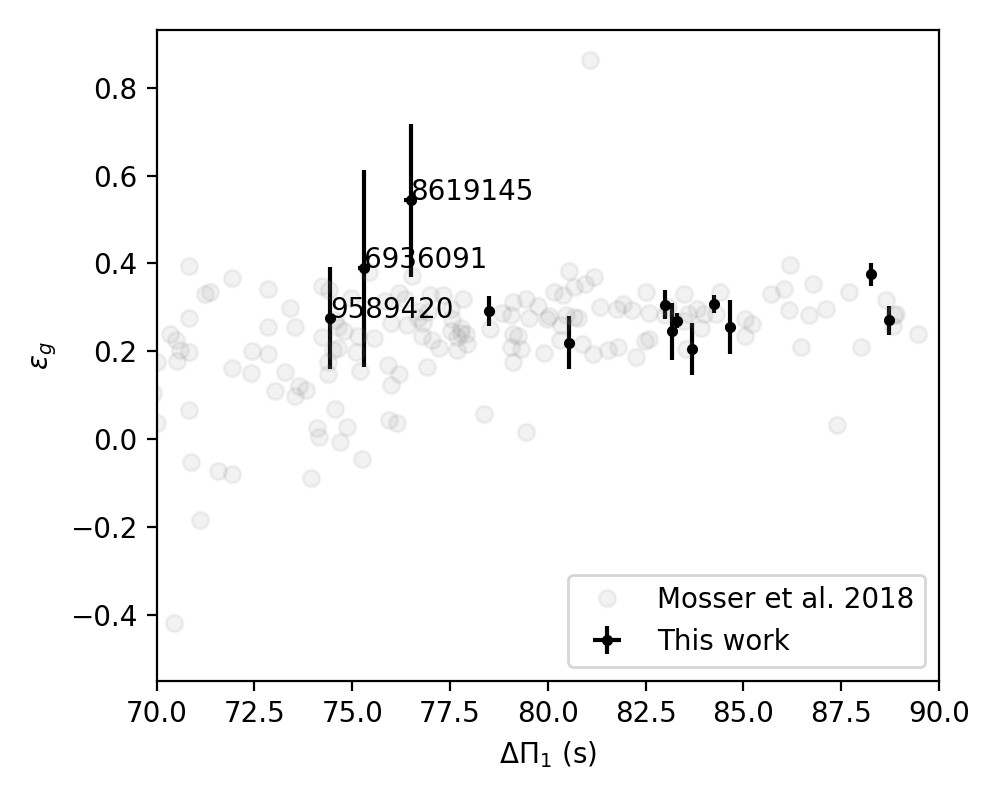}
    \caption{The g-mode phase $\varepsilon_\mathrm{g}$ as a function of $\Delta \Pi_1$. }
    \label{fig:g_mode_epsilon_g}
\end{figure}

The g-mode phases $\varepsilon_\mathrm{g}$ are displayed in Fig.~\ref{fig:g_mode_epsilon_g}. We only applied a uniform prior for $\varepsilon_\mathrm{g}$, but we still find that most of the stars have the typical $\varepsilon_\mathrm{g}$ values of red giant stars, which is $0.28\pm0.08$ reported by both observations \citep{mosser18} and theoretical predictions that do not include magnetic perturbations \citep{takata16}. Therefore, the magnetism-induced perturbation does not affect the distribution of g-mode phase. KIC\,8619145 has a large $\varepsilon_\mathrm{g}$ with large uncertainty, and we cannot provide strong constraint on $\varepsilon_\mathrm{g}$ of KIC\,6936091 either.

\begin{figure}
    \centering
    \includegraphics[width=\linewidth]{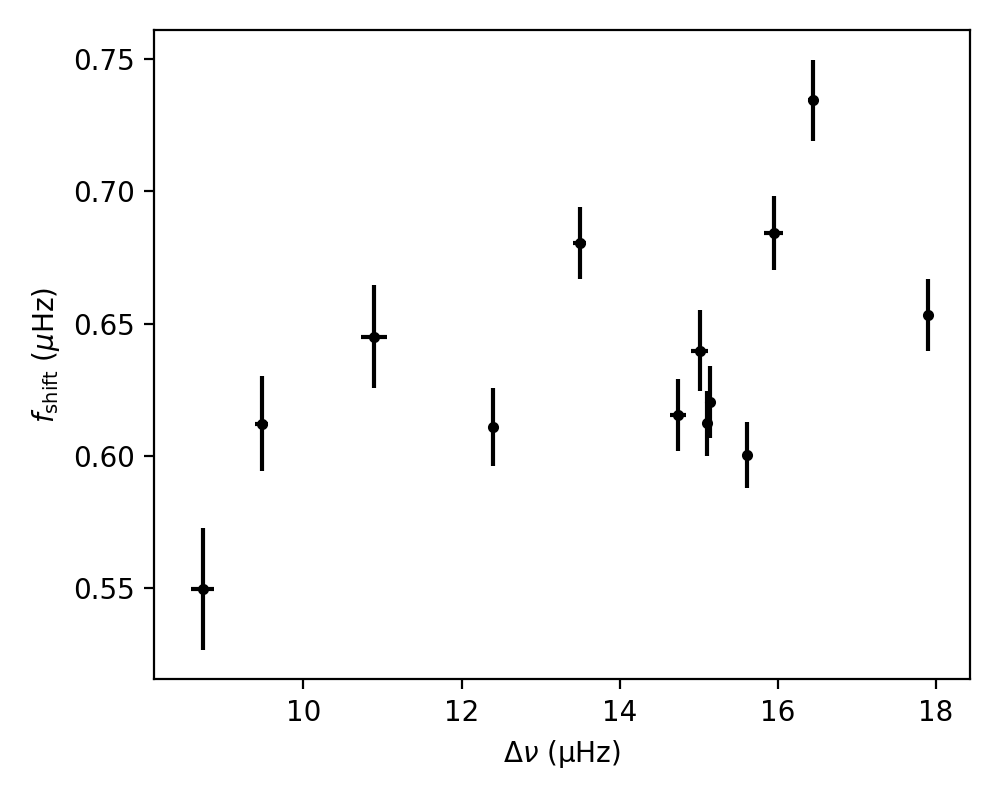}
    \caption{$f_\mathrm{shift}$ as a function of $\Delta \nu$. }
    \label{fig:f_shift}
\end{figure}

The frequency correction of $l=1$ pure p modes $f_\mathrm{shift}$ in Eq.~\ref{eq:p_mode_asymptotic_relation} is a new parameter that has not been studied in previous work. We plot $f_\mathrm{shift}$ as function of $\Delta \nu$ in Fig.~\ref{fig:f_shift}. The best values of $f_\mathrm{shift}$ lie between $\sim0.5$ to $\sim0.75\,\mathrm{\mu Hz}$. It seems that there is a rapid drop when $\Delta \nu \lesssim 10\,\mathrm{\mu Hz}$, but this feature needs more stars to confirm.

\section{Linear relation between $\mathcal{I}$ and $\Delta \Pi_1$}\label{appendixsec:relation_between_I_and_Delta_Pi_1}
\begin{figure}
    \centering
    \includegraphics[width=\linewidth]{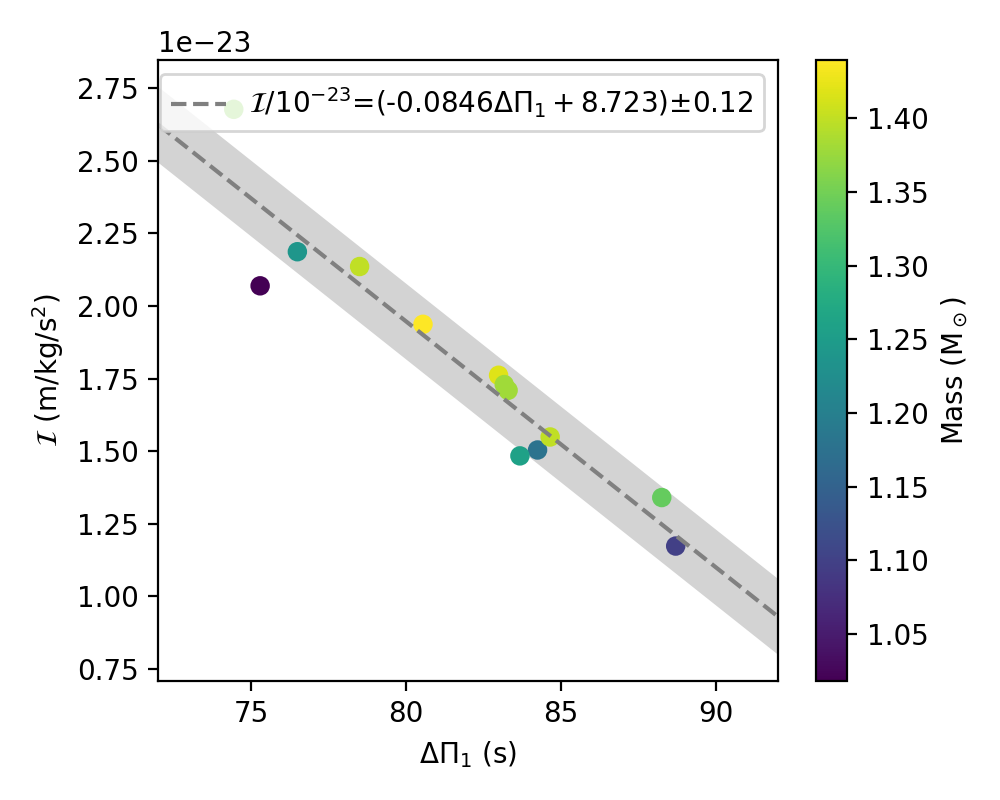}
    \caption{The core factor $\mathcal{I}$ as a function of $\Delta\Pi_1$, colour coded by the model-inferred masses. The grey dashed line shows the linear fit and the shaded area gives the $1\sigma$ range. }
    \label{fig:delta_pi_vs_factor}
\end{figure}

Using the best-fitting models, we calculated the core factors $\mathcal{I}$ and derived the field strengths, listed in table~\ref{tab:strength_table}. The core factor $\mathcal{I}$ is vital to calculate the field strength in Eq.~\ref{eq:square_field_strength}, but it is time-consuming to run a stellar evolution code to derive it. Hence, we report an empirical relation between $\mathcal{I}$ and the period spacing $\Delta \Pi_1$. As shown in Fig.~\ref{fig:delta_pi_vs_factor}, there is a linear relation between $\mathcal{I}$ and $\Delta \Pi_1$, which is characterised as
\begin{equation}
    \frac{\mathcal{I}\left(\mathrm{\frac{m}{s^2kg}}\right)}{10^{-23}}=\left(-0.0864\Delta\Pi_1+8.914\right)\pm0.11. \label{eq:linear_relation_of_core_factor}
\end{equation}

\end{appendix}

\end{document}